\documentclass[aps,prd,twocolumn,nofootinbib,showpacs,superscriptaddress]{revtex4-1}

\usepackage[T1]{fontenc} % if needed

\usepackage{amsmath}
\usepackage{amssymb}
\usepackage{graphicx}
\usepackage{bm}
\usepackage{epsfig}
\usepackage{amsfonts}
\usepackage{color}
\usepackage{mathtools}
\usepackage[section]{placeins}

\newcommand{\beq}{\begin{equation}}
\newcommand{\eeq}{\end{equation}}
\newcommand{\bea}{\begin{eqnarray}}
\newcommand{\eea}{\end{eqnarray}}
\newcommand{\beas}{\begin{eqnarray*}}
\newcommand{\eeas}{\end{eqnarray*}}
\newcommand{\chibar}{\ensuremath{\overline\chi}}

\newcommand{\re}{\operatorname{\mathfrak{Re}}}
\newcommand{\tr}{\operatorname{Tr}}

\begin{document}

\title {Finite temperature gluon spectral functions from $N_f=2+1+1$ lattice QCD}

\author{Ernst-Michael Ilgenfritz}
\affiliation{Bogoliubov Laboratory of Theoretical Physics, Joint Institute for Nuclear Research,
Joliot--Curie Str. 6, 141980 Dubna, Russia}
\author{Jan M.~Pawlowski}
\affiliation{Institute for Theoretical Physics, Universit\"at
  Heidelberg, Philosophenweg 12, D-69120 Germany}
\affiliation{ExtreMe Matter Institute EMMI, GSI, Planckstr. 1,
  D-64291 Darmstadt, Germany}
\author{Alexander Rothkopf}
\affiliation{Institute for Theoretical Physics, Universit\"at
  Heidelberg, Philosophenweg 12, D-69120 Germany}
\author{Anton Trunin}
\affiliation{Bogoliubov Laboratory of Theoretical Physics, Joint Institute for Nuclear Research,
Joliot--Curie Str. 6, 141980 Dubna, Russia}

\date{\today}

\begin{abstract}
  We investigate gluon spectral functions at finite temperature in
  Landau gauge, based on a subset of lattice QCD ensembles with
  $N_f=2+1+1$ dynamical twisted mass quarks flavors, generated by the
  tmfT collaboration. Our study uses a novel Bayesian approach for the
  extraction of non-positive definite spectral functions, which for
  each binned spatial momentum takes into account the gluon
  correlation functions at all available discrete imaginary
  frequencies. The spectral functions are extracted at three different
  lattice spacing, where for each of them, a scan of temperatures
  around the crossover transition is carried out at fixed scale. We
  find indications for the existence of a well defined quasi-particle
  peak. Due to a relatively small number of imaginary frequencies
  available, we focus on the momentum and temperature dependence of
  the position of this spectral feature. This dispersion relation
  reveals different in-medium masses for longitudinal and transversal
  gluons at high temperatures, qualitatively consistent with weak
  coupling expectations.
\end{abstract}

\maketitle

\flushbottom

%%%%%%%%%%%%%%%%%%%%%%%%%%%%%%%%%%%%%%%%%%%%%%%%%%%

%ToDo:
%
%  Benoetigen FRG Refernzen in der Einleitung 

%%%%%%%%%%%%%%%%%%%%%%%%%%%%%%%%%%%%%%%%%%%%%%%%%%%

%\FloatBarrier
\section{Introduction}
\label{sec:intro}

Understanding the evolution of strongly interacting matter in a
heavy-ion collision is one of the most demanding tasks in current
theoretical physics \cite{Muller:2013dea}. Not only do we need to
describe the real-time dynamics of matter in the high temperature
phase of QCD, the quark-gluon plasma, but also its transition to the
low temperature domain of hadrons seen and measured in experiment. In
particular around the chiral crossover transition, estimated on the
lattice \cite{Borsanyi:2010bp,Bazavov:2011nk} to occur at
$T_c=155\pm9$MeV, it is vital to uncover how the breaking of chiral
symmetry and the onset of confinement proceeds and changes the
relevant degrees of freedom in the system.

Transport and thermal properties of strongly interacting matter have been extracted from the correlation functions
of mesons, i.e. hadronic observables. For computations based on lattice simulations see e.g. \cite{Aarts:2014nba,Mages:2015rea,Ding:2016hua,Ding:2012sp,Aarts:2014cda,Kim:2014iga,Borsanyi:2014vka,Burnier:2015tda}. On the other hand such real-time properties can also be accessed via the spectral functions of
the fundamental constituents of QCD, gluons and quarks, see e.g. \cite{Rothkopf:2016luz,Christiansen:2014ypa,Dudal:2013yva,Haas:2013hpa,Qin:2013ufa,Strauss:2012dg,Mueller:2010ah,Karsch:2007wc}.

In the present study we focus on the gluonic sector. The study of gluon correlation functions in gauge fixed QCD has
garnered interest for quite some time, both with lattice simulations and with functional approaches, for finite temperature results see e.g. \cite{Mendes:2014gva,Aouane:2012bk,Cucchieri:2012nx,Aouane:2011fv,Maas:2011ez,Maas:2011se}
and \cite{Reinosa:2016iml,Fischer:2014ata,Fischer:2013eca,Fister:2013bh,Fister:2011uw,Fischer:2011mz,Fischer:2010fx,Braun:2009gm} respectively. 
The extraction of the corresponding
spectral functions is however hampered by the fact that gauge fixed
gluon spectra contain non positive-definite contributions, see e.g.\ 
\cite{Alkofer:2000wg}. This in turn defies standard approaches, based
on Bayesian inference, such as the Maximum Entropy Method (MEM)
\cite{Asakawa:2000tr}, to which then direct extensions
\cite{Hobson:1998bz,Qin:2013ufa}, modifications of the prior
\cite{Dudal:2013yva}, as well as of the data part, such as the
introduction of shift functions \cite{Haas:2013hpa}, have been applied
in the literature. Over the last few years progress has been made in
developing new Bayesian approaches independent from the MEM
\cite{Burnier:2013nla}, which recently have also been generalized to
non-positive definite spectra \cite{Rothkopf:2016luz}. Some
non-Bayesian approaches, such as the Backus-Gilbert method \cite{Brandt:2015sxa}
or the Sumudu transformation \cite{Pederiva:2014qea} also allow the treatment of spectra
with negative contributions.

Investigating the spectral properties of gluons serves several
complementary purposes, first and foremost it provides a direct and
intuitive handle on phenomena, such as the generation of a mass gap in
the context of confinement
\cite{Alkofer:2000wg,Cornwall:2013zra}, as well as the
emergence of thermal masses related to Debye screening.

Secondly gluon spectral functions play a vital role in the
self-consistent computation of transport coefficients in functional
approaches to QCD \cite{Haas:2013hpa,Christiansen:2014ypa}, such as
the functional renormalization group or Dyson-Schwinger approaches. If
the gluon spectral function is known, it may serve as input to a
closed set of real-time evolution equations for quark- and gluon
degrees of freedom, from which relevant quantities, such as
energy-momentum correlation functions may be computed. Thus, in turn,
transport properties become accessible.

From a practical point of view we are also interested in using lattice
gluon spectral functions to validate phenomenological models used in
the description of heavy-ion collisions. Some of these rely on a
quasi-particle picture for the fundamental constituents of strongly
interacting matter, which at high temperature is matched to resummed
perturbative predictions from hard-thermal loops. One example in this
regard is the parton-hadron string dynamics model
\cite{Bratkovskaya:2011wp,Hartnack:2011cn,Berrehrah:2016vzw}. Elucidating
the non-perturbative behavior of the gluon spectral function may
therefore lead to more refined approximations and a better
understanding of the validity of currently used model assumptions.

Single particle properties in QCD are conceptually more difficult to
capture than those of e.g. mesons. Quarks and gluons represent color
charged fields and thus their correlation functions are not gauge
invariant. Hence gauge fixing becomes necessary and one has to
carefully understand which of the observed properties is truly
physical and which depends on the choice of gauge. Dismissing all
together the study of gauge dependent correlators, however, is a too
narrow point of view, as they may still contain gauge independent
information. One example is the extraction of the heavy-quark
potential from Wilson-line correlators in Coulomb gauge
\cite{Rothkopf:2009pk,Rothkopf:2011db,Burnier:2014ssa}. In that case
the gauge independent spectral feature encoding the potential is
embedded in a gauge dependent background, which may be cleanly
separated.

In this first study of gluon properties in thermal lattice QCD with
$N_f=2+1+1$ dynamical twisted mass flavors, we use the conventional
choice of Landau gauge $\partial^\mu A_\mu^a=0$, which is manifestly
Lorentz invariant and retains global $SU(N_c)$ gauge symmetry. In the
following the computations are carried out in the Euclidean domain,
where time has been Wick-rotated to $\tau=it$. The gluon correlator is
defined from the Fourier transformed gauge fields
$\widetilde{A}^a_{\mu}(q)$ \beq D^{ab}_{\mu\nu}(q) =\left\langle
  \widetilde{A}^a_{\mu}(q)\widetilde{A}^b_{\nu}(-q) \right\rangle.
\label{eq:gluonzero}
\eeq
In the presence of a thermal bath, Lorentz invariance and hence Euclidean rotational invariance is broken and one may split the correlator into a transversal $D_T$ (``chromomagnetic'') as well as a longitudinal $D_L$ (``chromoelectric'') component
\beq
D^{ab}_{\mu\nu}(q)=\delta^{ab} 
            \left (P^{T}_{\mu\nu} D_{T}(q_{4}^{2},\vec{q}^{\,2})+ P^{L}_{\mu\nu} D_{L}(q_{4}^{2},\vec{q}^{\,2}) \right ) \; .
\eeq
For the particular choice of Landau gauge the projectors $P^{T,L}_{\mu\nu}$ 
are aligned transversally or longitudinally with the imaginary frequency $(\mu=4)$ direction:
\bea
P^{T}_{\mu\nu}&=&(1-\delta_{\mu4})(1-\delta_{\nu4}) 
       \left(\delta_{\mu\nu}- \frac{q_{\mu}q_{\nu}}{\vec{q}^{\;2}}\right), \\
P^{L}_{\mu\nu}&=&\left(\delta_{\mu\nu}-\frac{q_{\mu}q_{\nu}}{\vec{q}^2}\right)
                -P^{T}_{\mu\nu}\;.
\eea
The explicit expressions for the propagators $D_{T,L}$ read
\beq
 D_T(q)=\frac{1}{2 N_g} 
        \left\langle \sum_{i=1}^3  
         \widetilde{A}^a_i(q) \widetilde{A}^a_i(-q)
        -\frac{q_4^2}{\vec{q}^{\;2}} 
        \widetilde{A}^a_4(q)\widetilde{A}^a_4(-q) \right\rangle\label{Eq:GPT}
\eeq
and
\beq
 D_L(q)= \frac{1}{N_g}\left(1 + \frac{q_4^2}{\vec{q}^{\;2}}\right) 
        \left\langle \widetilde{A}^a_4(q) \widetilde{A}^a_4(-q) \right\rangle\label{Eq:GPL}
\; ,
\eeq
with $N_g=N_c^2-1$ and $N_c=3$. This fixes also the dressing functions $Z_{T,L}(q)=q^2 D_{T,L}(q)$, which are
often considered in functional computations. Taking consistent limits as momenta are sent to zero leaves us with 
\bea
\label{eq:zeromomprop}
D_T(0) &=& \frac{1}{3 N_g}
     \sum_{i=1}^3 \left\langle \widetilde{A}^a_i(0) \widetilde{A}^a_i(0)
\right\rangle, 
\\
D_L(0) &=& \frac{1}{N_g}
       \left\langle \widetilde{A}^a_4(0) \widetilde{A}^a_4(0) \right\rangle \;. 
\eea

We may relate the gluon correlators in imaginary frequencies $q_4$ to their spectral function via the K\"allen-Lehmann representation
\begin{align}
\nonumber D_{T,L}(q_4,{\mathbf q})&=\int_{-\infty}^\infty \frac{1}{iq_4+\omega} \rho_{T,L}(\omega,{\mathbf q})\\ &= \int_0^\infty \frac{2\omega}{q_4^2+\omega^2} \rho_{T,L}(\omega,{\mathbf q}), \label{Eq:KaellenRel}
\end{align}
with the spectral function being antisymmetric around the real-time frequencies origin $\rho(-\omega)=-\rho(\omega)$. Inverting this relation using the simulated correlator data represents 
a well known ill-posed problem, which we attack via the use of Bayesian inference as laid out in detail in the next section.

\section{Numerical methods}
\label{sec:nummethods}

 \subsection{Lattice simulations and gauge fixing}

The present study works with lattices that feature four dynamical quark 
flavors, u,d,s and c. They represent a subset of configurations, which 
were generated by the tmfT collaboration originally for the study of QCD 
thermodynamics \cite{Burger:2013hia} in the presence of a heavy doublet of 
quarks (third and fourth quark species). The dynamical quarks are implemented using so called twisted mass actions, 
which take the following form for the light sector
\begin{align}
\nonumber S^l_f[U,\chi_l,\chibar_l] = \sum_{x,y} \chibar_l(x) 
 &[\delta_{x,y} -\kappa D_\mathrm{W}(x,y)[U] \\
 &+ 2 i \kappa a \mu_l \gamma_5 \delta_{x,y}  \tau_3 ]
 \chi_l(y) \;,
\label{tmaction}
\end{align}
and the heavy sector 
\begin{align}
  S^h_f[U,\chi_h, \overline{\chi}_h] =
  &&&\sum_{x,y} \overline{\chi}_h(x) [ \delta_{x,y} - \kappa D_W(x,y)[U] \label{eq_heavyaction}\\
 \nonumber &&&+ 2 i \kappa a \mu_{\sigma} \gamma_5 \delta_{x,y} \tau_1 + 2 \kappa a \mu_{\delta} \delta_{x,y} \tau_3 ] \chi_h(y) \;
\end{align}
respectively, where the $\tau_i$ are the Pauli matrices in doublet (flavor) 
space. The term $D_{W}[U]$ denotes the standard gradient term for Wilson 
fermions 
\begin{align}
\nonumber D_{W}[U] = \frac{1}{2 a} [ \gamma_\mu ( \nabla_\mu + \nabla^{*}_\mu )
- \nabla^{*}_\mu \nabla_\mu ] \;
 \label{Eq:DiracWilson}
\end{align}  
and $\kappa_l=(2 a m_{0,l} + 8 r)^{-1}$ represents the usual hopping term 
with $r=1$.

The gauge field degrees of freedom are governed -- apart from the fermion backreaction -- by an improved 
Iwasaki action
\begin{align}
\nonumber S_g[U] =\beta  \Big{(} &c_0 \sum_{P} \lbrack 1 - \frac{1}{3} \re \tr \left ( U_{P} \right ) \rbrack  \\
 + &c_1 \sum_{R} \lbrack  1 - \frac{1}{3} \re \tr  \left ( U_{R} \right ) \rbrack \Big{)} \;.
\end{align}
with ($c_0 = 3.648$ and $c_1 = -0.331$), where the sum ($P$) contains all 
plaquettes and the sum  ($R$) all planar rectangles.
 
The light doublet $\chi_l = (\chi_u,\chi_d)$ in the twisted basis is related
by a chiral rotation to the doublet $\psi^{phys} = (\psi_u,\psi_d)$ in the 
physical basis
\begin{align}
\nonumber \psi^{phys}_l = e^{i \omega_l \gamma_5 \tau_3/2} \chi_l \qquad
\overline{\psi}^{phys}_l = \overline{\chi}_l e^{i \omega_l \gamma_5 \tau_3/2} 
 \label{Eq:light-Rotation}
\end{align}
with the twisting angle $\omega_l$. Twisted mass light fermions are taken
at maximal twist, if the bare untwisted mass $m_{0,l}$ is tuned to its 
critical value $m_{\rm crit}$. When $|m_{0,l} - m_{\rm crit}| \to 0$,
the twisting angle $\omega_l \to \frac{\pi}{2}$ (maximal twist). This fixes 
the twisted basis.

A similar rotation 
\begin{align}
\nonumber \psi^{phys}_h = e^{i \omega_h \gamma_5 \tau_3/2} \chi_h \qquad
\overline{\psi}^{phys}_h = \overline{\chi}_h e^{i \omega_h \gamma_5 \tau_3/2} 
\end{align}
relates the two bases in the heavy sector. $\kappa_h=(2 a m_{0,h} + 8 r)^{-1}$ 
(with $r=1$) is the hopping parameter for heavy quarks.
Again, $\omega_h \to \frac{\pi}{2}$ if $|m_{0,h} - m_{\rm crit}| \to 0$.

An economic procedure dealing with the $N_f=2+1+1$ case consists in the 
choice $a m_{0,l} = a m_{0,h} =\frac{1}{2 \kappa} -4 $ with a common 
hopping parameter. Tuning to maximal twist means tuning 
$\kappa = \kappa_{\rm crit}(\beta)$. The critical $\kappa$ corresponds
to the vanishing of the PCAC light quark mass $m_{\rm PCAC}$ and is
determined as function of $\beta$ at zero temperature \cite{Baron:2010bv}.

The bare light-quark ($\mu_l$) twisted-mass parameter (in the first doublet) 
and the two bare heavy-quark twisted-mass parameters $\mu_\sigma$ and 
$\mu_\delta$ (in the second doublet) also need to be tuned (as functions of 
$\beta$) at zero temperature to stay on a line of constant physics, defined 
by the ``pion mass'' and by matching masses of hadrons containing strange 
and charm quarks. For light hadrons this has been performed for the first
time for $\beta=1.90$ and $\beta=1.95$ in Ref. \cite{Baron:2010bv}.

The bare twisted-mass parameters $\mu_{\sigma}$ and $\mu_{\delta}$ are
related to the renormalized strange and charm quark masses
\begin{align}
\nonumber(m_s)_R =Z_P^{-1} ( \mu_{\sigma} - \frac{Z_P}{Z_S} \mu_{\delta} )
(m_c)_R =Z_P^{-1} ( \mu_{\sigma} + \frac{Z_P}{Z_S} \mu_{\delta} )
\end{align}
with the renormalization constants $Z_P$ and $Z_S$ of the pseudoscalar
and scalar quark densities.

For a more detailed description of the simulation setup see 
Ref. \cite{Baron:2010bv,Baron:2011sf}.

The tmfT collaboration has adopted three parameter sets for their finite 
temperature studies from the zero-temperature ensembles used by the ETMC 
collaboration (under the names A60.24, B55.32 and D45.32 defined in 
Ref. ~\cite{Baron:2011sf}). In Ref. \cite{Alexandrou:2014sha}
these ensembles have been calibrated with the help of the baryon spectrum,
and we adopt these results for the lattice spacing.
The set of $\beta$ values is fixed (according to the fixed-scale approach) 
and has been extended to include $\beta=1.90$ (A), $\beta=1.95$ (B) and 
$\beta=2.10$ (D). For example, the $T=0$ nomenclature ``A60.24'' indicates, 
besides the $\beta$ value, a lattice size $24^3\times 48$ for zero temperature
and a light twisted-mass parameter $a \mu_l = 0.0060$. 
The corresponding physical lattice spacings and pion masses $m_{\pi^{\pm}}$, 
together with the resulting deconfinement crossover temperatures are listed 
in Tab.~\ref{tab:tc}. The tmfT nomenclature refers to (apart from the $\beta$ 
value) to the approximate pion mass only. The temperature is varied by changing
$N_\tau$.

\begin{table} [h]
\centering      
\begin{tabular}{r@{\quad}c@{\quad}c@{\quad}c}
\hline\hline       
ETMC ens. ($T=0$)    & A60.24      & B55.32      & D45.32 \\       
\hline
tmfT ens. ($T\ne0$)  & A370        & B370        & D370   \\     
\hline
$\beta$                   & 1.90        & 1.95        & 2.10   \\  
\hline       
$a \, [\text{fm}]$        & $0.0936$    & $0.0823$    & $0.0646$  \\       
$m_\pi\, [\text{MeV}]$    & $364(15)$   & $372(17)$   & $369(15)$ \\       
$T_\mathrm{deconf}\, [\text{MeV}]$ &202(3)(0)& $201(6)(0)$ & $193(13)(2)$ \\       
$N_\tau=N_{q_4}\,{\rm range}$ & 4-14 & 10-14& 4-20\\
\hline\hline     
\end{tabular} 
\caption{Properties of the three sets of finite-temperature ensembles used in our study, 
among them the deconfinement cross-over temperature $T_\mathrm{deconf}$ (defined by the Polyakov loop susceptibility).}
\label{tab:tc} 
\end{table} 

To compute the gluon correlation functions \eqref{Eq:GPT} and \eqref{Eq:GPL}, each
generated configuration needs to be fixed to Landau gauge. This corresponds to the following
discretized local condition
\beq 
 \nabla_{\mu}A_{\mu}=
\sum_{\mu=1}^{4} \left(A_{\mu}(x+\hat{\mu}/2) - A_{\mu}(x-\hat{\mu}/2) \right) = 0
\label{eq:gaugecondition}
\eeq
on the gauge fields defined from the link variables as
\beq
 A_{\mu}(x+\hat{\mu}/2)=
               \frac{1}{2iag_{0}}(U_{x\mu}-U_{x\mu}^{\dagger})\mid_{traceless}\,.
\label{eq:potential}
\eeq

This condition may be fulfilled by iteratively applying local gauge transformations $g_{x}$
\beq
U_{x\mu} \stackrel{g}{\mapsto} U_{x\mu}^{g}
= g_x^{\dagger} U_{x\mu} g_{x+\mu} \,,
\qquad g_x \in SU(3) \,,
\label{gaugetraf}
\eeq
in order to maximize the functional 
 \beq
F_{U}[g]=
\dfrac{1}{3} \sum_{x,\mu} \re \tr \left( g_{x}^{\dagger} U_{x\mu} g_{x+\mu} \right).
\label{eq:functional}
\eeq
We consider a configuration to have reached a (local) extremum if the the global deviation 
is less than
\beq
\max_{x}\re\tr[\nabla_{\mu}A_{x\mu}\nabla_{\nu}A_{x\nu}^{\dagger}]<
10^{-13} \,.
\eeq
This procedure has been carried out by means of the cuLGT library \cite{Schrock:2012fj}, which we have adapted for the use with lattice configurations in the ILDG format.

Subsequently we transform the gauge fields \eqref{eq:potential} into Fourier space, where the lattice momenta are defined as 
\begin{align}
k_\mu a=\frac{\pi n_{\mu}}{N_{\mu}}, \quad n_\mu\in \left(-N_{\mu}/2, N_{\mu}/2\right].
\end{align}
They are related to physical momenta via
\beq
q_{\mu}(n_{\mu}) = \frac{2}{a} \sin\left(\frac{\pi
 n_{\mu}}{N_{\mu}}\right)\,.
\eeq

\subsection{Bayesian spectral reconstruction}
\label{Sec:Bayes}

The extraction of gluon spectral functions from simulated correlation functions poses an inherently ill-defined problem. Our task is to select a unique continuous function to reproduce a finite and noisy set of datapoints. In order to be able to resolve the anticipated peaked features in the spectrum one discretizes it along real-time frequencies $\omega$ with $O(1000)$ bins, while as shown in Tab.\ref{tab:tc} the number of available correlator points ranges over $N_{q_4}\in[4\ldots20]$. Hence inverting a discretized Eq.\eqref{Eq:KaellenRel}
\begin{align}
D_i^\rho=\sum_{l=1}^{N_\omega} \Delta\omega_l K_{il} \rho_l,\quad i\in[0,N_{q_4}], \quad {N_\omega \gg N_{q_4}}
\end{align}
via a naive $\chi^2$ fit of the $\rho_l$ parameters would yield an infinite number of degenerate solutions. The severity of the ill-posed-ness of the inverse problem in the case of gluon spectra is even stronger than for the reconstruction of e.g. mesonic spectra. The space of possible reconstructions for the latter is restricted to positive definite spectra, while already perturbative computations at high energy show that gluon spectra do contain both positive and negative contributions.

Just as in the positive definite case, Bayes theorem can provide us with a viable strategy to regularize the otherwise ill-defined problem. It states that the probability of a test spectral function $\rho$ to be the correct spectrum, given measured data and further, so called prior information $(I)$ is proportional to the product of two terms
\begin{align}
P[\rho|D,I]\propto P[D|\rho,I]P[\rho|I].
\end{align}
This expression follows from the multiplication theorem of conditional probabilities and formally allows prior information $I$ to influence both terms on the right hand side. The first $P[D|\rho,I]={\rm exp}[-L]$ refers to the likelihood probability, which in our case is related to the $\chi^2$ fitting functional. The likelihood $L$ measures the quadratic distance between the correlator points corresponding to the test function $\rho$ and the simulated data $D_i$
\begin{align}
L=\frac{1}{2}\sum_{i,j=1}^{N_{q_4}}(D_i-D^\rho_i)C_{ij}^{-1}(D_j-D^\rho_j),
\end{align}
where $C_{ij}$ refers to the usual covariance matrix of the simulated $D_i$'s. Prior information enters implicitly, as we will only accept spectra for which $L=N_{q_4}$, which corresponds to our prior knowledge that the correct spectrum, sampled randomly with Gaussian noise, would on average lead to such a value of the likelihood. Note that $L$ by itself possesses $N_\omega-N_{q_4}$ flat directions, which in the Bayesian approach are regularized by introducing the prior probability $P[\rho|I]={\rm exp}[-\alpha S (\omega)]$.

This second term encodes further information we possess about the spectrum, beyond the simulation data, which may take the form of a smoothness condition, a sum rules or in the case of hadronic spectra refer to positive definiteness. Prior information enters in two ways: on the one hand the functional form of $S$ itself encodes part of that information, on the other hand $S[m]$ conventionally depends on a function $m(\omega)$ called the default-model. By definition $m$ corresponds to the correct spectrum in the absence of data, i.e. it represents the unique extremum of $S$. In this study we will use a sum-rule for gluon spectra, which requires that the area under the spectrum integrated over positive (or equivalently due to antisymmetry negative) frequencies vanishes, by choosing the default model to vanish identically.

Since for gluon spectra we may not assume positive definiteness, standard regulators, such as those of the MEM $S_{\rm SJ}$ \cite{Asakawa:2000tr} or the standard BR method \cite{Burnier:2013nla} 
\begin{align}
S_{\rm BR}=\int d\omega \big( 1- \frac{\rho(\omega)}{m(\omega)} + {\rm log}\big[ \frac{\rho(\omega)}{m(\omega)} \big]\big)
\end{align}
are not applicable. While the Shannon Jaynes entropy of the MEM  has been generalized to treat non-positive definite spectra \cite{Hobson:1998bz}, that approach requires us to apriori choose a decomposition of the spectrum in positive and negative components. Since the information of where the spectrum starts to become negative is among those we wish to learn from such a reconstruction, we refrain from following this strategy. On the other hand other regulators, such as quadratic ones have been proposed \cite{Dudal:2013yva}. Unfortunately we have found in previous studies that these very strongly pull the final result towards the default model, especially if only a relatively small number of correlator points are available. This does not allow the information encoded in the simulation data to manifest itself in a satisfactory manner.

Instead we will deploy here a recently developed regulator $S_{\rm BR}^g$ \cite{Rothkopf:2016luz}, which shares many of the advantageous analytic properties of that used in the standard BR method
\begin{align}
S_{\rm BR}^g=\int d\omega \Big(  \frac{|\rho(\omega)-m(\omega)|}{h(\omega)} + {\rm log}\Big[ \frac{|\rho(\omega)-m(\omega)|}{h(\omega)}+1 \Big]\Big).
\end{align}
Since now $\rho$ and $m$ can both take on the value zero, one uses a different measure for deviation between the default model and the spectrum $r_l=|\rho_l-m_l|/h_l$. The function $h_l$, absent in the standard BR regulator, corresponds to an additional default model like function, which encodes the confidence we have in $m_l$. $S_{\rm BR}^g$ does not require us to choose a decomposition apriori, since the role of $m(\omega)$ is unchanged and furthermore only imprints itself relatively weakly onto the end results compared to $S_{SJ}$ or the quadratic prior.

Both the choice of $m$ and $h$ contribute to the systematic uncertainties of the reconstructed spectrum. Thus their values need to be varied  to ascertain, which parts of the spectrum are fixed predominantly by the correlator data. Note also that a hyper-parameter $\alpha$ has been introduced in the definition of the prior probability, taking into account that we may weight the influence of data and prior information independently from each other. The analytic form of $S_{\rm BR}^g$ allows us to integrate $\alpha$ out in a straight forward fashion, assuming full ignorance about its values $P[\alpha]=1$
\begin{align}
P[\rho|D,I,m]\propto P[D|\rho,I]\int_0^{\infty} d\alpha P[\rho | m,\alpha] P[\alpha]\label{Eq:IntOutAlpha}.
\end{align}

Once $m(\omega)$ and $h(\omega)$ are specified we have to carry out a numerical search for the most probable Bayesian spectrum according to 
\begin{align}
\left. \frac{\delta P[\rho|D,I]}{\delta \rho} \right|_{\rho=\rho^{\rm Bayes}} = 0,
\end{align}
which, as laid out above, consists of a competition between reproducing the simulation data and conforming to prior information. Even though we do not restrict the space of basis functions for this search, i.e. the optimization is performed in $N_\omega\gg N_{q_4}$ degrees of freedom, we have confirmed that different starting points lead to the same extremum. The underlying reason is that $S_{\rm BR}^g$ fulfills the conditions required in the proof of uniqueness of the Bayesian solution given in \cite{Asakawa:2000tr}.

For the reconstructions to be presented in Sec.\ref{sec:numspecs} we deploy the generalized BR method on a frequency grid $\omega\in[10^{-3},100]$GeV divided in $N_\omega=2000$ bins. To ensure that the evaluation of the convolution in \eqref{Eq:KaellenRel} evaluated over such a relatively large frequency interval does not suffer from numerical precision losses, the computations are carried out using $512$bit arithmetic. In order to further improve the stability of the numerical optimization task, we deploy the regularization prescription for the kernel $K(q_4,\omega)=2{\rm ArcTan}(\omega)\omega/(q_4^2+\omega^2)$. It takes into account that the spectrum vanishes at $\omega=0$ and that the UV region of the Matsubara frequencies on the lattice is affected by the finite lattice spacing. 

In anticipation of the zero area sum rule for gluon spectra, our default model $m(\omega)$ is set to zero, while the confidence function $h(\omega)$ is set to unity. We have checked through a ten bin Jackknife that the dominant uncertainties do not arise from statistical fluctuations but instead from the variation of $m$ and $h$. The robustness of the reconstruction is thus probed via deviating from the sum-rule by varying the value of the constant default model between $m=\{-2,0,2\}$, as well as also varying the confidence function $h=\{1,2\}$. We find that for all temperatures, except those where less than six datapoints are available the results remain robust. The spread of reconstructed values will be shown as errorbars and errobands.

\section{Correlation functions}
\label{sec:numcorrs}

We begin by presenting the computed gluon correlation functions in Landau gauge on $N_f=2+1+1$ tmfT lattices. In order to both connect to the preceding literature, as well as prepare for our study of gluon spectra, we provide several different visualizations. The most common is that of zero Matsubara frequency correlators $D(0,|\vec{q}|^2)$ evaluated at different spatial momenta. These can be directly compared to the same quantities routinely calculated in functional approaches to QCD. On the other hand for our inverse problem we utilize the correlators at a fixed spatial momentum $D(q_4,|\vec{q}|)$ along the different available imaginary frequencies $q_4$ on the lattice. 

In the continuum, the finite extent of the Euclidean time axis leads to equidistantly spaced Matsubara frequencies $\mu_n=2\pi T n$, at which the imaginary frequency correlator is routinely evaluated. In the presence of a finite lattice spacing the discrete Fourier transform connecting Euclidean times and imaginary frequencies leads to a artifacts close to the edge of the Brillouin zone, so that the physical momenta $q_4$ are not any longer spaced at the same distance. As we carry out the reconstruction using the kernel based on $q_4$ we also plot the correlators in these physical imaginary frequencies.

In the following we will discuss the qualitative features of the correlation functions using the ensembles of $D370$, which is both closest to the continuum and spans the broadest temperature range. The figures for the two other sets of ensembles can be found in Appendix A.  Since temperature scans are carried out using a fixed scale prescription, the maximum available imaginary frequency remains constant and it is the spacing between values of $q_4$ that grows with increasing $T$. 

\begin{figure*}[th]
\includegraphics[scale=0.3]{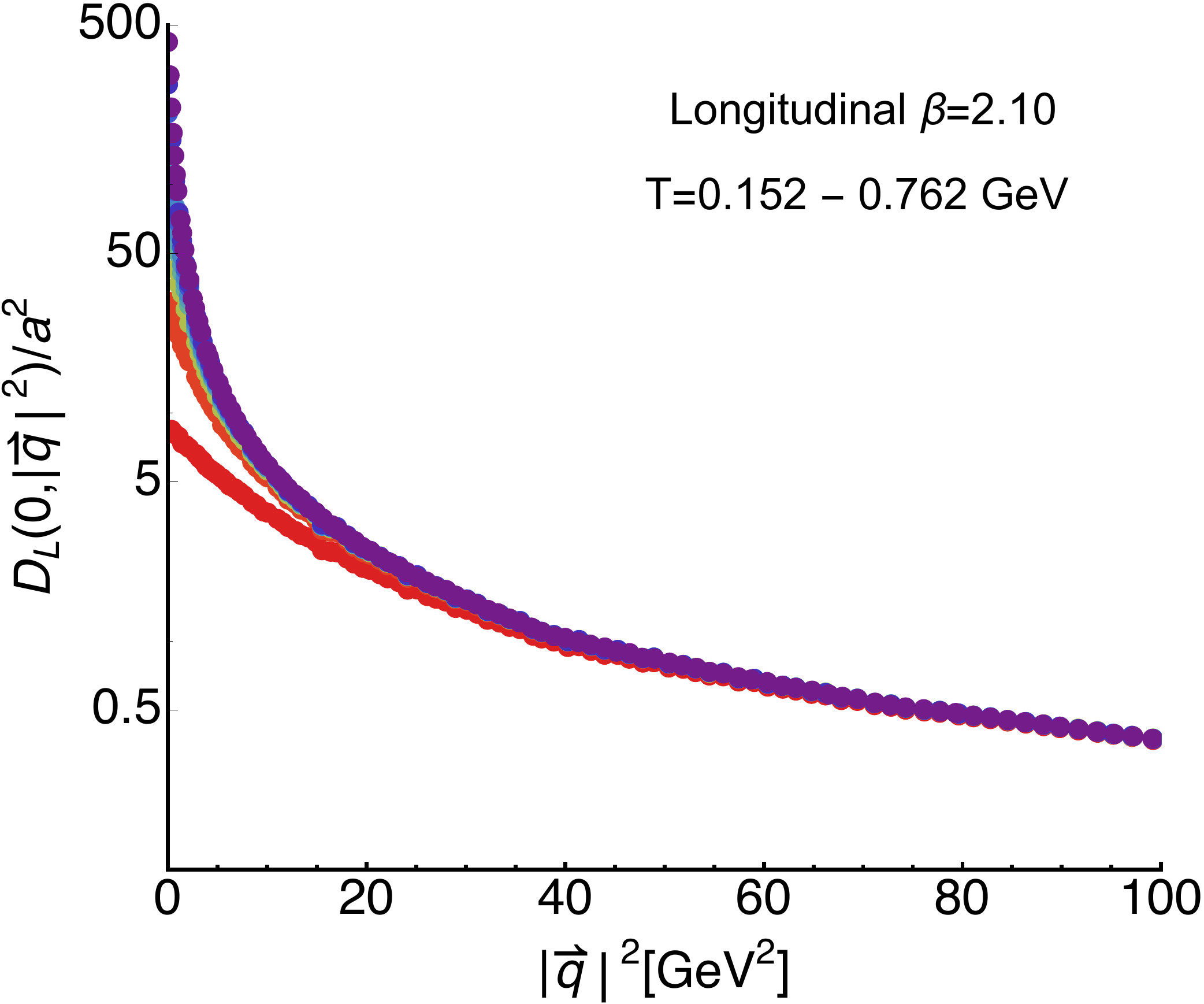}
\includegraphics[scale=0.3]{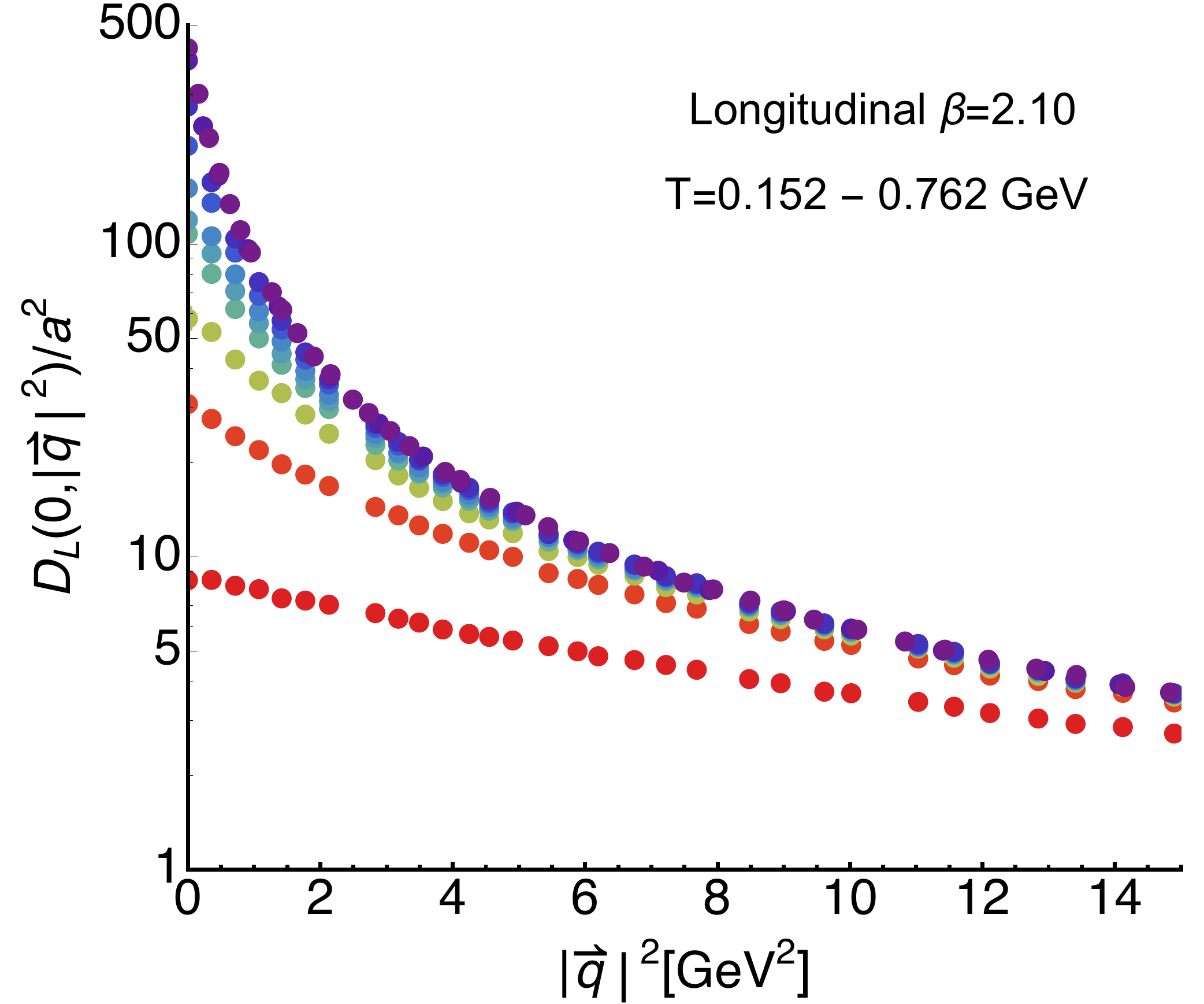}
\caption{The longitudinal gluon correlators at $\beta=2.10$ evaluated at vanishing imaginary frequency $q_4=0$ but finite spatial momenta $|\vec{q}|^2$ for different temperatures $T=152\ldots762$MeV. The left panel shows values up to momenta of $100{\rm GeV}^2$, while the right panel contains a zoomed in interval around the origin. }\label{Fig:LongCorr210VsPSqr}
\end{figure*}
\begin{figure*}
\includegraphics[scale=0.3]{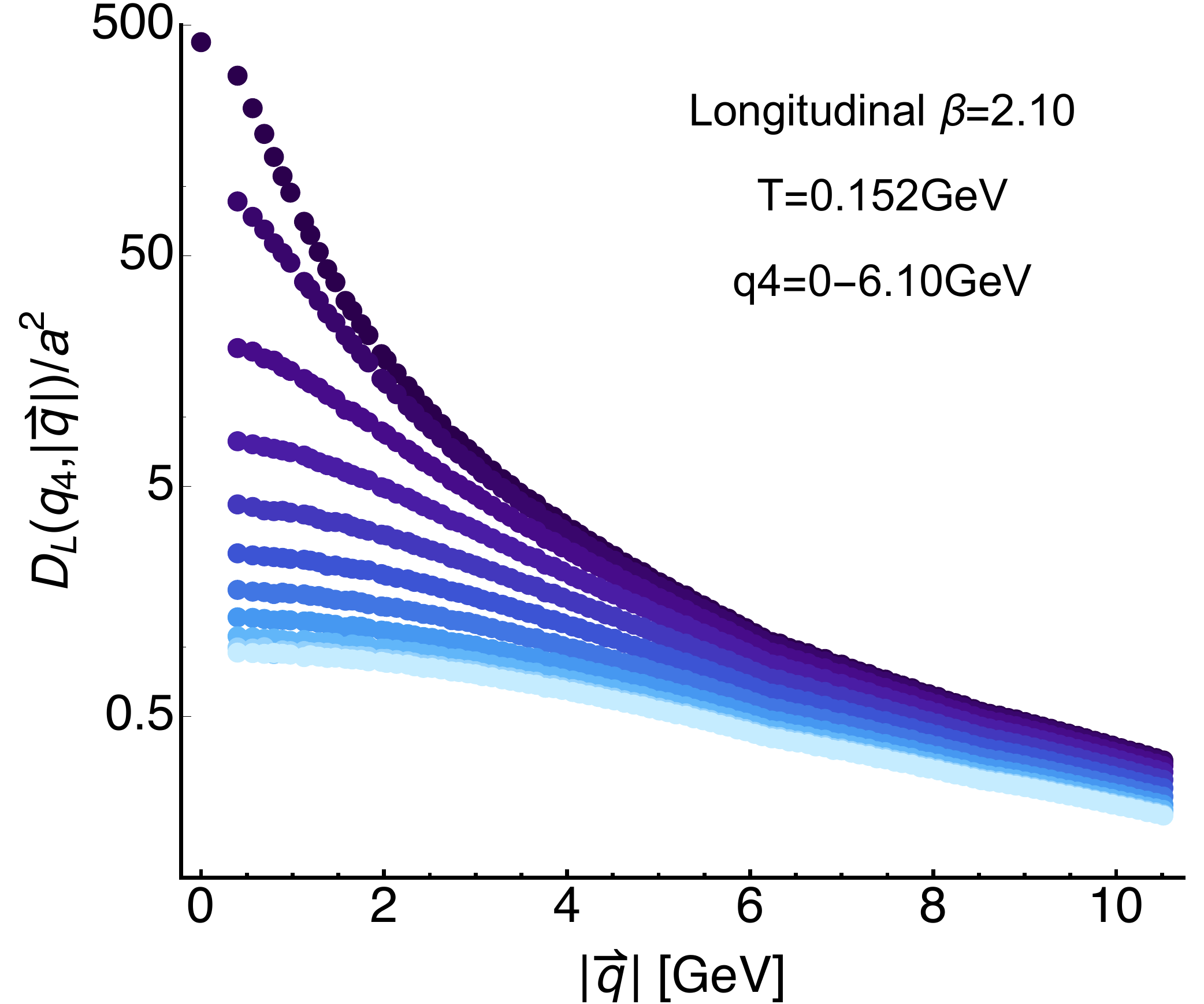}\vspace{0.5cm}
\includegraphics[scale=0.3]{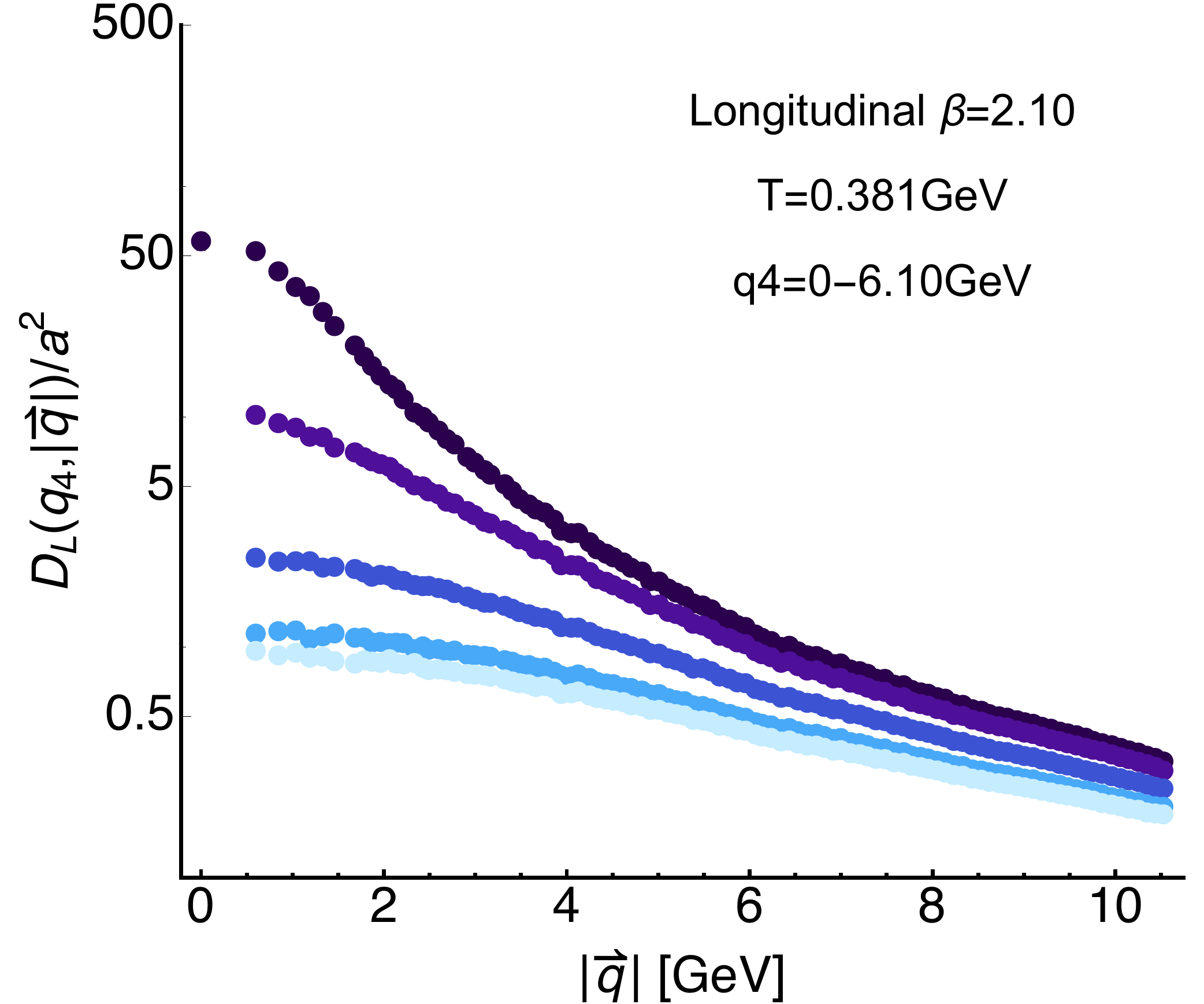}\vspace{0.5cm}
\includegraphics[scale=0.3]{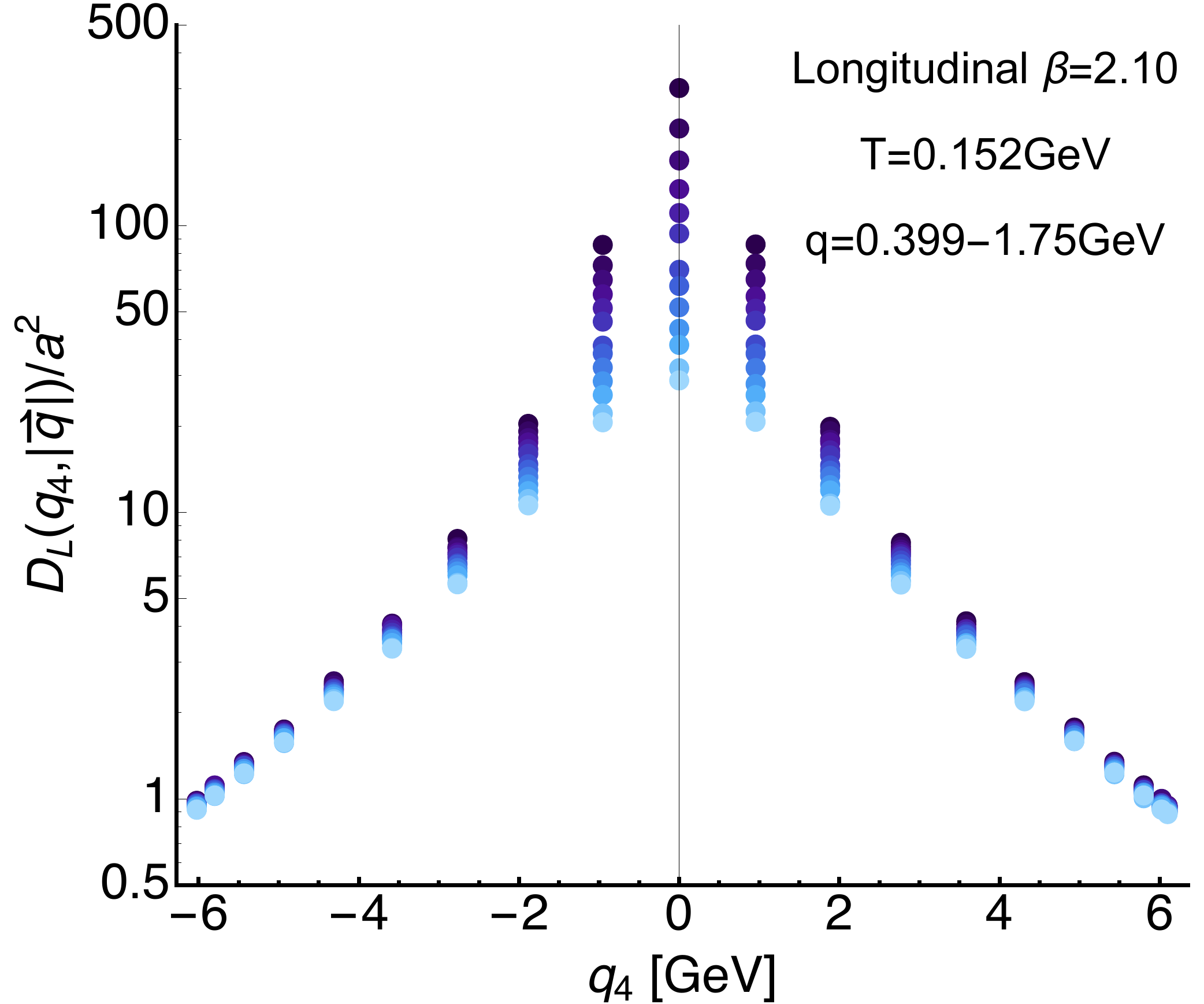}
\includegraphics[scale=0.3]{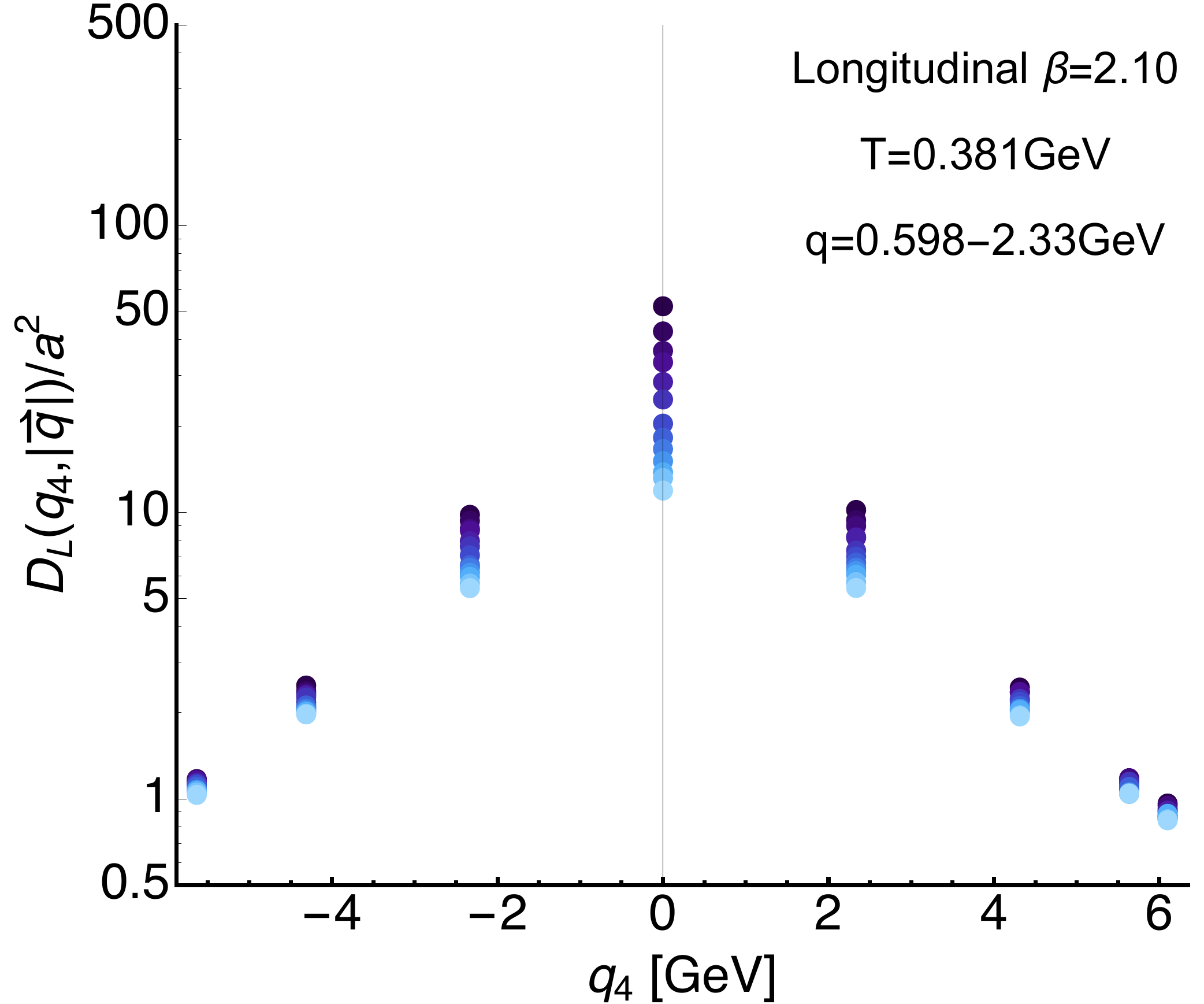}
\caption{The longitudinal gluon propagators at $\beta=2.10$ evaluated at both finite imaginary frequency $q_4$ and spatial momentum $|\vec{q}|$.  The top row shows the $|\vec{q}|$ dependence at fixed $q_4$, while the bottom row we present the $q_4$ dependence for the fourteen lowest $|\vec{q}|$ values that will be used in the spectral reconstruction subsequently. The color coding assigns darkest colors to the lowest value of the corresponding parameter. }\label{Fig:LongCorr210VsMu}
\end{figure*}

The difficulty of the inverse problem will be immediately apparent, as we find that down to the first finite $q_4$ the correlators are nearly indistinguishable (except at the highest temperatures considered). Thus the most important contributions to the physics of gluons at finite temperature is hidden in the interval between $q_4=0$ and $q_4\approx\mu_1=2\pi T$. 

\subsection{Longitudinal Correlators}

\begin{figure*}[th!]
\includegraphics[scale=0.3]{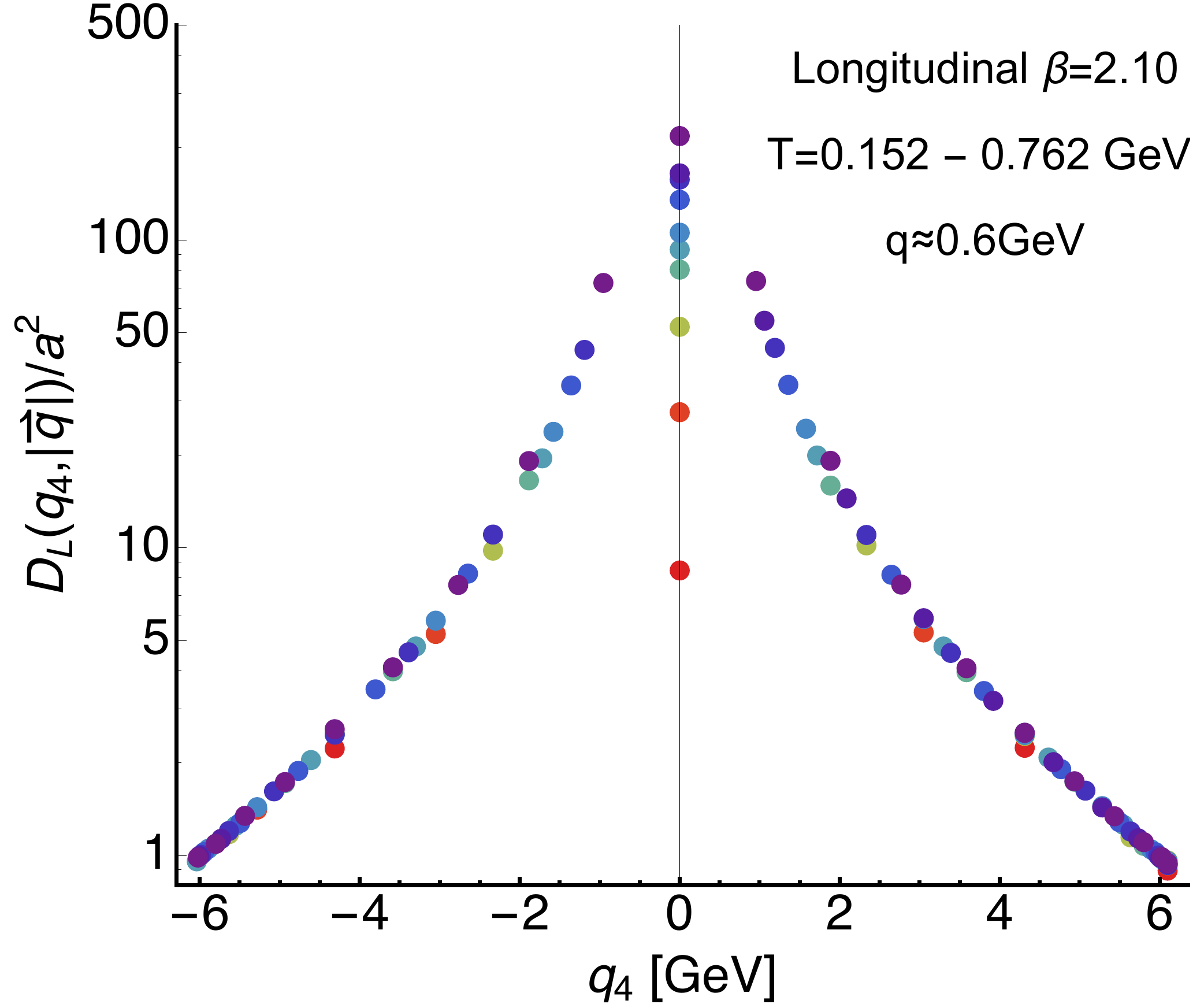}\hspace{0.0cm}
\includegraphics[scale=0.3]{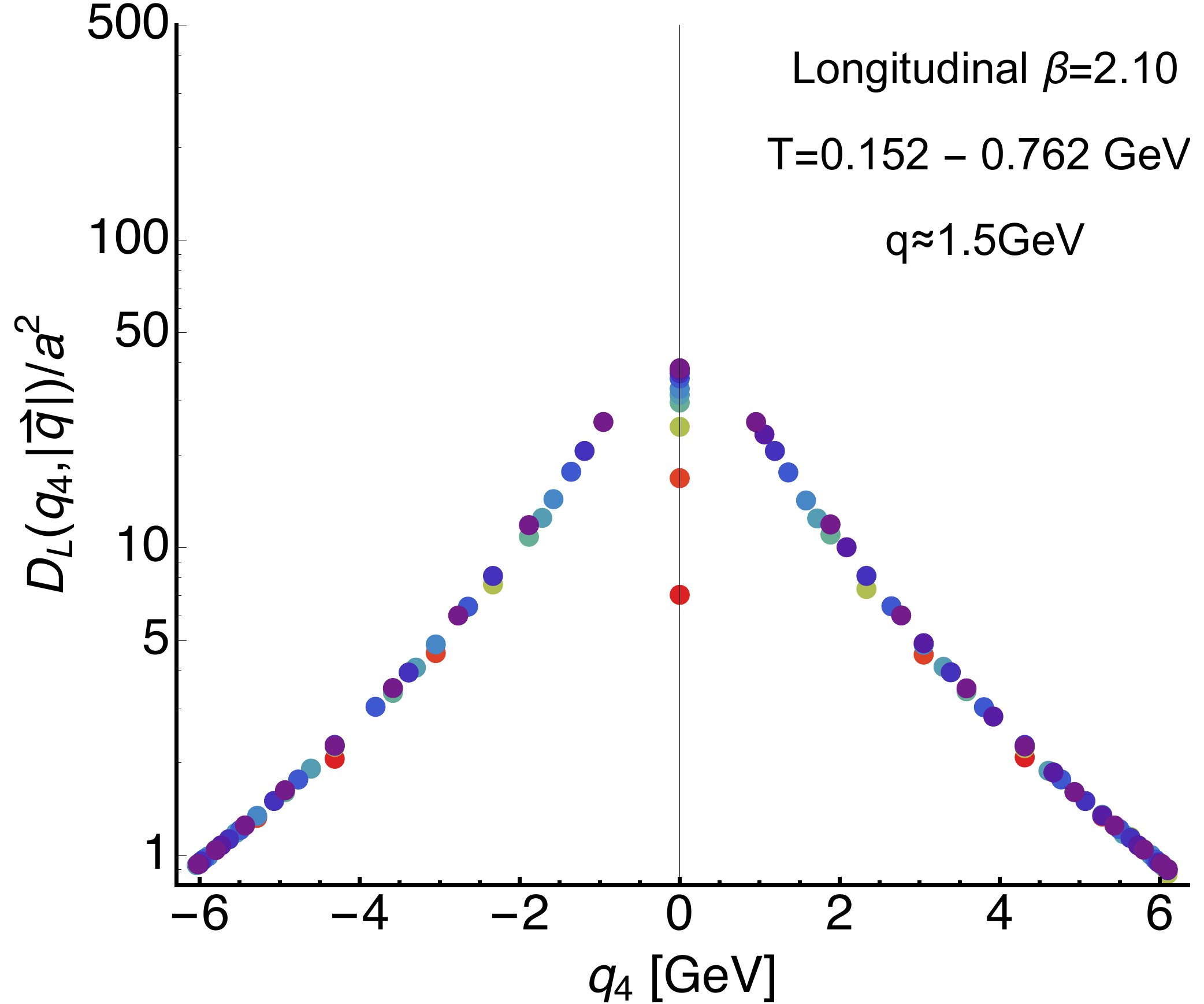}\hspace{0.0cm}
\caption{The longitudinal gluon propagators at $\beta=2.10$ evaluated along imaginary frequencies for the ten available temperatures among the ensembles of $D370$, i.e. $T=152\ldots762$MeV. The left panel contains correlators at $|\vec{q}|\approx0.6$GeV while the right panel shows those at $|\vec{q}|\approx1.5$GeV. }\label{Fig:LongCorr210VsMuDiffTmp}
\end{figure*}
\begin{figure*}[th!]
\includegraphics[scale=0.3]{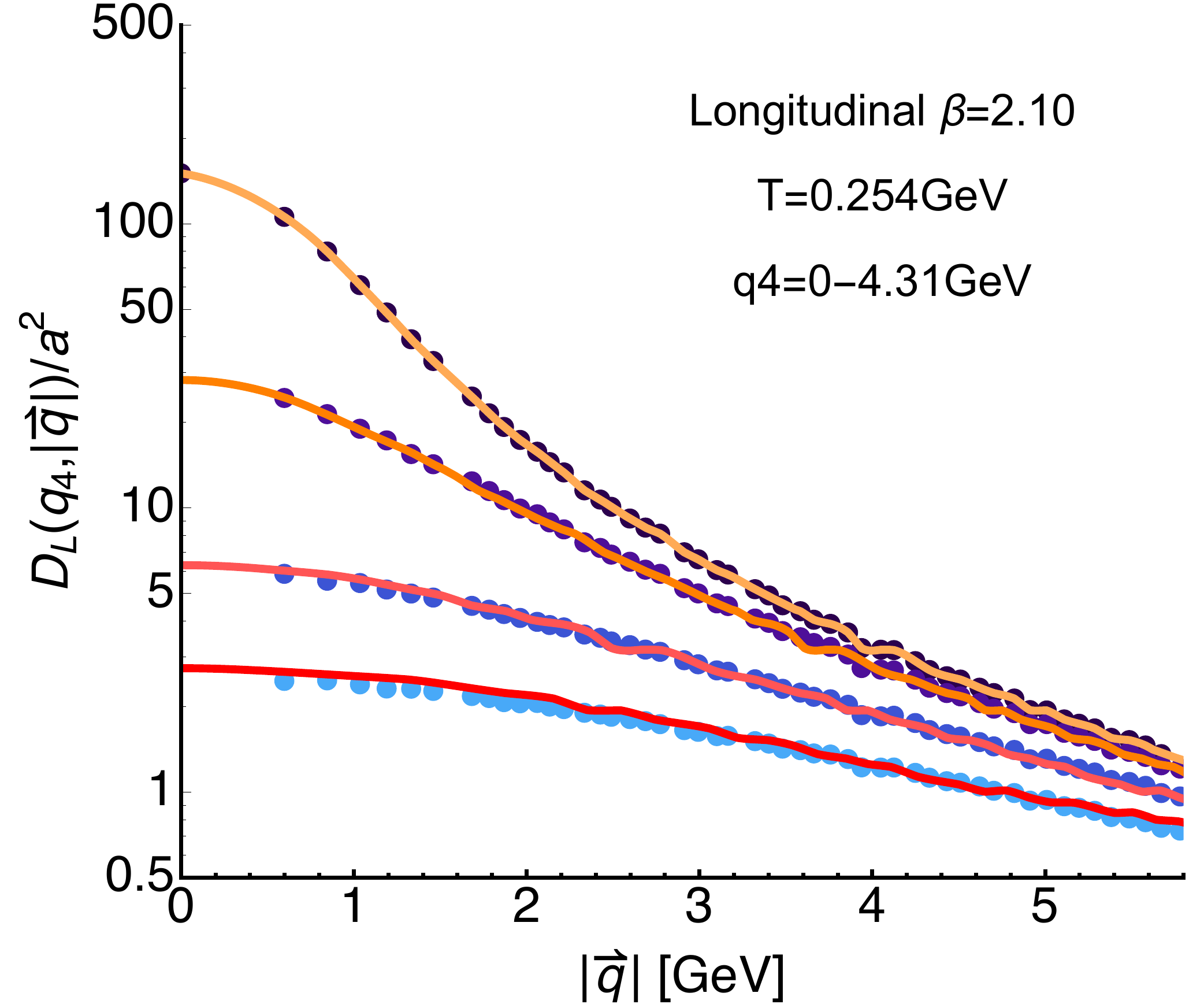}\hspace{0.0cm}
\includegraphics[scale=0.3]{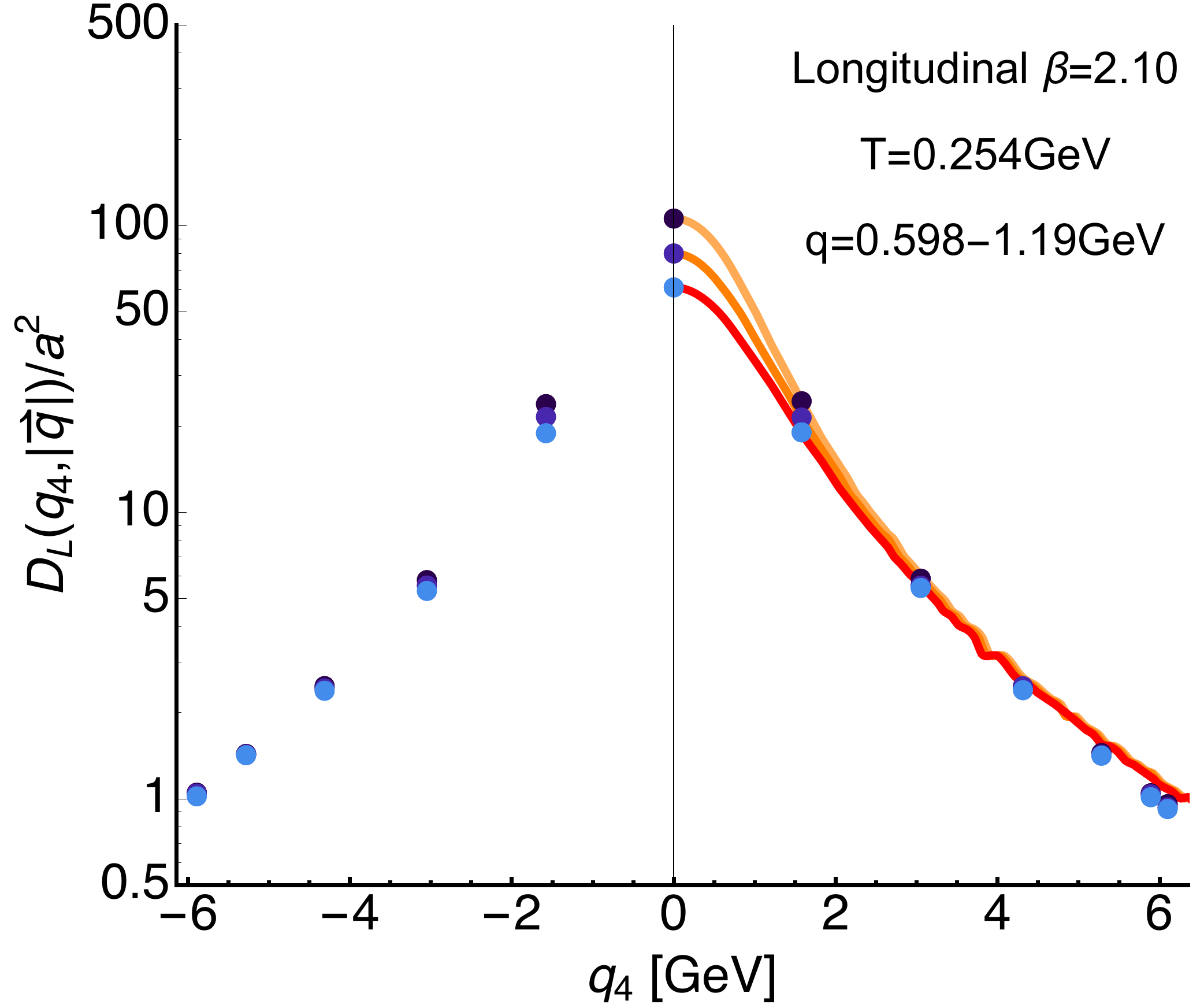}\hspace{0.0cm}
\caption{The longitudinal gluon propagators for $\beta=2.10$ evaluated at $T=254$MeV. (left) Shown are correlators vs. spatial momenta at the first four available imaginary frequencies $q_4=0-4.31$GeV. The orange curve represents a naive spline interpolation $\tilde{D}_L(|\vec{q}|)$ of $D_L(q_4=0,|\vec{q}|)$. The dark orange and red curve corresponds to the interpolation evaluated using the assumption of $O(4)$ invariance: $\tilde{D}_L(\sqrt{q_4^2 + |\vec{q}|^2})$. (right) Use of the interpolation $\tilde{D}_L(|\vec{q}|)$ in order to reproduce the finite imaginary frequency behavior of the propagator (blue points) . The solid curves show the $O(4)$ evaluation $\tilde{D}_L(\sqrt{q_4^2 + |\vec{q}|^2})$. We can see that while for small imaginary frequencies the $O(4)$ ansatz works quite well, it starts to degrade as one approaches the boundary of the Brillouin zone due to breaking of rotational symmetry on the lattice.}\label{Fig:LongCorr210VerifyInterpolation}
\end{figure*}

In Fig.\ref{Fig:LongCorr210VsPSqr} we show the longitudinal gluon correlators \eqref{Eq:GPL} evaluated at vanishing Matsubara frequency and along finite spatial momentum for the ten available temperatures among the ensembles of $D370$ (see Tab.\ref{tab:D370}). While the left panel shows a large momentum range up to $|\vec{q}|^2=100{\rm GeV}^2$, the right panel contains a zoomed in interval around the origin. From the former we find that as expected \cite{Pawlowski:2016eck} for $|\vec{q}|^2\gg T^2$ the correlators take on the same values and at the highest momenta shown are virtually indistinguishable. From the zoom-in we see that the longitudinal correlators are ordered in amplitude according to temperature. $D_{L}$ at the lowest $T$ (dark violet) takes on the largest values and exhibits the smallest values at the highest $T$ (red).

\begin{table} [h]
\centering      
\begin{tabular}{c|c|c|c|c|c|c|c|c|c|c|}
\hline\hline       

$D370$ $N_{\tau}$  & 4& 6& 8& 10& 11& 12& 14& 16& 18& 20    \\     
\hline
$T$  MeV & 762& 508& 381& 305& 277& 254& 218& \
191& 170& 152    \\
\hline
$N_s$  & 32& 32& 32& 32& 32& 32& 32& 32& 40& 48    \\     
\hline
$N_\mathrm{meas}$  & 310 & 400  & 120& 410& 420& 380&790&610&590&280 \\   
\hline\hline     
\end{tabular} 
\caption{Grid sizes and temperatures in the D370 ensembles used for the computation of the correlation functions below. $N_{\rm meas}$ refers to the number of available correlator measurements.} 
\label{tab:D370} 
\end{table} 

Now let us have a look at the correlators evaluated instead at finite Matsubara frequencies and finite spatial momentum as shown in Fig.\ref{Fig:LongCorr210VsMu}. Here we find another ordering, i.e. the correlators vary in magnitude according to spatial momenta or imaginary frequencies at a fixed temperature. Those corresponding to lowest momenta or imaginary frequencies take on the largest values. 

We continue with plotting the longitudinal correlators along finite imaginary frequencies at a fixed momentum for different temperatures in Fig.\ref{Fig:LongCorr210VsMuDiffTmp}. The left panel corresponds to a low value of momentum $|\vec{q}|\approx0.6$GeV, while the right panel to $|\vec{q}|\approx1.5$GeV. We see that at $q_4=0$ there is a clear difference between the values visible, while already at $q_4\approx 2\pi T$ the correlators appear to lie very close to a general trend which does not depend on temperature. At higher momenta (right) the agreement of the data points above $q_4\approx 2\pi T$, as expected, is even more pronounced than at low spatial momenta.

\begin{figure*}[th!]
\includegraphics[scale=0.3]{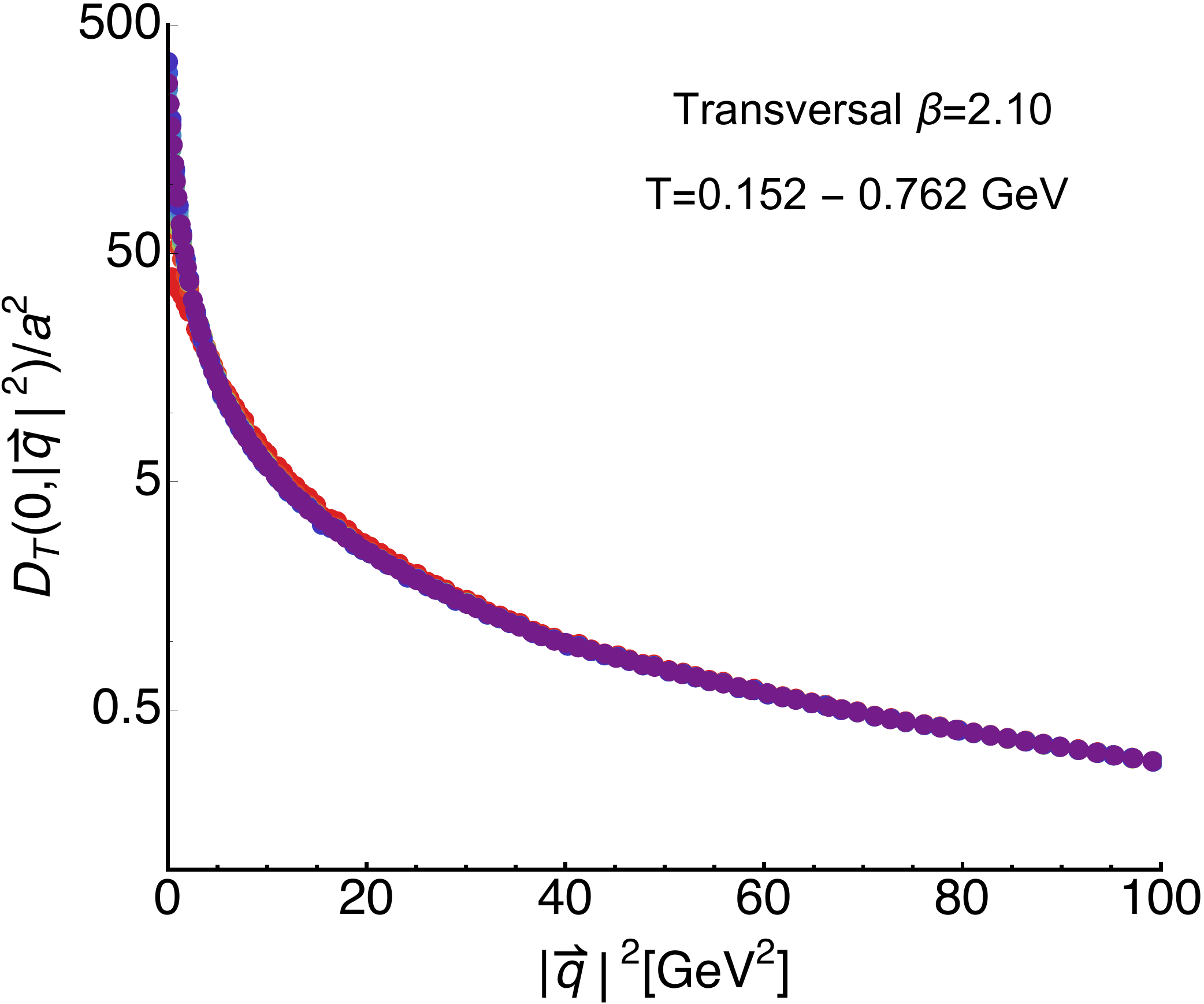}
\includegraphics[scale=0.3]{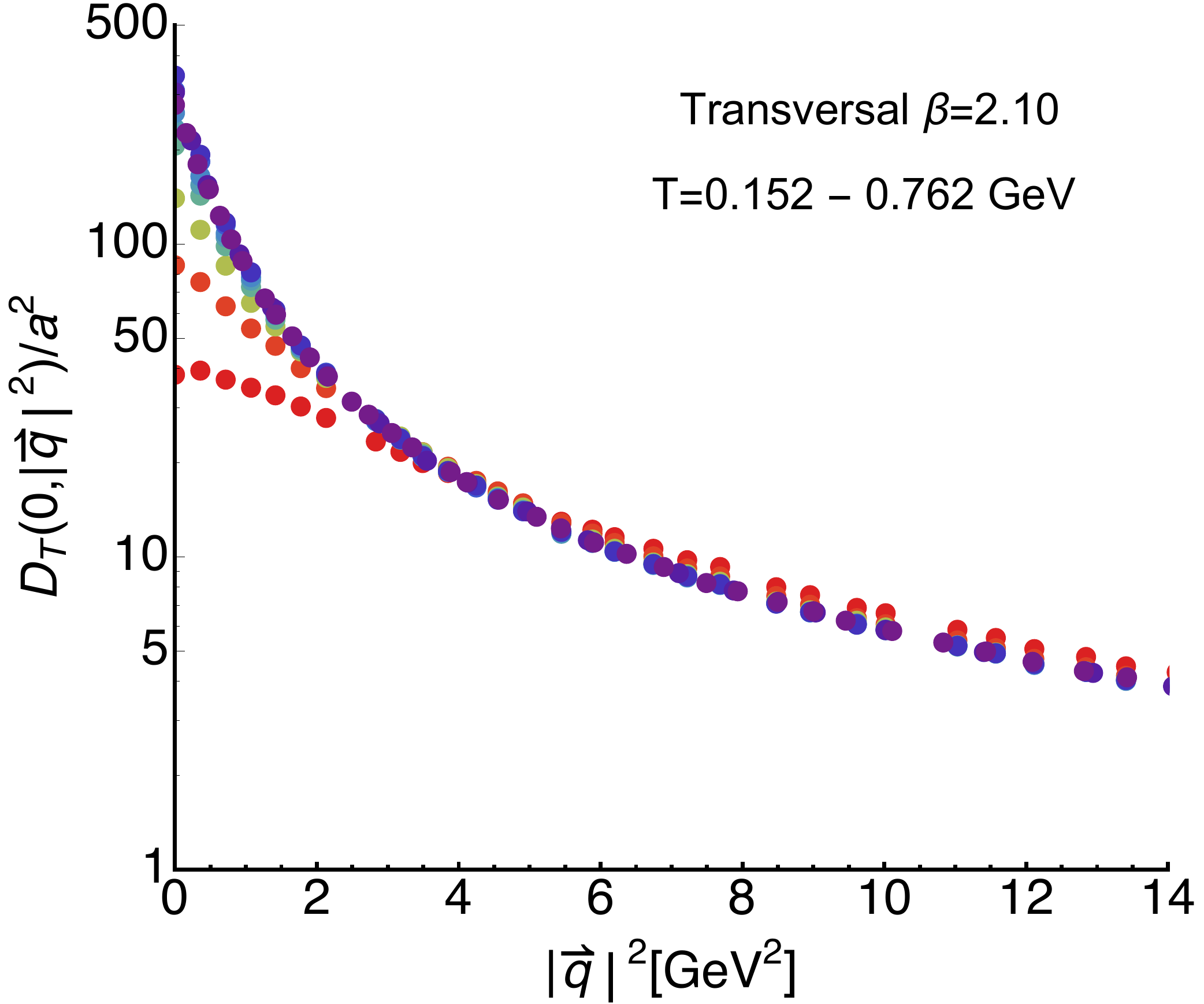}
\caption{The transversal gluon correlators at $\beta=2.10$ evaluated at vanishing imaginary frequency $q_4=0$ and finite spatial momenta $|\vec{q}|^2$ for different temperatures $T=152\ldots762$MeV. The left panel shows their values up to momenta of $100{\rm GeV}^2$, while the right panel contains a zoomed in interval around the origin. }\label{Fig:TransCorr210VsPSqr}
\end{figure*}
\begin{figure*}[th]
\includegraphics[scale=0.3]{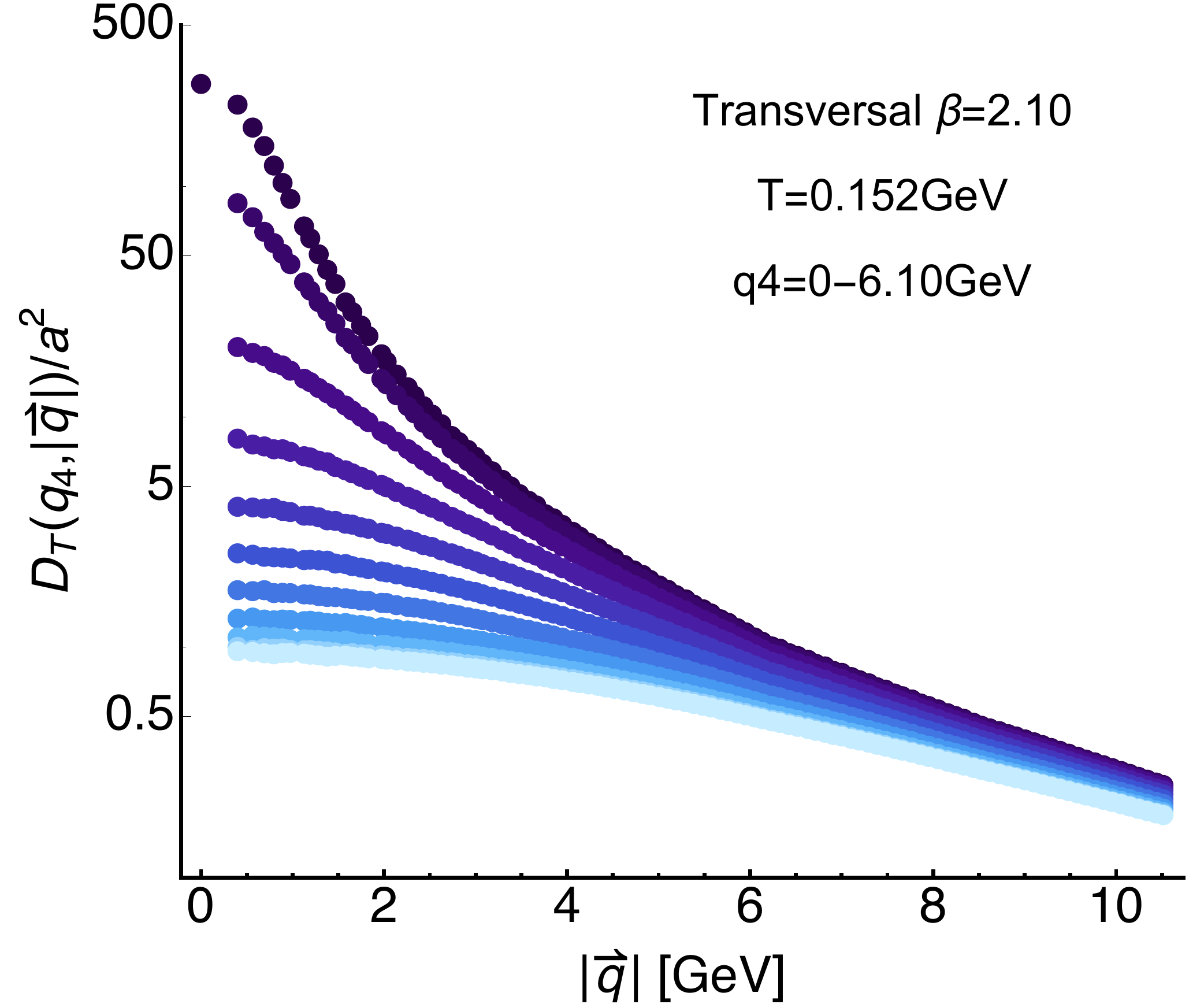}\vspace{0.5cm}
\includegraphics[scale=0.3]{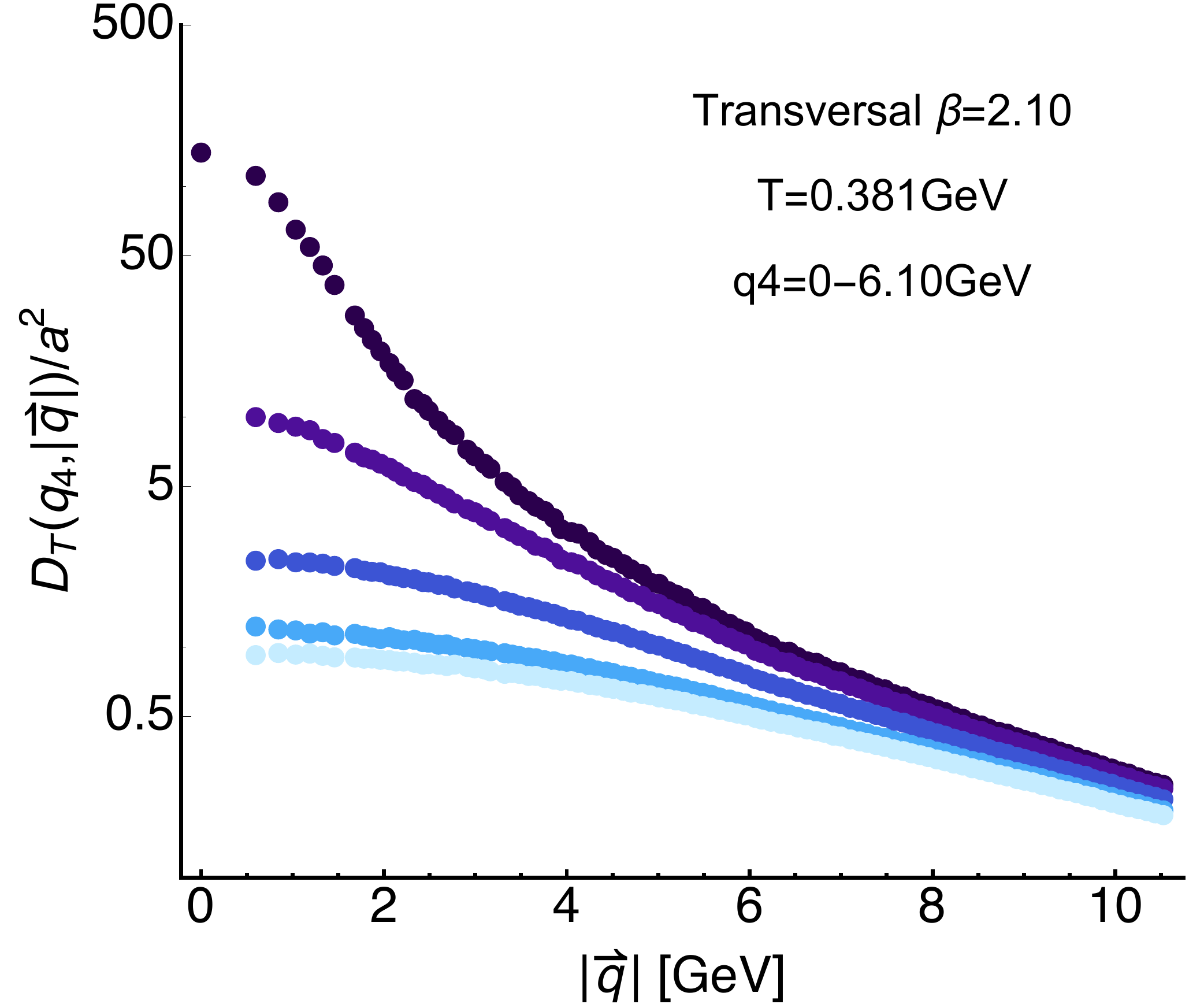}\vspace{0.5cm}
\includegraphics[scale=0.3]{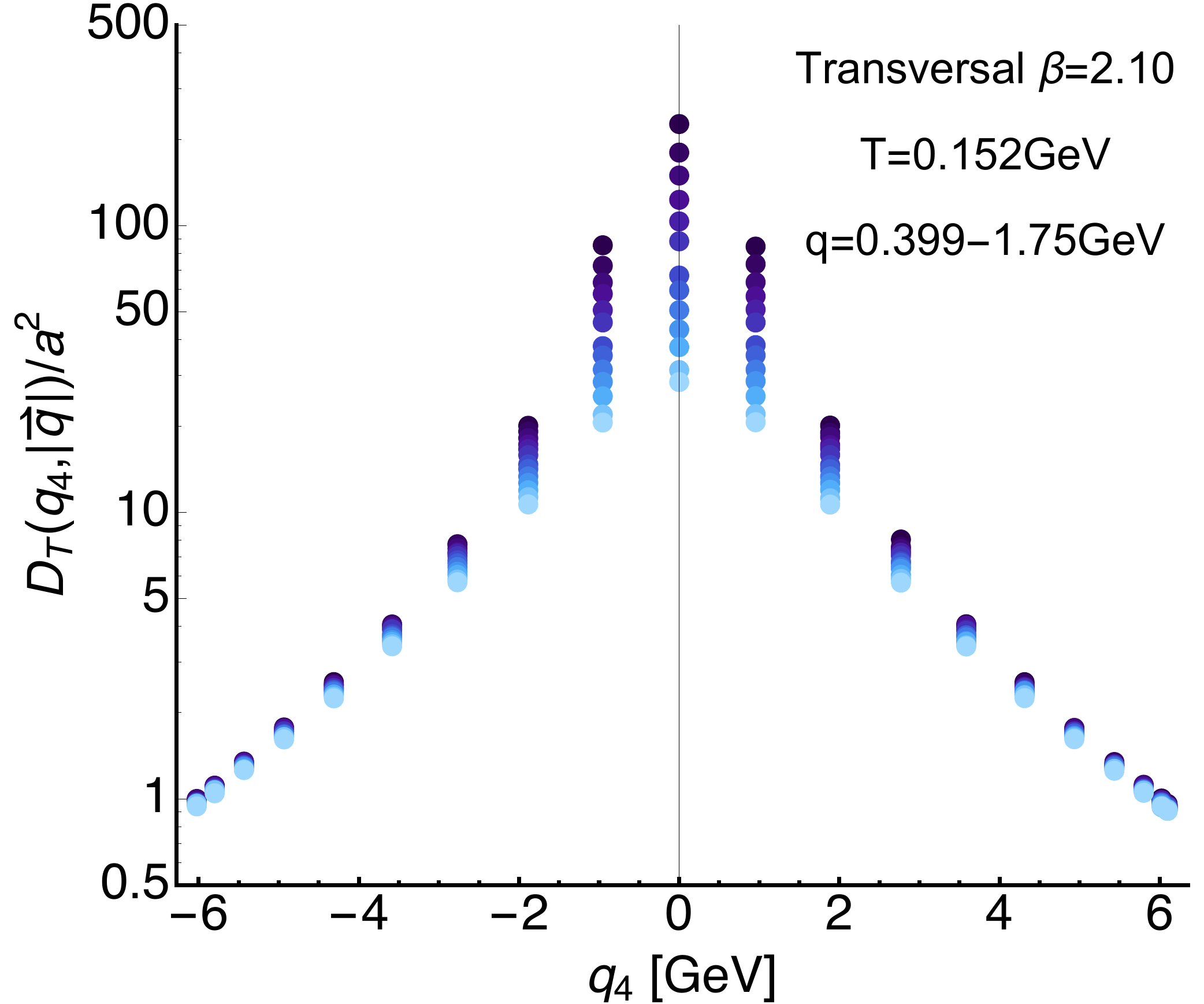}
\includegraphics[scale=0.3]{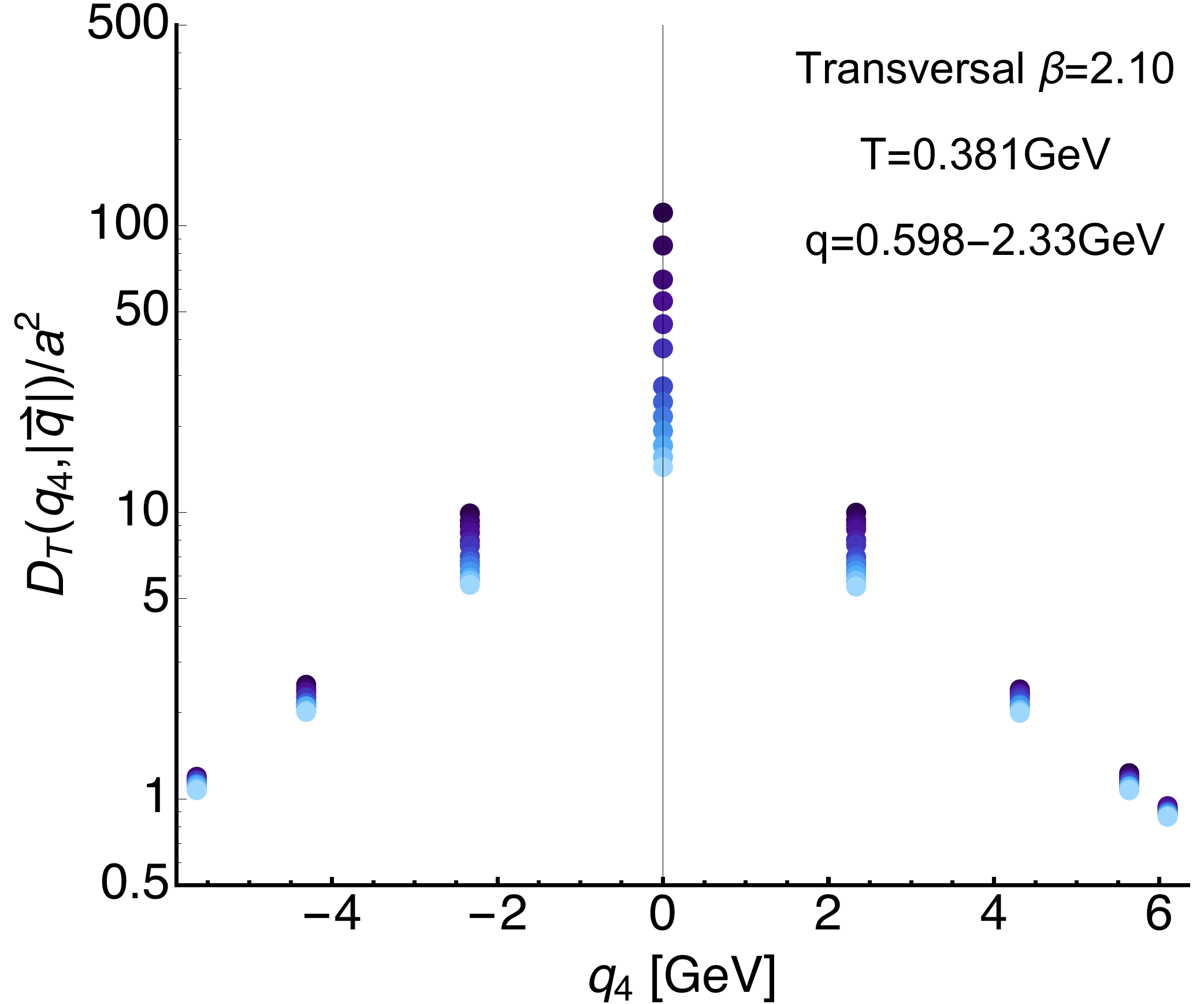}
\caption{The transversal gluon propagators at $\beta=2.10$ evaluated at both finite imaginary frequencies $q_4$ and spatial momentum $|\vec{q}|$. The top row shows the $|\vec{q}|$ dependence at fixed $q_4$, while the bottom row we present the $q_4$ dependence for the fourteen lowest $|\vec{q}|$ values that will be used in the spectral reconstruction subsequently. The color coding assigns darkest colors to the lowest value of the corresponding parameter.}\label{Fig:TransCorr210VsMu}
\end{figure*}

\begin{figure*}[th!]
\includegraphics[scale=0.3]{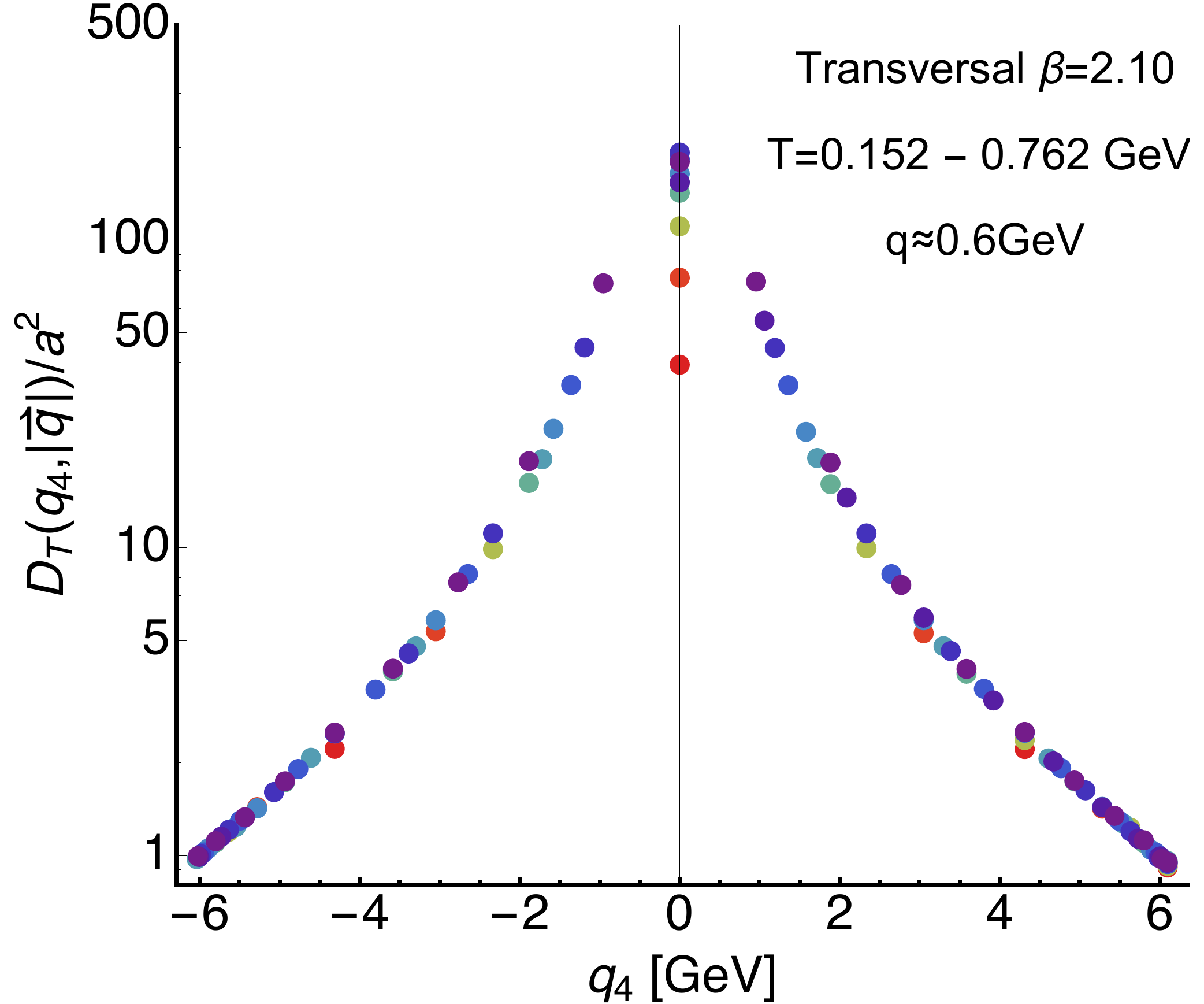}\hspace{0.0cm}
\includegraphics[scale=0.3]{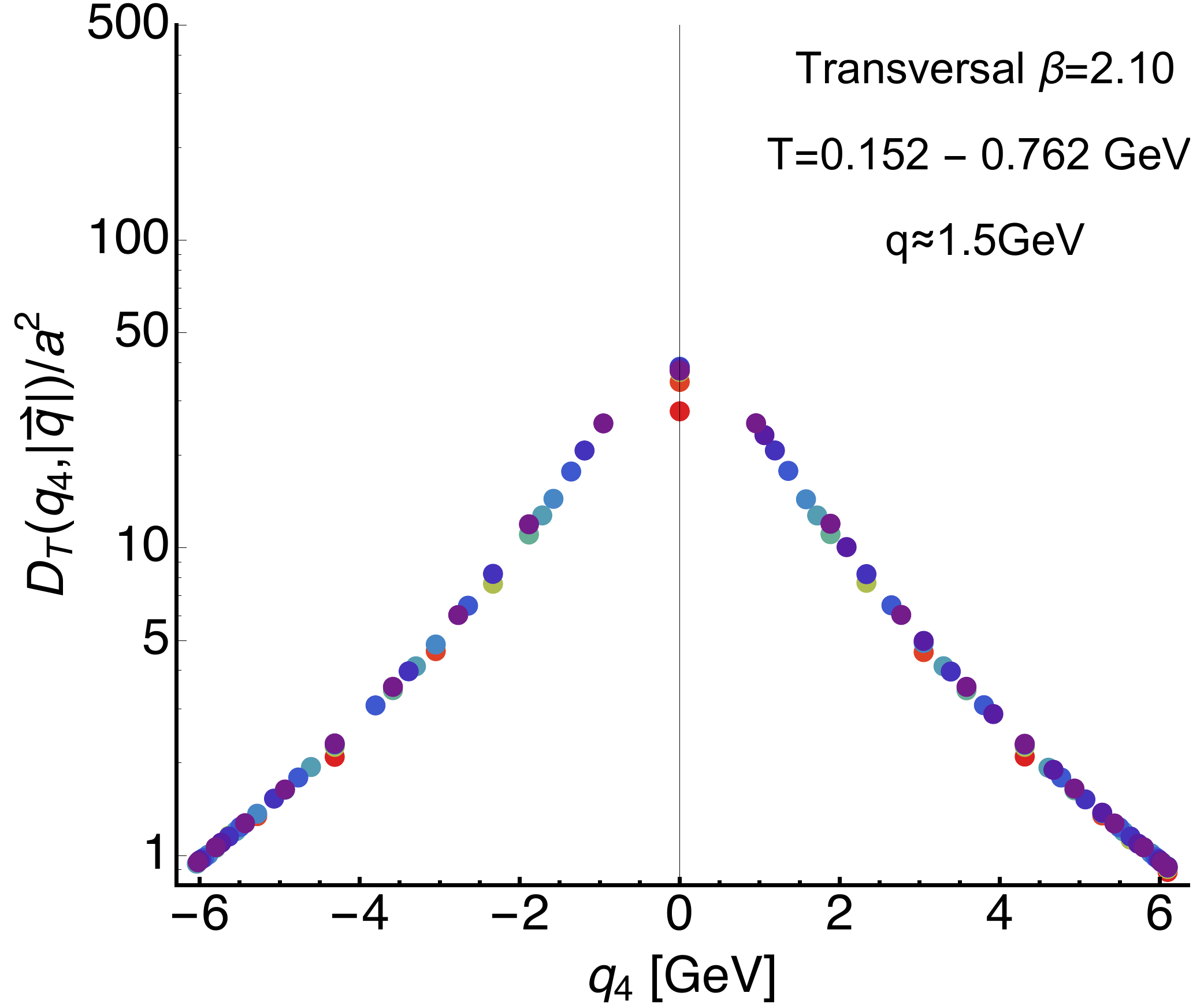}\hspace{0.0cm}
\caption{The transversal gluon propagators at $\beta=2.10$ evaluated along imaginary frequencies for the ten available temperatures among the ensembles of $D370$, i.e. $T=152\ldots762$MeV. The left panel contains correlators at $|\vec{q}|\approx0.6$GeV, while the right panel shows those at $|\vec{q}|\approx1.5$GeV.}\label{Fig:TransCorr210VsMuDiffTmp}
\end{figure*}
\begin{figure*}[th!]
\includegraphics[scale=0.3]{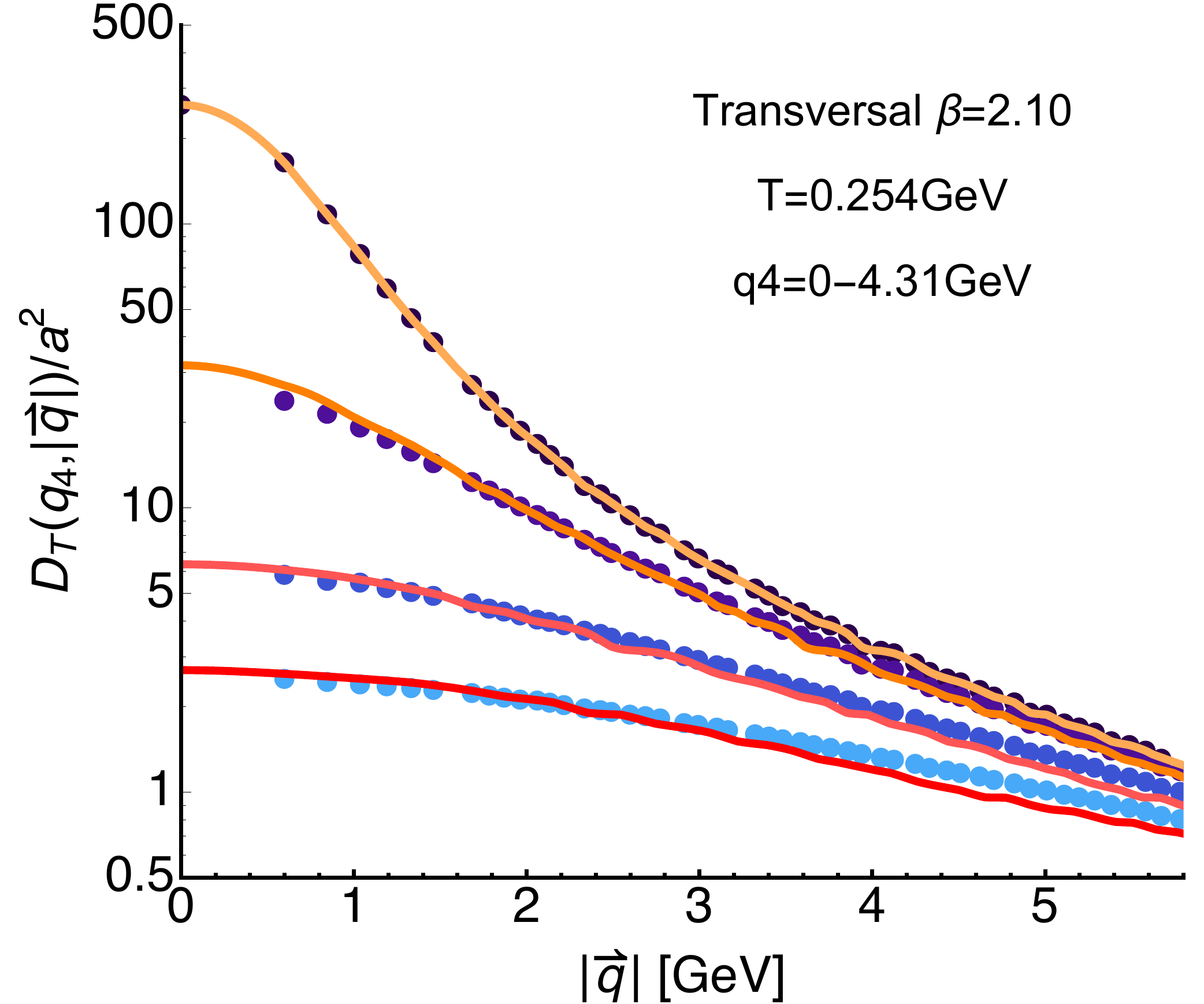}\hspace{0.0cm}
\includegraphics[scale=0.3]{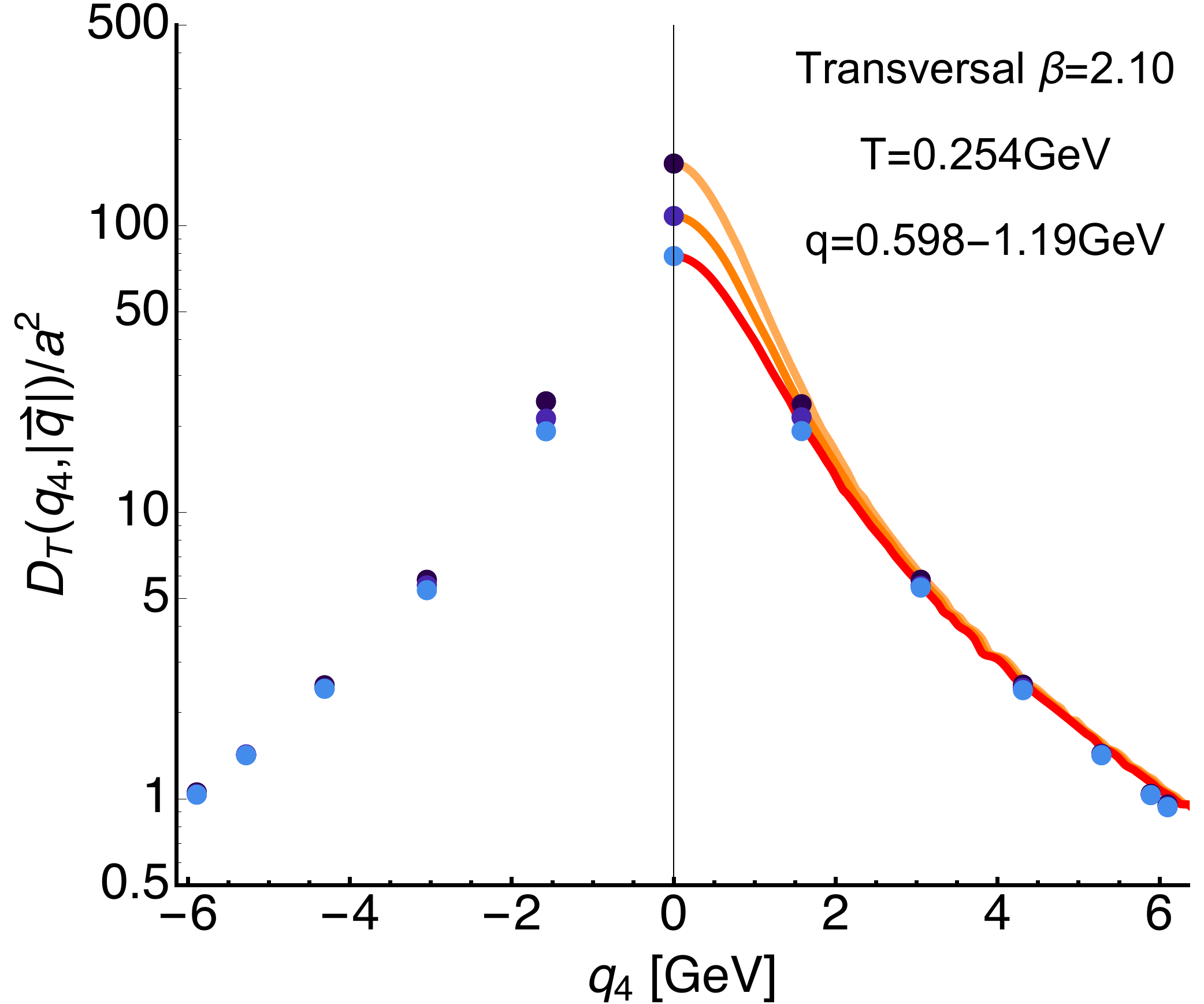}\hspace{0.0cm}
\caption{The transversal gluon propagators for $\beta=2.10$ evaluated at $T=254$MeV. (left) Shown are correlators vs. spatial momenta at the first four available imaginary frequencies $q_4=0-4.31$GeV. The orange curve represents the naive interpolation $\tilde{D}_T(|\vec{q}|)$ of $D_T(q_4=0,|\vec{q}|)$. The dark orange and red curve correspond to the interpolation evaluated using the assumption of $O(4)$ invariance $\tilde{D}_T(\sqrt{q_4^2 + |\vec{q}|^2})$. (right) Use of the interpolation $\tilde{D}_T(|\vec{q}|)$ in order to reproduce the finite imaginary frequency behavior of the propagator (blue points) . The solid curves show the naive $O(4)$ evaluation $\tilde{D}_T(\sqrt{q_4^2 + |\vec{q}|^2})$. We can see that while for small imaginary frequencies the $O(4)$ ansatz works quite well it starts to degrade as one approaches the boundary of the first Brilloin zone due to breaking of rotational symmetry on the lattice.}\label{Fig:TransCorr210VerifyInterpolation}
\end{figure*}

As last step we investigate the validity of the $O(4)$ scaling assumption for the longitudinal correlator. It states that the values of the correlator at finite imaginary frequencies may be obtained by evaluating the correlator at zero imaginary frequencies while appropriately shifting the finite spatial momentum $D(q_4,|\vec{q}|)\approx D(0,\sqrt{q_4^2+|\vec{q}|^2})$. Using a spline interpolation along spatial momenta, this ansatz has been verified in continuum computations at zero and finite temperature to apply with less than 10\% error up to the first Matsubara frequency and with even less error at the higher frequencies. This experience in continuum has motivated use of the $O(4)$ ansatz also for reconstructions of lattice spectral functions in previous studies \cite{Dudal:2013yva}. 

In the left panel of Fig.\ref{Fig:LongCorr210VerifyInterpolation} we set up a spline interpolation $\tilde{D}_L$ (solid yellow line) of the longitudinal correlator $D_L$ for $\beta=2.10$ at $T=254$MeV at vanishing imaginary frequencies (topmost points) along spatial momenta $|\vec{q}|$. The orange and red curves then correspond to this interpolation, evaluated according to $\tilde{D}_L(0,\sqrt{q_4^2+|\vec{q}|^2})$ at the first three finite available imaginary frequencies. We find that for $q_4\approx2 \pi T$ the $O(4)$ ansatz works well, starting from $|\vec{q}|\approx 1.5$GeV$^2$. Since contrary to the continuum, the finite lattice spacing manifests itself in breaking of rotational symmetry, which affects the edge of the Brillouin zone most severely, it is exactly at high spatial momenta, where we see deviations from the $O(4)$ behavior to appear. In the right panel we use $\tilde{D}_L$ to evaluate the correlator along imaginary frequencies. We find that it provides a smooth interpolation, which for the four lowest finite imaginary frequencies available on the lattice lies quite close to the actual datapoints, while it starts to deviate again close to the edge of the Brillouin zone.

Since for the spectral reconstruction very precise correlator data is required and we find that systematic uncertainties due to finite lattice spacing artifacts are manifest in the application of the $O(4)$ ansatz to finite temperature lattice gluon correlators, we will only use the actual computed correlator values for the spectral reconstructions in the next section.

\subsection{Transversal Correlators}

The transversal gluon correlators \eqref{Eq:GPT} evaluated at vanishing imaginary frequency and along finite spatial momentum for the ten available temperatures among the ensembles $D370$ are shown in Fig.\ref{Fig:TransCorr210VsPSqr}. As in the longitudinal case, we present in the left panel the momentum range up to $|\vec{q}|^2=100{\rm GeV}^2$, while in the right panel a zoomed-in interval around the origin is plotted. Again for $|\vec{q}|^2\gg T^2$ the correlators take on similar values and at the highest momenta shown are virtually indistinguishable. In the zoomed-in plot on the left we can see a first difference to the longitudinal case, in that the spread of values is much smaller here. Except for the highest temperature all $D_T$'s fall together on a single line already at around $4$GeV$^2$, while in the longitudinal case even at $15$GeV$^2$ clear differences between temperatures are visible. In addition the ordering in magnitude with temperature is not as clear. At small $p^2$ the values of the correlator at the lowest temperature seem to lie below those close to the deconfinement transition.

If instead evaluated at finite imaginary frequencies and finite spatial momentum as shown in Fig.\ref{Fig:TransCorr210VsMu}, we find similar ordering, as in the longitudinal case. I.e. the correlators vary in magnitude according to spatial momenta or imaginary frequencies at a fixed temperature, where those corresponding to lowest momenta or Matsubara frequencies take on the largest values. 

Let us plot the transversal correlators along finite imaginary frequencies at a fixed momentum for different temperatures in Fig.\ref{Fig:TransCorr210VsMuDiffTmp}. The left panel contains $D_T$ at a low momentum value $|\vec{q}|\approx0.6$GeV, while the right panel at $|\vec{q}|\approx1.5$GeV. Its behavior is qualitatively similar to that of the longitudinal correlator at finite $q_4>2\pi T$. Already at $q_4\approx 2\pi T$ the correlators appear to lie very close to a general trend which does not depend on temperature. At higher momenta (right) the agreement of the data points above $q_4\approx 2 \pi T$, as expected, is even better than at low spatial momenta. At vanishing $q_4$ we again see a clear difference between the values of $D_T$. However the ordering here now differs from that of the longitudinal case. At the two lowest temperatures we find that $D_T(q_4=0,|\vec{q}|)$ lies actually below that of the higher lying temperatures. A monotonous ordering is not present, the values below the deconfinement crossover transition behave qualitatively different at $q_4=0$ than those in the deconfined phase.

We also check the applicability of the assumption of $O(4)$ invariance. To this end in the left panel of Fig.\ref{Fig:TransCorr210VerifyInterpolation} we again set up a spline interpolation $\tilde{D}_T$ (solid yellow line), now of the transversal correlator $D_T$ for $\beta=2.10$ at $T=254$MeV at vanishing imaginary frequencies (topmost points) along spatial momenta $|\vec{q}|$. The orange and red curves then correspond to this interpolation evaluated according to $\tilde{D}_T(0,\sqrt{q_4^2+|\vec{q}|^2})$ at the first three finite available imaginary frequencies. We find, similarly to the longitudinal case, that for $q_4\approx2 \pi T$ the $O(4)$ ansatz works well, starting from $|\vec{q}|\approx 1.5$GeV$^2$. As expected from experience with continuum computations the $O(4)$ scaling does not work as well for the transversal part at higher imaginary frequencies and starts to deviate from the simulated data at around $|\vec{q}|=3$GeV. As in the previous subsection, we use $\tilde{D}_T$ to evaluate the correlator $D_T$ along imaginary frequencies. We find that it provides a smooth interpolation, which for the four lowest finite imaginary frequencies available on the lattice lies quite close to the actual datapoints, while it starts to deviate again close to the edge of the Brillouin zone.

\section{Reconstructed spectral functions}
\label{sec:numspecs}

In this section we present the reconstructed gluon spectral functions, based on the correlator data discussed above. 

\subsection{Longitudinal Spectra}

\begin{figure*}[th!]
\includegraphics[scale=0.3]{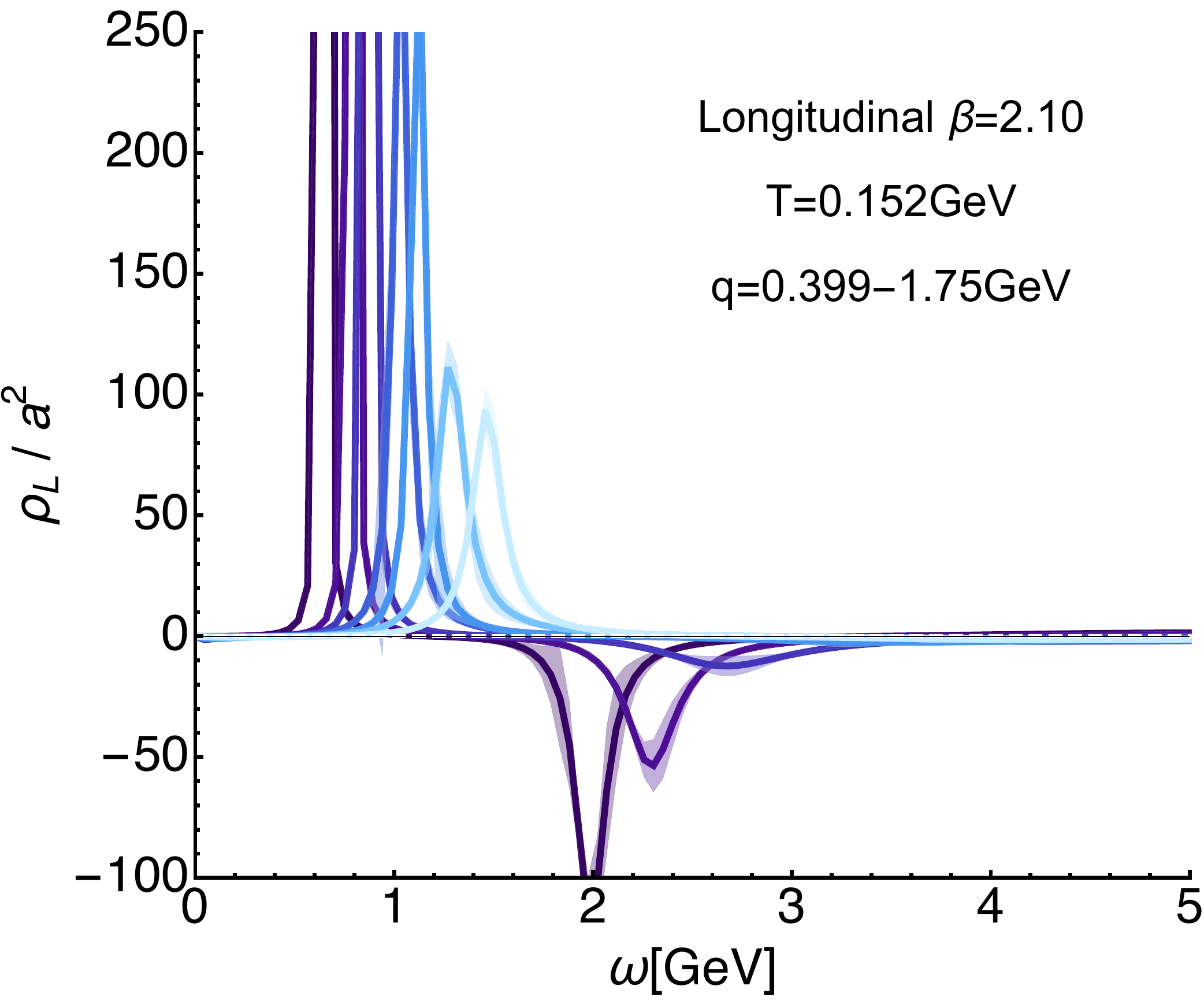}\hspace{0.5cm}
\includegraphics[scale=0.3]{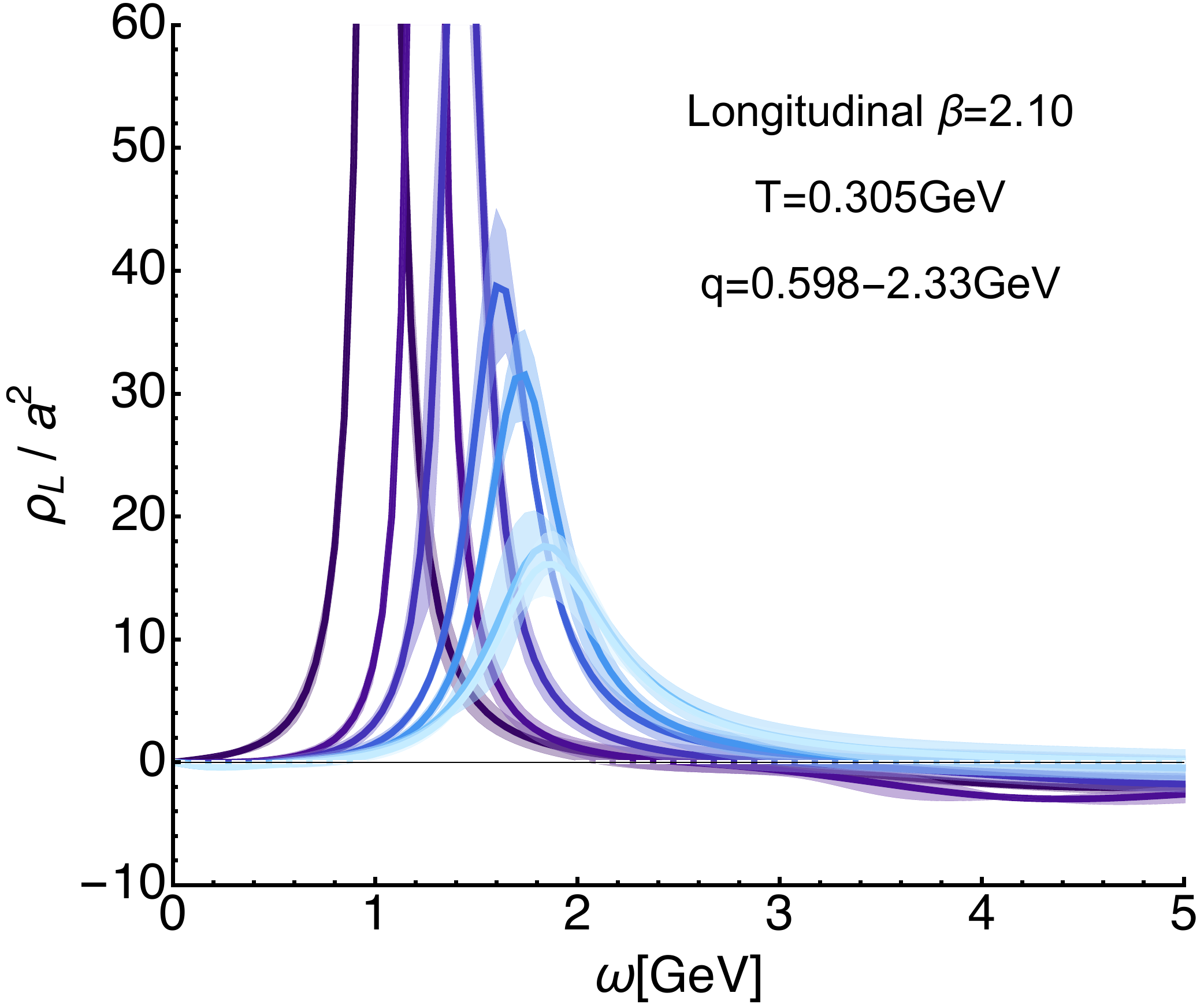}
\caption{Reconstructed longitudinal gluon spectra on the $\beta=2.10$ ensembles for (left) $T=152$MeV and (right) $T=305$MeV. The different curves denote seven of the fourteen lowest spatial momenta at which correlator data is available. The y-axis is cut off to showcase the existence of negative contributions. One can clearly see a well defined lowest lying positive peak with a subsequent trough, which dies out towards higher frequencies. Errorbands arise from varying the default model functions $m$ and $h$.}\label{Fig:LongRecMultiMom2d}
\end{figure*}

\begin{figure*}[th!]
\includegraphics[scale=0.4]{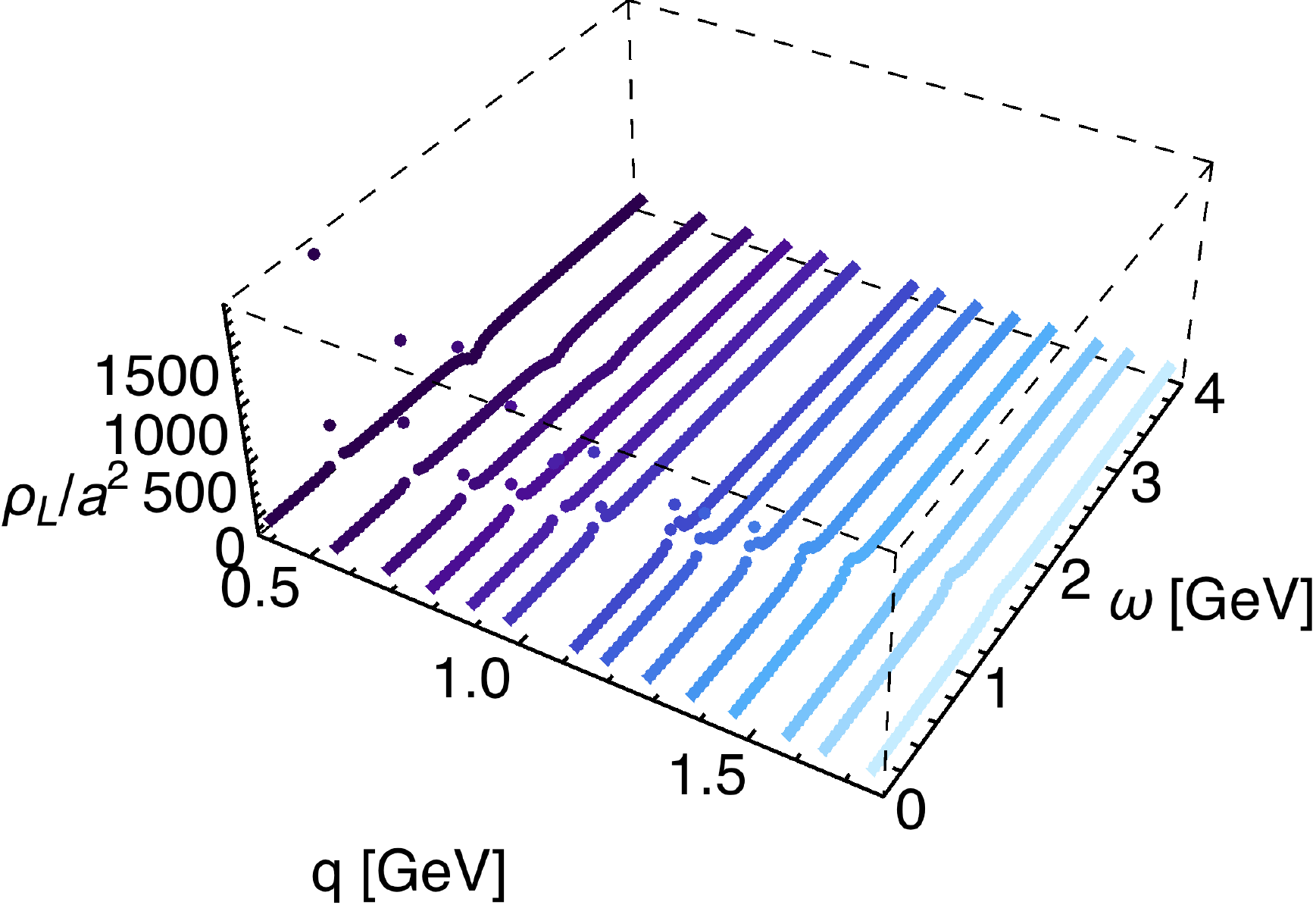}\hspace{0.5cm}
\includegraphics[scale=0.4]{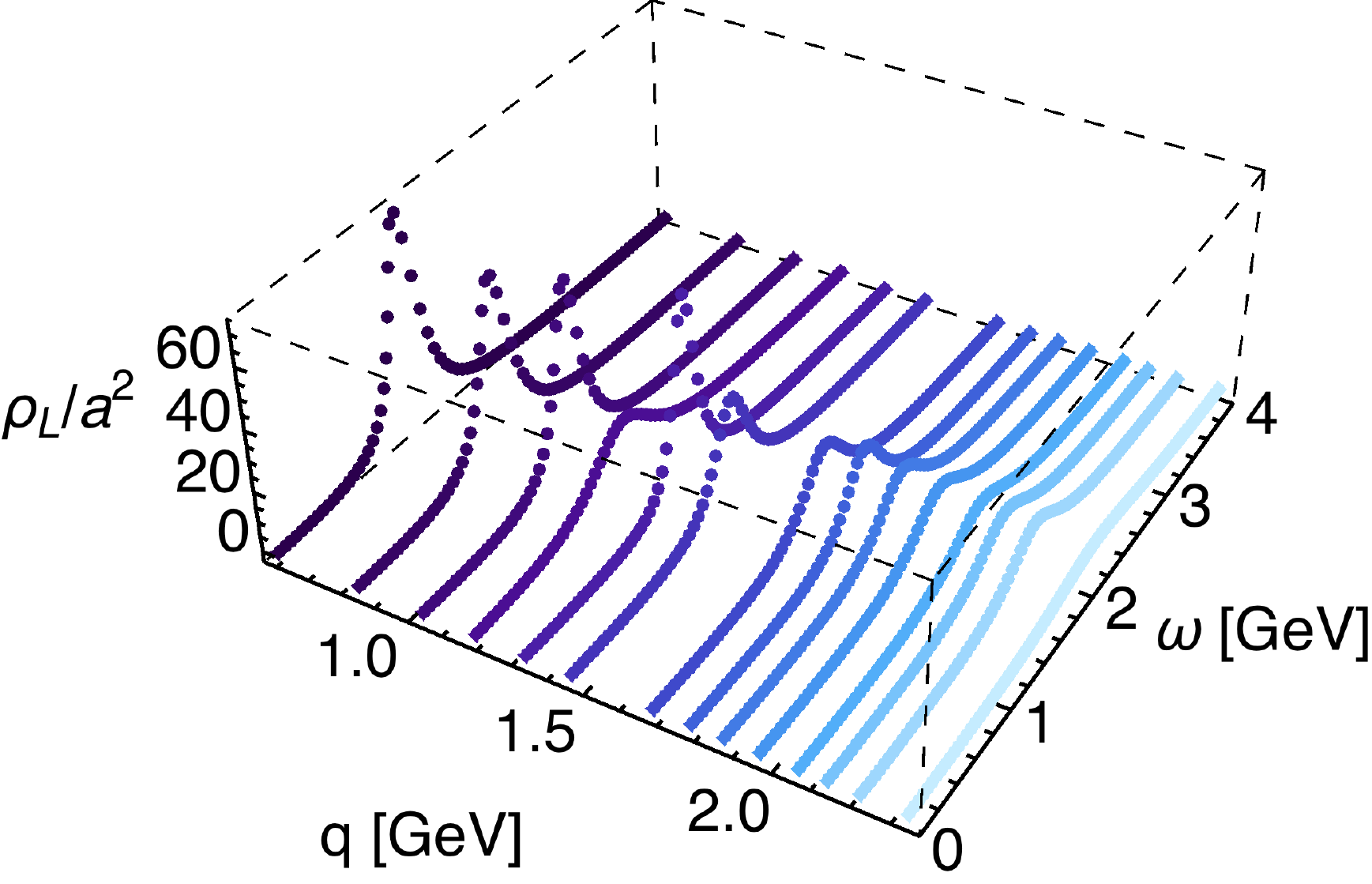}
\caption{Three dimensional visualization of the momentum dependence of the reconstructed longitudinal gluon spectra on the $\beta=2.10$ ensembles for (left) $T=152$MeV and (right) $T=305$MeV. The different curves denote the fourteen lowest spatial momenta at which correlator data is available. A clear dependence of the quasi-particle peak on $q=|\vec{q}|$ is visible.}\label{Fig:LongRecMultiMom3d}
\end{figure*}
\begin{figure*}[th!]
\includegraphics[scale=0.4]{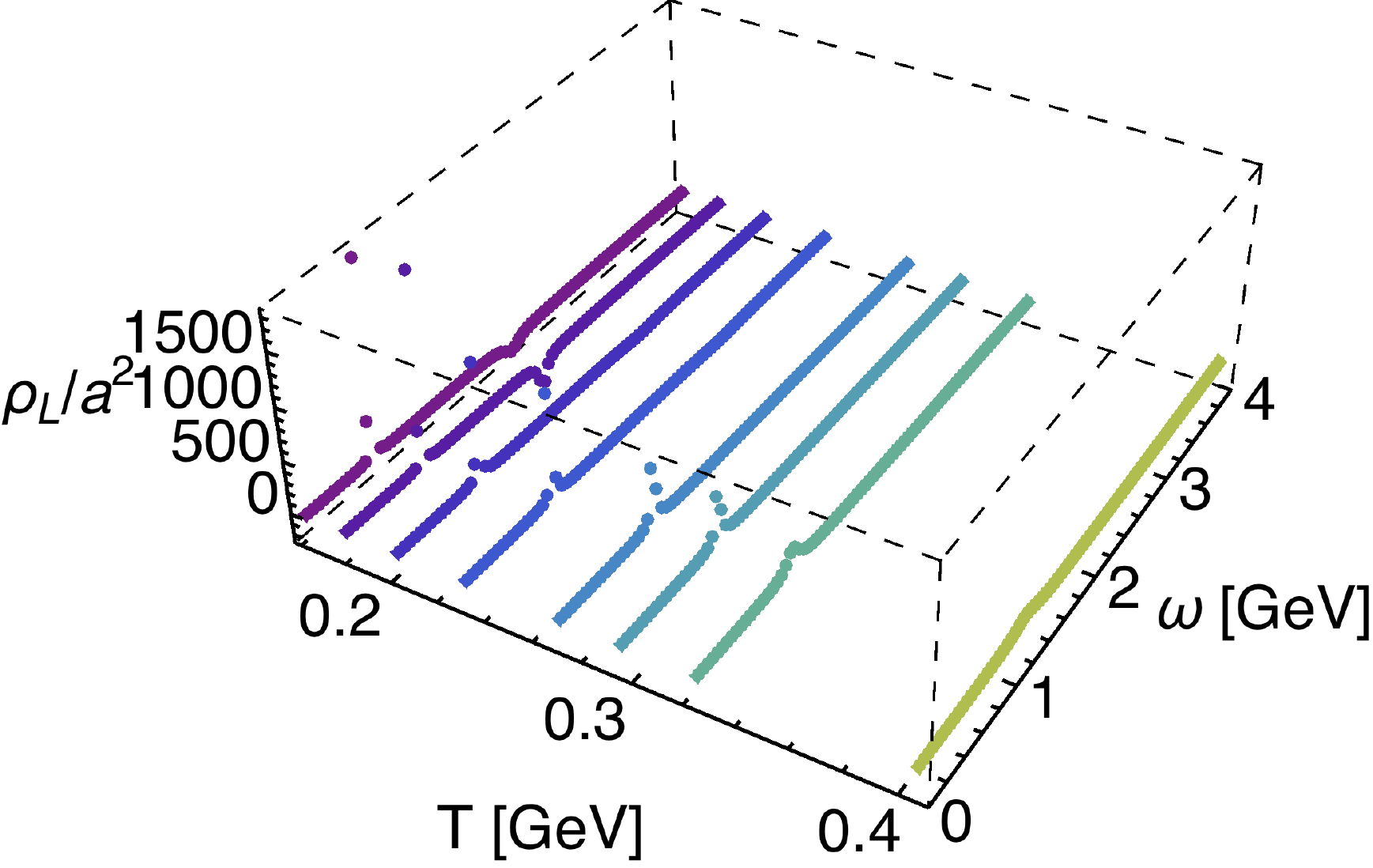}\hspace{0.5cm}
\includegraphics[scale=0.4]{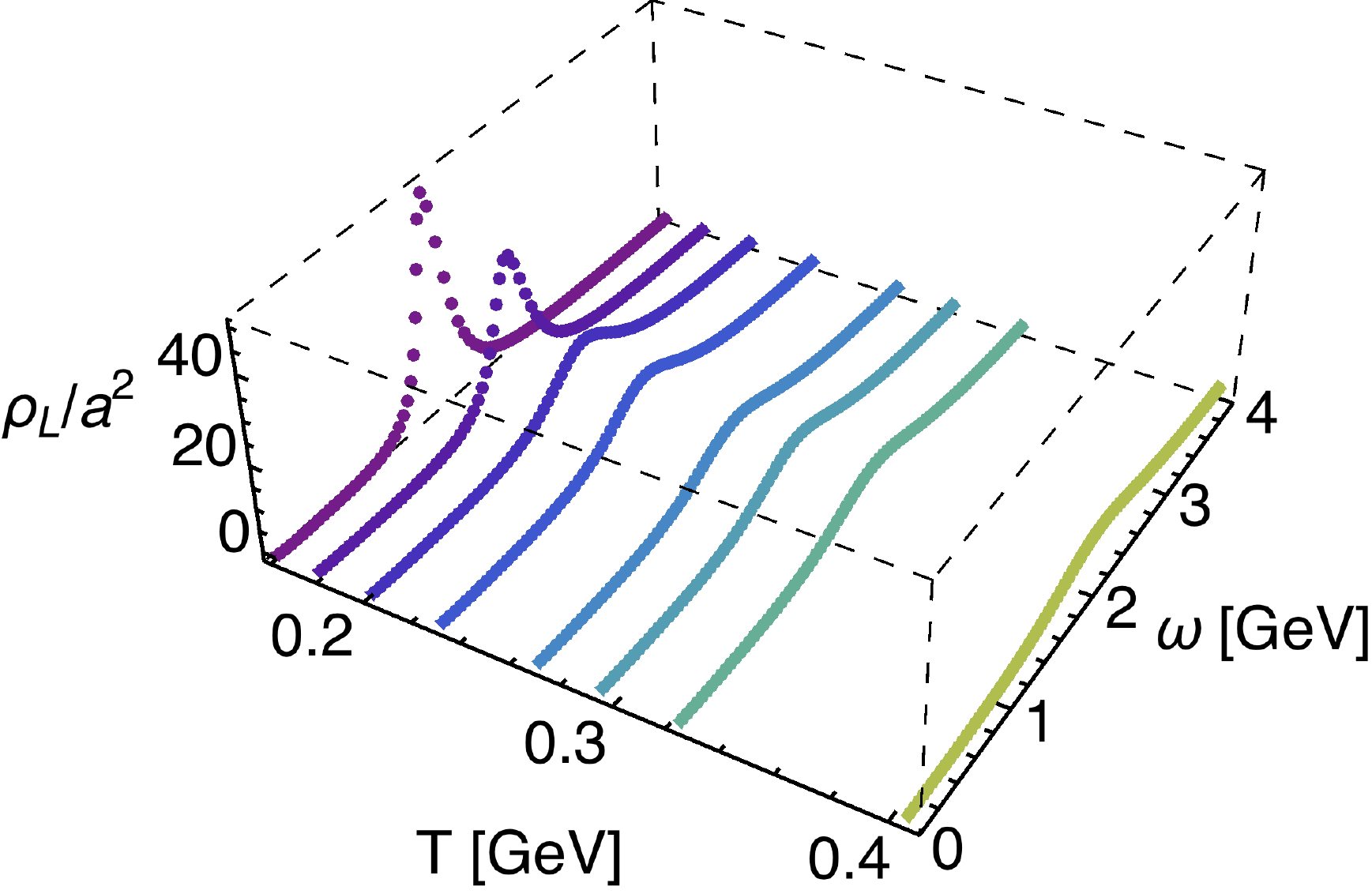}
\caption{Three dimensional visualization of the temperature dependence of the reconstructed longitudinal gluon spectra on the $\beta=2.10$ ensembles for (left) the lowest and (right) highest available spatial momentum. The different curves denote the eight lowest temperatures at which the reconstruction exhibited reasonably small default model dependence.}\label{Fig:LongRecMultiT3d}
\end{figure*}

In Fig.\ref{Fig:LongRecMultiMom2d} the reconstructed longitudinal gluon spectra at low (left, $T=152$MeV) and high (right, $T=305$MeV) temperature are shown with each curve corresponding to a different value of spatial momentum. Since the extent of the Brillouin zone depends on the physical lattice spacing, the available spatial momenta $q=|\vec{q}|$ are denoted in the figure. At the lowest momenta available, the spectrum exhibits a characteristic peak-trough structure. A well defined lowest lying positive peak is followed by a negative valley, which approaches the x-axis from below. The amplitude of both structures diminishes as one increases momenta, as well as temperature. Their position is clearly correlated with spatial momentum, which we will study quantitatively below.

The large frequency behavior is qualitatively consistent with the asymptotic form obtained in continuum computations. We however note that due to the ill-posed nature of the reconstruction task the spectra can actually show oscillations around the x-axis with a monotonously decreasing amplitude.

The magnitude of the negative contributions to the longitudinal gluon spectra appear to show a mild dependence on the lattice spacing. On coarser lattices, e.g. at $\beta=1.90$ for which spectra are shown in Appendix \ref{sec:AppACorrAndRecSpec} in Fig.\ref{Fig:SpectraLong190}, the trough in the confined phase is less strongly pronounced than for $\beta=2.10$ (keeping in mind the slightly different temperatures and momenta). On the other hand in the quark-gluon plasma phase the positivity violation seems to remain stronger the farther away we are from the continuum limit. 

To obtain a better impression of the momentum dependence of the positive peak and its height, we plot in Fig.\ref{Fig:LongRecMultiMom3d} a three-dimensional visualization of the reconstructions. They are given along both frequencies and spatial momenta at the same two temperatures as in Fig.\ref{Fig:LongRecMultiMom2d}, i.e. on the left $T=152$MeV and on the right $T=305$MeV. An increase of the peak position to higher frequencies, as well as its decreasing amplitude with rising momenta is clearly visible.

\begin{figure*}[th!]
\includegraphics[scale=0.55]{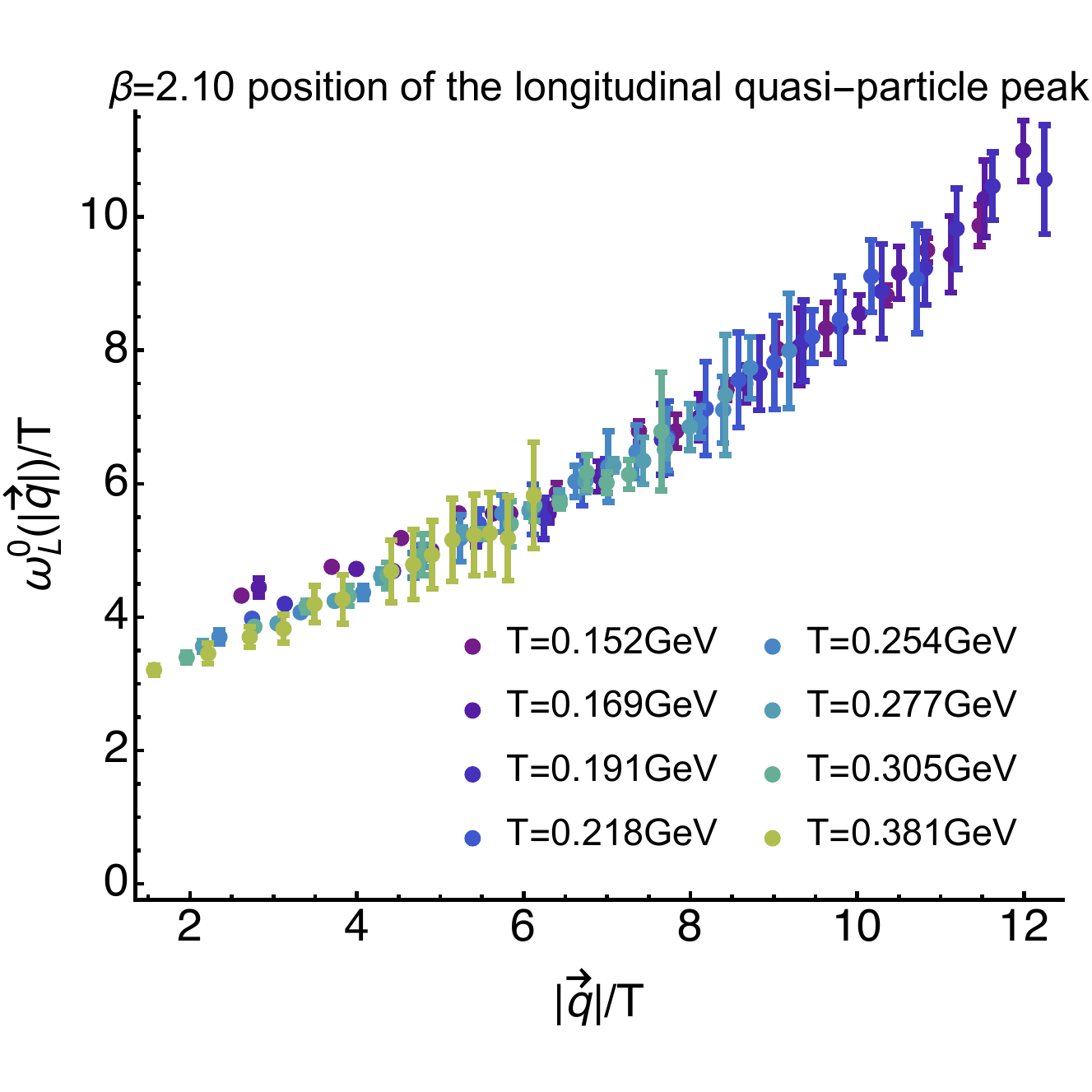}
\includegraphics[scale=0.55]{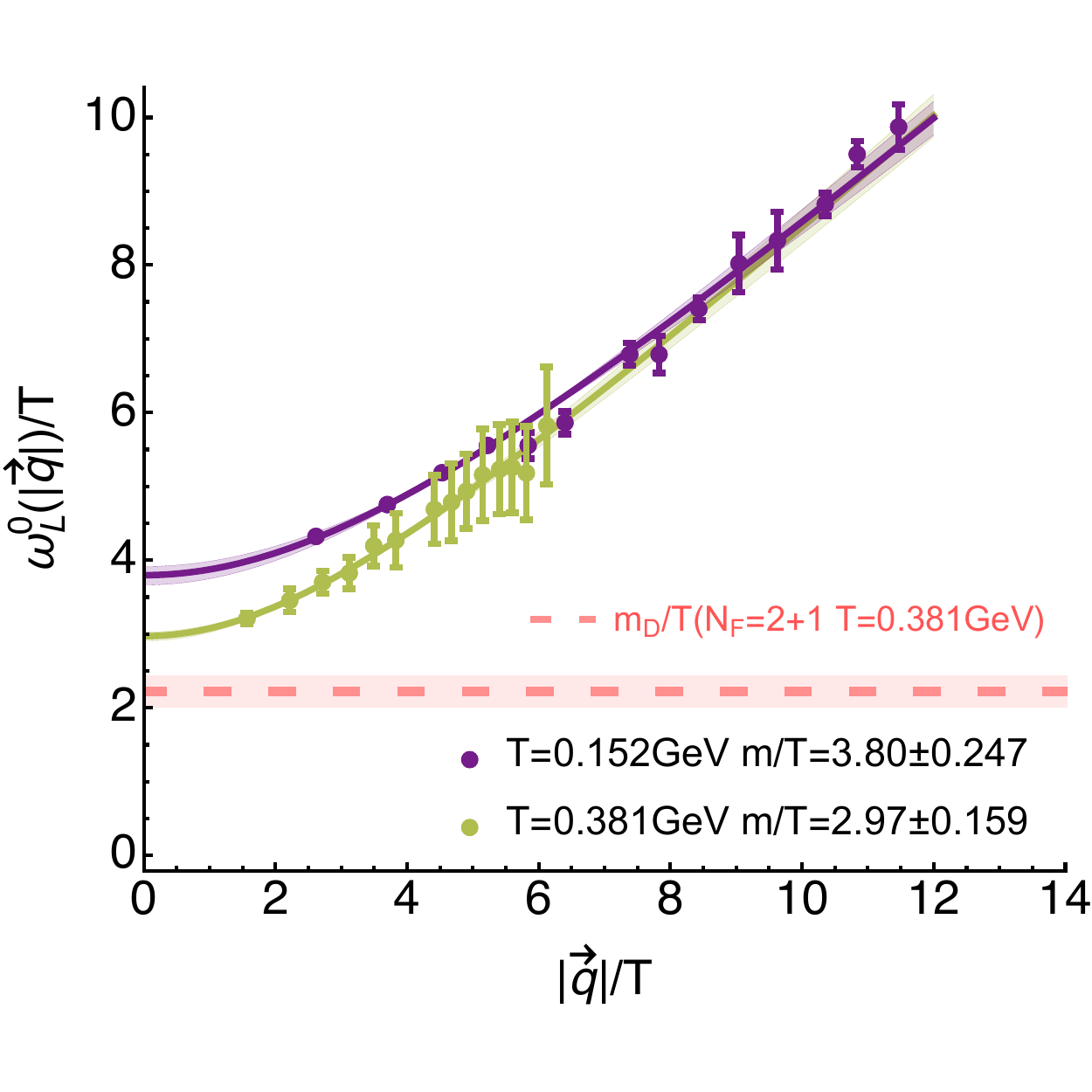}
\caption{(left) Momentum dependence of the longitudinal quasi-particle peak position at $\beta=2.10$, which for small momenta takes on a non-zero value. The reconstructions at the two lowest temperatures (within the hadronic phase) seem to exhibit a slightly larger intercept than the curves in the deconfined phase. (right) Fit of the lowest and highest temperature curves with the ansatz $\omega_L^0(|\vec{q}|)=A\sqrt{B^2+|\vec{q}|^2}$. The mass of the quasiparticle excitation is defined as the value of the intercept $m=AB$. The Debye mass evaluated from the heavy-quark potential in $N_f=2+1$ lattice QCD is given as reference as red dashed line. We note that the behavior at small $T$ is compatible not only with a finite intercept but also with a diverging functional form. This uncertainty needs to be resolved in a future study.}\label{Fig:LongQuasiParticlePos}
\end{figure*}

For completeness we give another visualization in Fig.\ref{Fig:LongRecMultiT3d}. There the reconstructions are plotted against frequencies and temperature. The left panel corresponds to the lowest available spatial momentum at each temperature, while the right panel to the highest of the fourteen considered ones.

Having inspected the behavior of the reconstructed spectra qualitatively, we proceed to quantitatively determine properties of the first positive peak. This structure is of interest, as its position has been interpreted as defining the dispersion relation $\omega^0_L(|\vec{q}|)$ of a gluon quasi-particle, which in turn may become part of a dynamical model of the quark-gluon plasma. In the left panel of Fig.\ref{Fig:LongQuasiParticlePos} we show the peak positions, defined naively via its topmost point, plotted against spatial momentum for those eight temperatures at which the default model dependence was mild enough for a robust determination. Both the peak position, as well as momenta are rescaled by the temperature, which allows a straight forward comparison of non-trivial differences in the behavior of $\omega_L^0$ at different temperatures. The errorbars arise from the variation of the results among changing both $m(\omega)$ and $h(\omega)$ as described in Sec.\ref{Sec:Bayes}.

We find that all curves approach the y-axis at a non-zero value and that above $|\vec{q}|/T\approx6$GeV all of them exhibit an identical behavior within the relatively large systematic errorbars. As the number of correlator points reduces with increasing temperature the reliability of the reconstruction also decreases concurrently. In turn we are not able to distinguish differences among the peak positions in the deconfined phase at low momenta. On the other hand in the confined phase, i.e. for the lowest two temperatures, the peak position seems to flatten off at a value above that in the deconfined phase. This difference is probed more quantitatively in the right panel of Fig.\ref{Fig:LongQuasiParticlePos}, where the peak positions for $T=0.152$GeV$<T_c$ and $T=0.381$GeV$>T_c$ are plotted together with a modified free theory fit $\omega_L^0(|\vec{q}|)=A\sqrt{B^2+|\vec{q}|^2}$. This simple fit ansatz manages to retrace the values reasonably well and leads to significantly different intercepts, i.e. quasi-particle masses between the lowest and highest temperature:
\begin{align}
&\left. m_L/T\right|_{T=0.152{\rm GeV}}=3.80\pm0.25\label{LMassConf}\\
&\left. m_L/T\right|_{T=0.381{\rm GeV}}=2.97\pm0.16\label{LMassDeConf}
\end{align}
For comparison purposes we also display (red dashed curve) a recent lattice QCD determination of the Debye mass with $N_f=2+1$ flavors of light HISQ quarks \cite{Burnier:2015tda,Burnier:2015nsa}, which at $T=0.381$GeV coincidentally agrees within errors with the HTL value of $m_D$ evaluated for four massless flavors at the scale $\mu=2\pi T$. Another possible behavior for the peak position in the confined phase that has been discussed in the literature is that of a divergence towards the origin. With the currently available lattice box sizes the momentum spacing, in particular for $T<T_c$, does not yet allow us to distinguish the two cases. A future study on larger physical box sizes will be necessary to resolve this question.

\begin{figure*}[th!]
\includegraphics[scale=0.3]{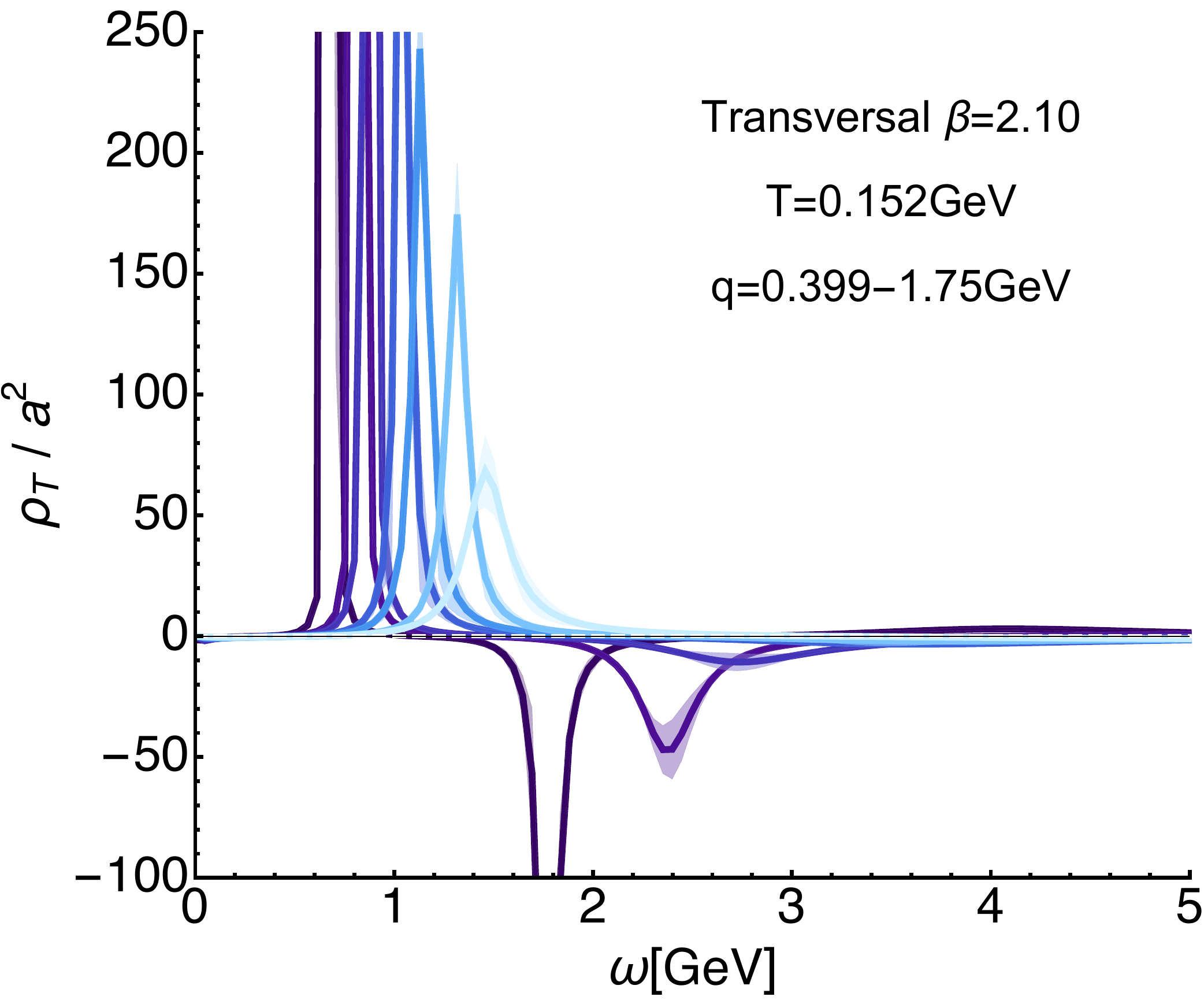}\hspace{0.5cm}
\includegraphics[scale=0.3]{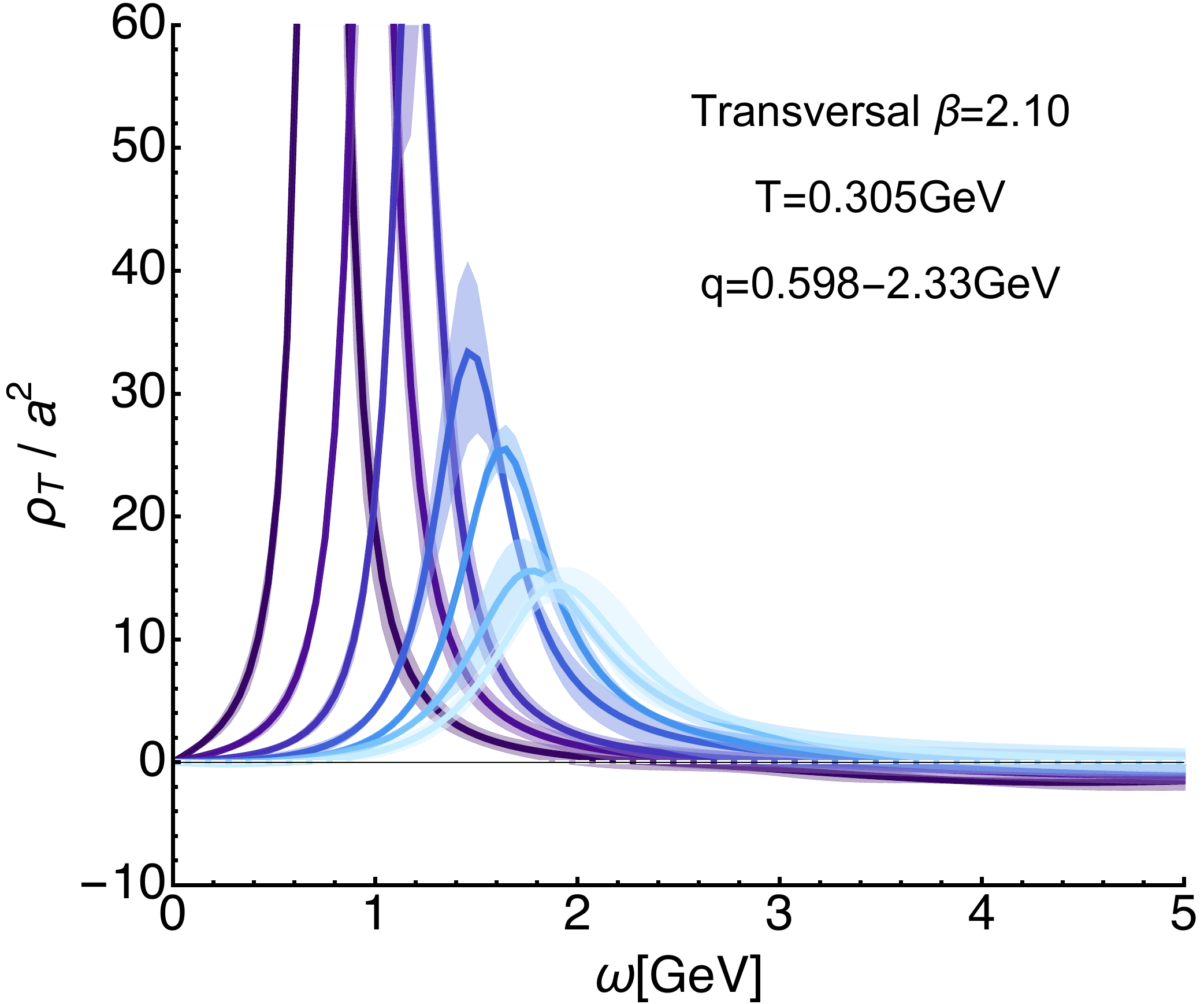}
\caption{Reconstructed transversal gluon spectra on the $\beta=2.10$ ensembles for (left) $T=152$MeV and (right) $T=305$MeV. The different curves denote seven of the fourteen lowest spatial momenta at which correlator data is available. The y-axis is cut off to showcase the existence of negative contributions. One can clearly see at well defined lowest lying positive peak with a subsequent trough, which dies out towards higher frequencies. Errorbands arise from varying the default model functions $m$ and $h$.}\label{Fig:TransRecMultiMom2d}
\end{figure*}

\begin{figure*}[th!]
\includegraphics[scale=0.4]{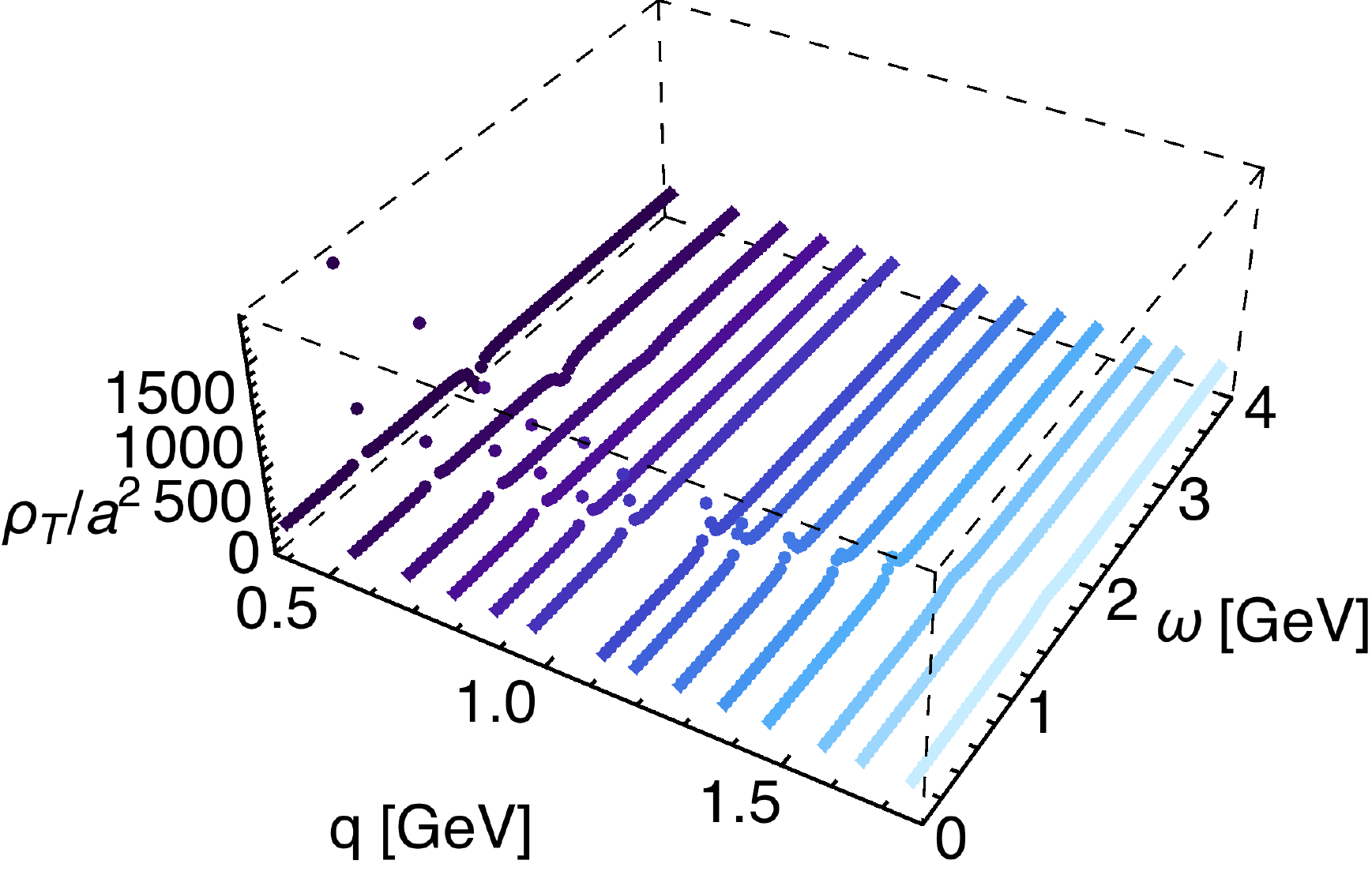}\hspace{0.5cm}
\includegraphics[scale=0.4]{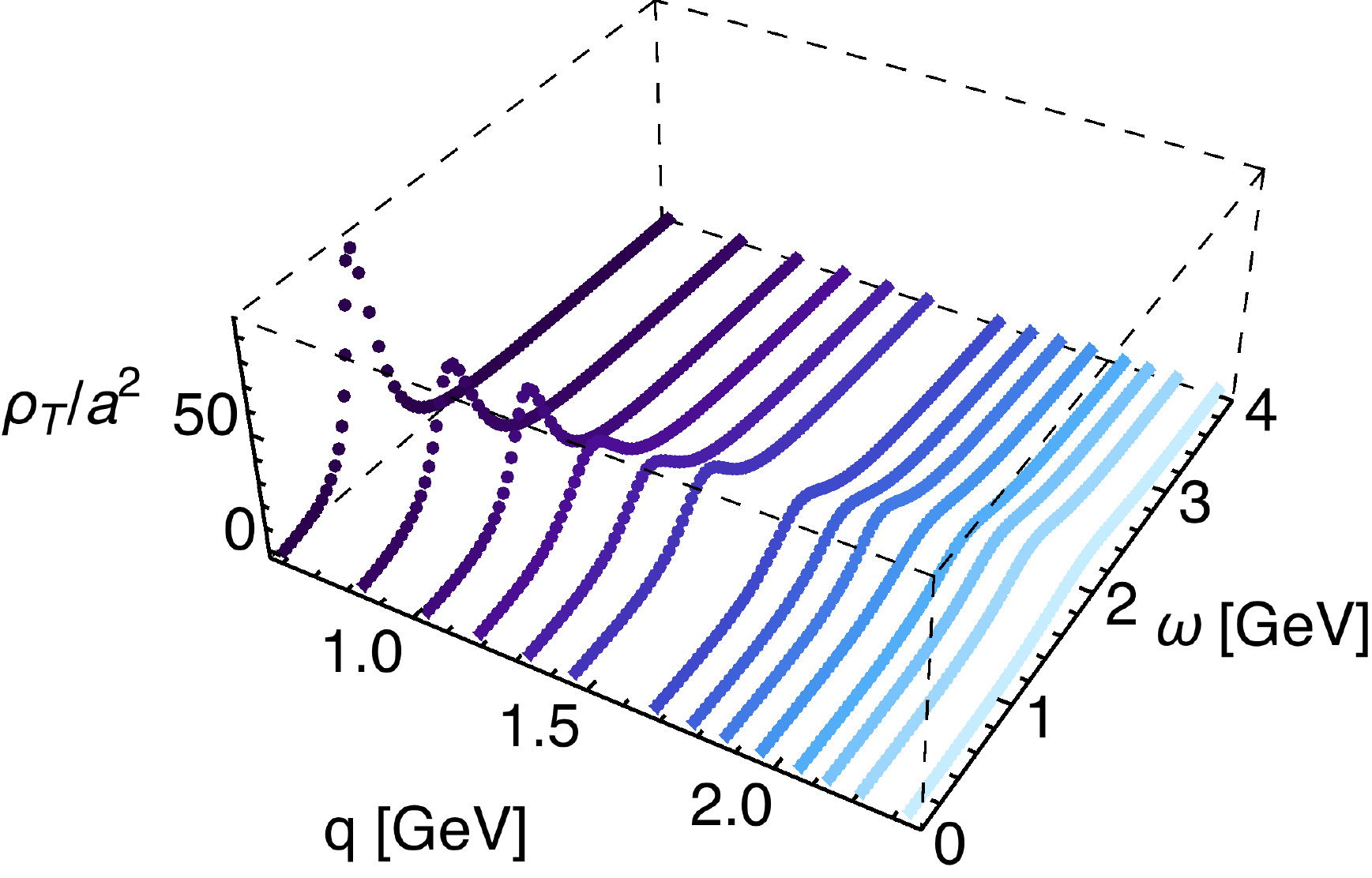}
\caption{Three dimensional visualization of the momentum dependence of the reconstructed transversal gluon spectra on the $\beta=2.10$ ensembles for (left) $T=152$MeV and (right) $T=305$MeV. The different curves denote the fourteen lowest spatial momenta at which correlator data is available. A clear dependence of the quasi-particle peak on $q=|\vec{q}|$ is visible.}\label{Fig:TransRecMultiMom3d}
\end{figure*}
\begin{figure*}[th!]
\includegraphics[scale=0.4]{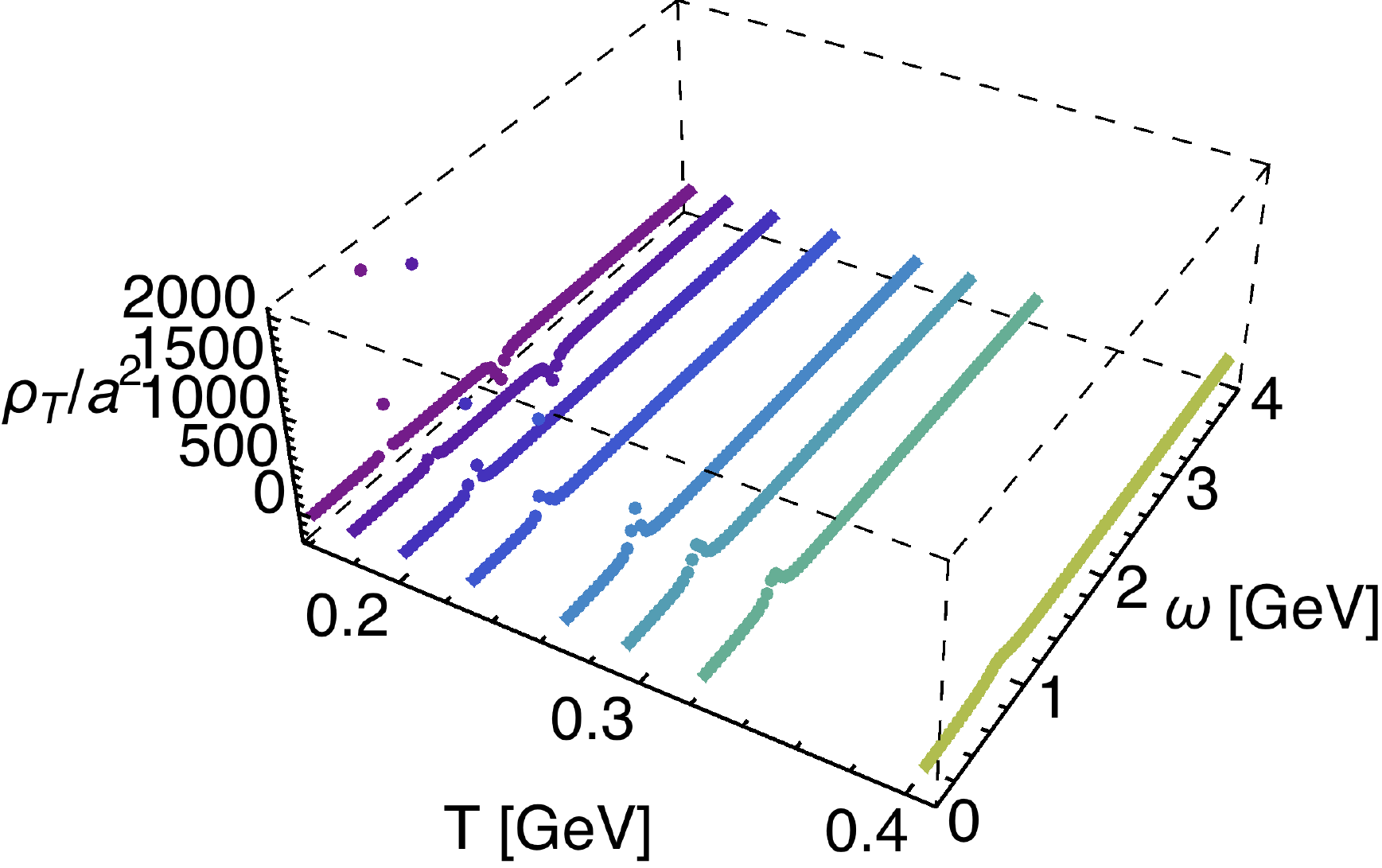}\hspace{0.5cm}
\includegraphics[scale=0.4]{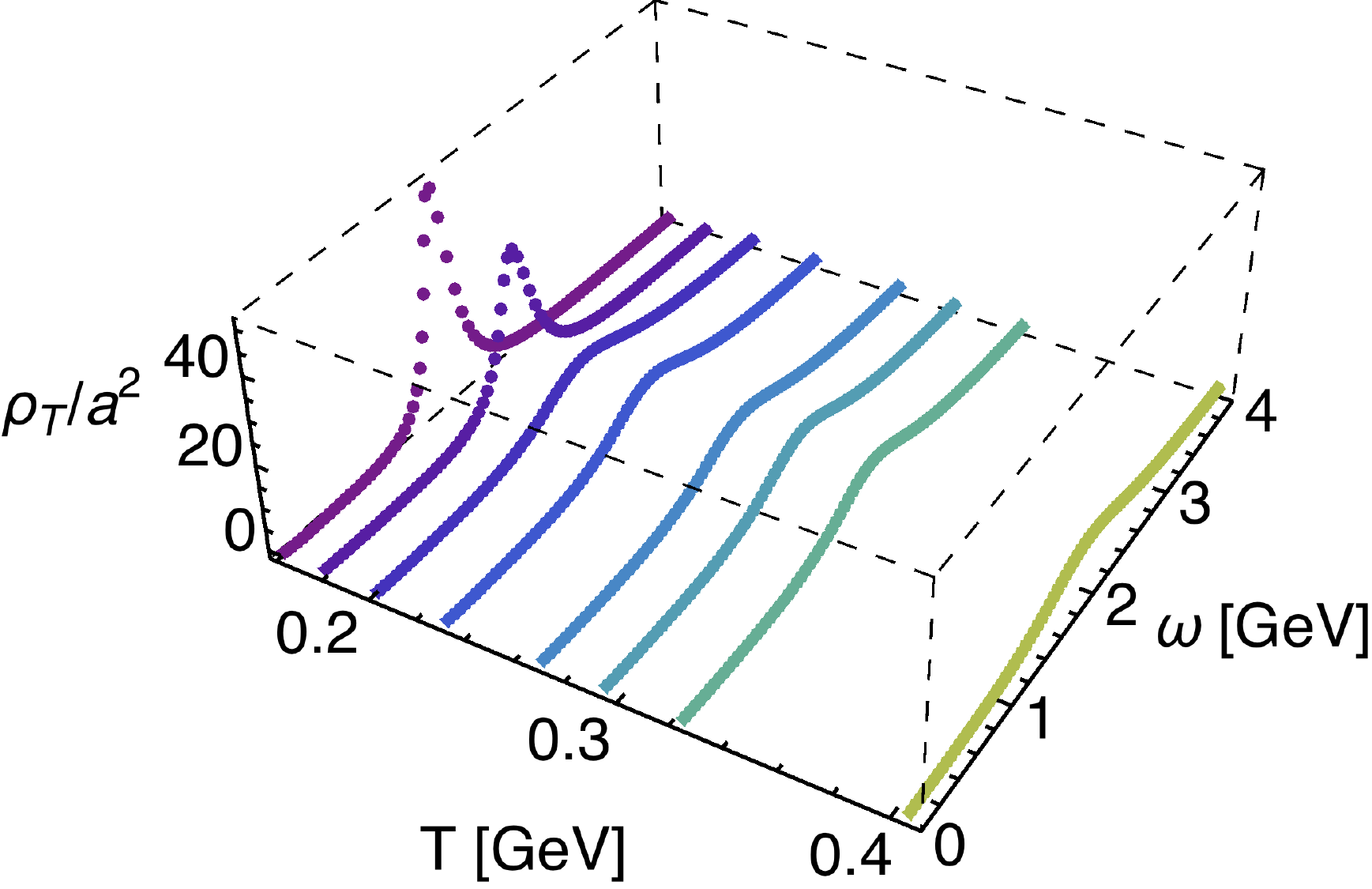}
\caption{Three dimensional visualization of the temperature dependence of the reconstructed transversal gluon spectra on the $\beta=2.10$ ensembles for (left) the lowest and (right) highest available spatial momentum. The different curves denote the eight lowest temperatures at which the reconstruction exhibited reasonably small default model dependence.}\label{Fig:TransRecMultiT3d}
\end{figure*}

\subsection{Transversal Spectra}

\begin{figure*}[th!]
\includegraphics[scale=0.55]{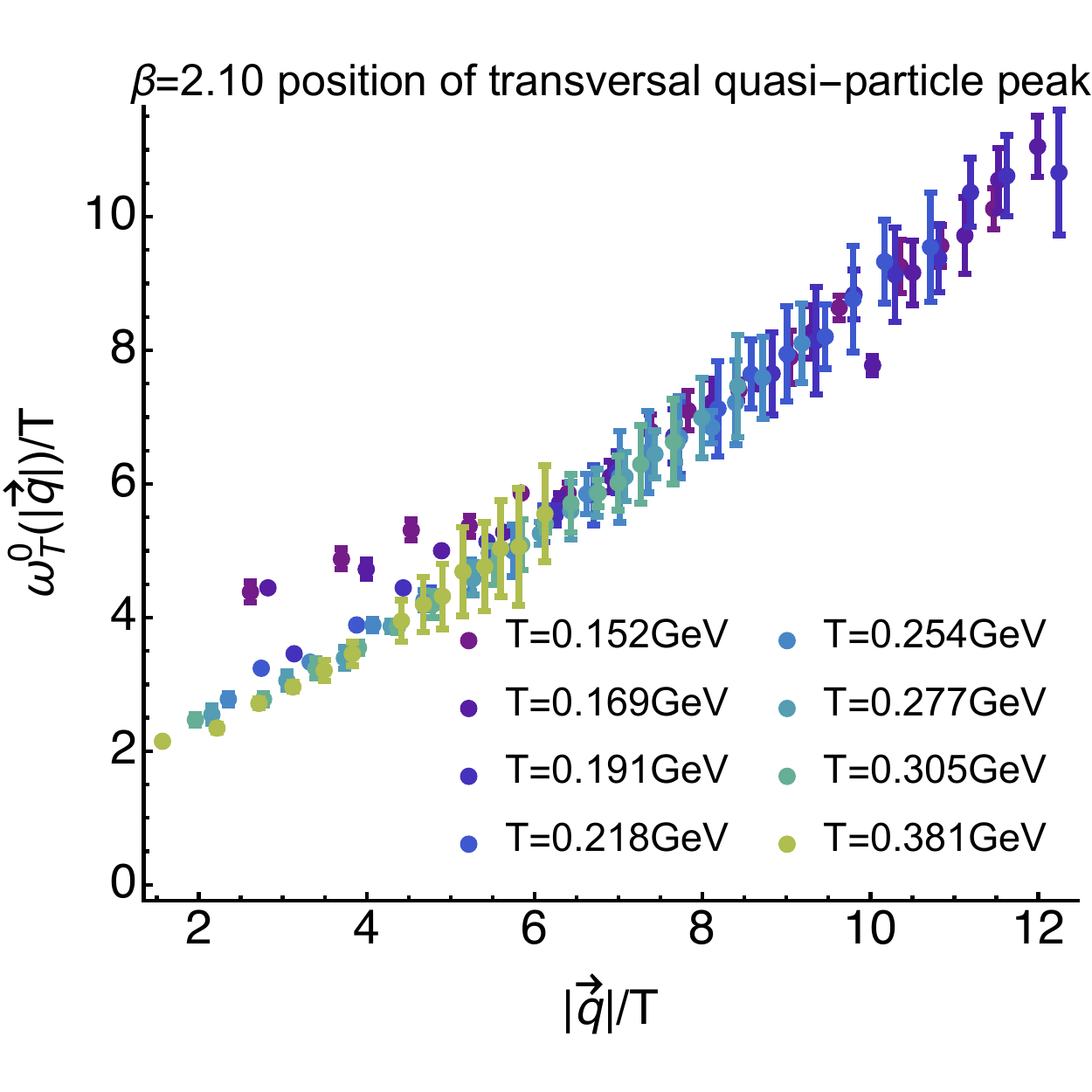}
\includegraphics[scale=0.55]{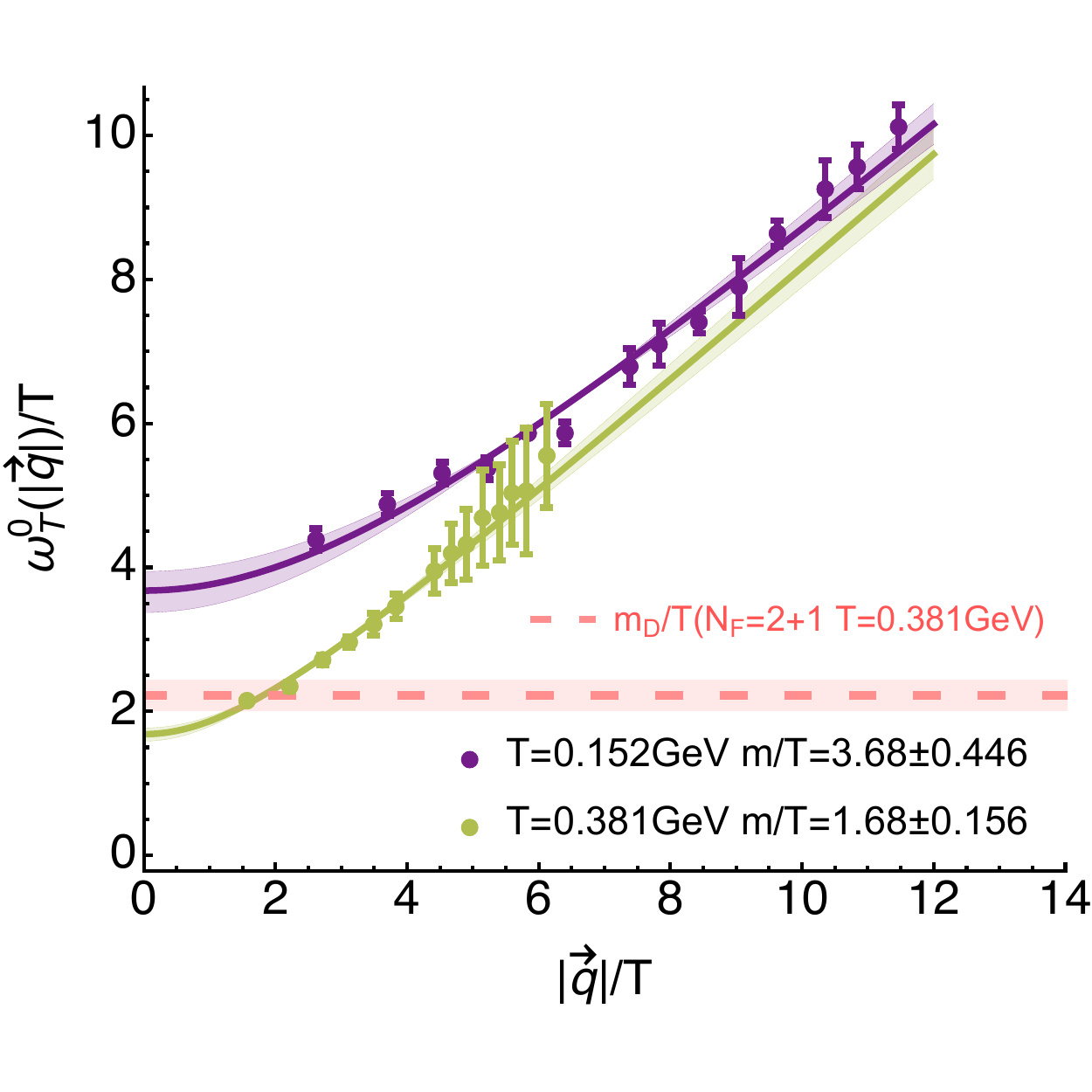}
\caption{(left) Momentum dependence of the transversal quasi-particle peak position at $\beta=2.10$, which for small momenta takes on a non-zero value. The reconstructions at the two lowest temperature within the hadronic phase seem to exhibit a larger intercept than the curves in the deconfined phase. (right) Fit of the lowest and highest temperature curves with the ansatz $\omega_L^0(|\vec{q}|)=A\sqrt{B^2+|\vec{q}|^2}$. The mass of the quasiparticle excitation is defined as the value of the intercept $m=AB$. The Debye mass evaluated from the heavy-quark potential in $N_f=2+1$ lattice QCD is given as reference as red dashed line. We note that the behavior at small $T$ is compatible not only with a finite intercept but also with a diverging functional form.}\label{Fig:TransQuasiParticlePos}
\end{figure*}
Let us now turn to the reconstructed transversal spectral functions, which we plot in Fig.\ref{Fig:TransRecMultiMom2d}. The left panel contains the spectra at low temperature $T=152$MeV, while the right panel features those at $T=305$MeV. Each curve corresponds to a different value of spatial momentum $q=|\vec{q}|$, the interval probed in each case is denoted in the legend of the corresponding panel. We again find at low momenta the characteristic peak-trough structure of a well defined lowest lying positive peak, followed by a negative valley, which approaches the x-axis from below. Compared to the longitudinal case the amplitude of the negative trough appears to be larger. Artificial numerical oscillations around the x-axis at high frequencies are unfortunately also present.

While the effect is less pronounced than in the longitudinal case, we again see indications for a mild dependence of the negative contributions on the lattice spacing. At the coarser $\beta=1.90$ for which spectra are shown in Fig.\ref{Fig:SpectraTrans190} of Appendix \ref{sec:AppACorrAndRecSpec}, the trough in the confined phase is slightly less strongly developed than for $\beta=2.10$ (keeping in mind the slightly different temperatures and momenta). In the quark-gluon plasma phase the positivity violation seems to remain stronger, farther away from the continuum limit. 

We give here the same three dimensional visualizations of the reconstructions as in the longitudinal case to obtain a better impression of the momentum dependence of the peak and its height. In Fig.\ref{Fig:TransRecMultiMom3d} the spectra are plotted along both frequencies and spatial momenta at the same two temperatures as in Fig.\ref{Fig:TransRecMultiMom2d}, i.e. on the left $T=152$MeV and on the right $T=305$MeV. An increase of the peak position to higher frequencies, as well as its decreasing amplitude with rising momenta is clearly visible.

On the other hand in  Fig.\ref{Fig:TransRecMultiT3d} the reconstructions are plotted against frequencies and temperature. The left panel corresponds to the lowest available spatial momentum at each temperature, while the right panel to the highest of the fourteen considered ones.

What remains is to determine the momentum and temperature dependence of the quasi-particle peak in the transversal sector. In the left panel of Fig.\ref{Fig:TransQuasiParticlePos} we give the corresponding peak positions plotted against spatial momentum for those eight temperatures at which the default model dependence was mild enough for a robust determination. Again both the peak position as well as momenta are rescaled by the temperature and the errorbars arise from the variation of the results among changing both $m(\omega)$ and $h(\omega)$.

We find that, qualitatively consistent with the longitudinal case, all curves approach the y-axis at a non-zero value. Also above $|\vec{q}|/T \approx 6$ they exhibit the same behavior within systematic errors. At low momenta similarities, as well as differences to the longitudinal case can be observed. On the one hand, the behavior in the confined phase appears to be consistent with that of the longitudinal spectra. On the other hand above $T_c$ the longitudinal peak positions move to significantly lower values. Thus for $|\vec{q}|/T<6$ the difference between the two regimes is here more pronounced. This impression from a visual inspection is confirmed by the fits carried out on the right panel of Fig.\ref{Fig:TransQuasiParticlePos}. There the peak positions for $T=0.152$GeV$<T_c$ and $T=0.381$GeV$>T_c$ are plotted together with a modified free theory fit $\omega_L^0(|\vec{q}|)=A\sqrt{B^2+|\vec{q}|^2}$. The values for the quasi-particle mass, given by the intercepts
\begin{align}
&\left. m_T/T\right|_{T=0.152{\rm GeV}}=3.68\pm0.45\label{TMassConf}\\
&\left. m_T/T\right|_{T=0.381{\rm GeV}}=1.68\pm0.16\label{TMassDeConf}
\end{align}
agree with the longitudinal masses of \eqref{LMassConf} in the confined phase but differ significantly in the deconfined regime. 

This difference in in-medium masses at high temperature is qualitatively consistent with the expectations from weak coupling considerations. In a perturbative setting the strong couling $\alpha$ is small and the electric scale of Debye screening $\sim \alpha T$ is expected to be well separated from the non-perturbative magnetic sector $\sim \alpha^2 T$. The corresponding magnetic in-medium mass therefore will be smaller than its electric Debye counterpart. 

Also here we display the lattice QCD determination of the Debye mass with $N_f=2+1$ flavors of light HISQ quarks \cite{Burnier:2015tda,Burnier:2015nsa} as red dashed curve.  The possibility that the $T<T_c$ dispersion relation will actually diverge as one approaches the y-axis again cannot be excluded from the momenta currently available. 

\section{Conclusion}
\label{sec:Conclusion}

We have presented the first computation of finite temperature gluon correlation functions in Landau gauge on full QCD ensembles with $N_f=2+1+1$ flavors of dynamical quarks, generated by the tmfT collaboration and gauge fixed using the cuLGT library on GPU's. Based on these data we carried out a Bayesian investigation of gluon spectral functions, the first being independent of the assumption of $O(4)$ invariance, as well as with a systematic error budget. Spectral function reconstructions were performed with a novel Bayesian approach, which generalizes the BR method to arbitrary, i.e. non positive-definite functions.

As expected, we found (see Fig.~\ref{Fig:LongCorr210VsMuDiffTmp} and Fig.~\ref{Fig:TransCorr210VsMuDiffTmp}) on the level of correlators that for a fixed momentum, at imaginary frequencies around the first Matsubara frequency $q_4\approx 2\pi T$ the computed values are already very close to the $T\approx0$ behavior observed at higher $q_4$. At $q_4=0$ on the other hand significant differences between the correlators are manifest. Unfortunately at $q_4=0$ the correlator will suffer most severely from the inherent finite extent of the Euclidean axis in standard lattice simulations. We have checked (see Fig.~\ref{Fig:LongCorr210VerifyInterpolation} and Fig.~\ref{Fig:TransCorr210VerifyInterpolation}) that while interpolating the correlators using the assumption of $O(4)$ invariance works reasonably well at small $q_4$, it degrades towards the boundaries of the Brillouin zone. The interpolation is found to work better on the longitudinal correlators than on the transversal ones.

The reconstructed spectral functions (see Fig.~\ref{Fig:LongRecMultiMom2d} and Fig.~\ref{Fig:TransRecMultiMom2d}) both in the longitudinal and transversal sector show clear signs of positivity violation at high frequencies. In general we find one well defined positive peak structure at low frequencies, followed by a negative trough at higher $\omega$. At higher frequencies, the spectrum approaches the frequency axis from below, qualitatively similar to the expected asymptotic behavior. Due to the imperfections of the reconstruction process at high frequencies the spectra can begin to artificially oscillate around the $\omega$-axis with a diminishing amplitude.

Since a well pronounced positive peak at low frequencies was identified in all reconstructed spectra, we use its tip as a naive definition of a quasi-particle dispersion relation. The corresponding values plotted against spatial momentum (see Fig.~\ref{Fig:LongQuasiParticlePos} and Fig.~\ref{Fig:TransQuasiParticlePos}) are compatible with a finite intercept at $|\vec{q}|=0$. Such an intercept, determined from a modified free-theory fit, can give a first rough estimate of the quasi-particle mass in the longitudinal and transversal sector (see \eqref{LMassDeConf} and \eqref{TMassDeConf}). We find that below $T_c$ the values of this mass appear to be consistent with each other for longitudinal and transversal gluons, while above the deconfinement transition a clear separation of values emerges. This difference is qualitatively consistent with the weak coupling expectation that the magnetic mass should be parametrically smaller than the Debye mass which screens the electric fields. The possibility that at $T\approx 0$ the dispersion relation diverges cannot be excluded with our current data, where the minimum available value of $|\vec{q}|/T$ remained relatively large.

Our findings are encouraging: it appears possible to reconstruct characteristic features of gluon spectral functions from Landau gauge lattice QCD correlators with a relatively small number of available frequency points, since the statistics of the ensembles is high. The position of the lowest lying peak structure is one example. To connect to the perturbative high momentum regime, where the signal to noise ratio in the correlators is still weak will require increasing the statistics further. The quasi-particle peak width on the other hand demands simulations with significantly larger number of temporal lattice points, i.e. a smaller lattice spacing. Connecting our lattice results on gluon spectra to e.g. the PHSD framework therefore needs to be postponed to future studies. Performing the full continuum extrapolation on the correlators and subsequent reconstructions has to be attempted in a future study, as we have seen indications that e.g. parts of the negative spectral contributions still depend quantitatively on the lattice spacing.

Already with the currently available data we may attempt to use the reconstructed spectra in a self consistent computation for transport coefficients in full QCD, which is work in progress.

We are confident that with a further increase of the statistics on the tmfT ensembles and subsequently on the computed correlators the determination of the quasi-particle dispersion relation can be brought to a more robust quantitative level, in particular that it will become possible to resolve more clearly the temperature dependence of its intercept at $|\vec{q}|=0$.

One open problem left is to estimate the influence of the number of light 
quark species and of the light quark mass on the properties of the gluon 
spectral functions at temperatures below and above the thermal crossover.
Considering the ongoing work on thermodynamics for $N_f=2+1+1$ flavors, 
we will be able in near future to investigate the case of more realistic 
light quark masses (pion masses of near to 200 MeV) for $N_f=2+1+1$ flavors
along the lines of the present paper.

In the further course of work we plan to compare these results with the cases 
of pure gluodynamics and two-flavor full QCD, adopting the fixed-scale approach
that has guided us in the investigations of $N_f=2+1+1$ QCD.

\vspace{-0.2cm}
\begin{acknowledgments}

  E.-M. I. and A. T. are grateful to the members of the tmfT
  collaboration for years of common work, in particular we thank
  Florian Burger who has run most of the simulations for the
  $N_f=2+1+1$ project. We thank Michel M\"uller-Preussker and
  Maria-Paola Lombardo for numerous discussions and continuous
  interest in gauge-fixed correlation functions and real-time
  approaches. Landau-gauge fixing and generation of part of the tmfT
  configurations have been performed on the ``Lomonosov''
  supercomputer of Moscow State University and on the ``HybriLIT''
  cluster of JINR. We are grateful to the MSU Supercomputer Center and
  the HybriLIT team for extensive computational resources and
  responsive support. This work is supported by EMMI, the grants
  ERC-AdG-290623, BMBF 05P12VHCTG. It is part of and supported by the
  DFG Collaborative Research Centre "SFB 1225 (ISOQUANT)".

\end{acknowledgments}

\newpage

\appendix

\begin{widetext}
\section{Correlators and Reconstructed Spectra at $\beta=1.90$ and $\beta=1.95$}
\label{sec:AppACorrAndRecSpec}

In this appendix we provide for completeness the figures for the
Landau gauge gluon correlation functions and spectral reconstructions
at the other two lattice spacings $\beta=1.90$ and $\beta=1.95$. These
exhibit a qualitatively similar behavior compared to those at
$\beta=2.10$. 
\begin{figure*}
\includegraphics[scale=0.125]{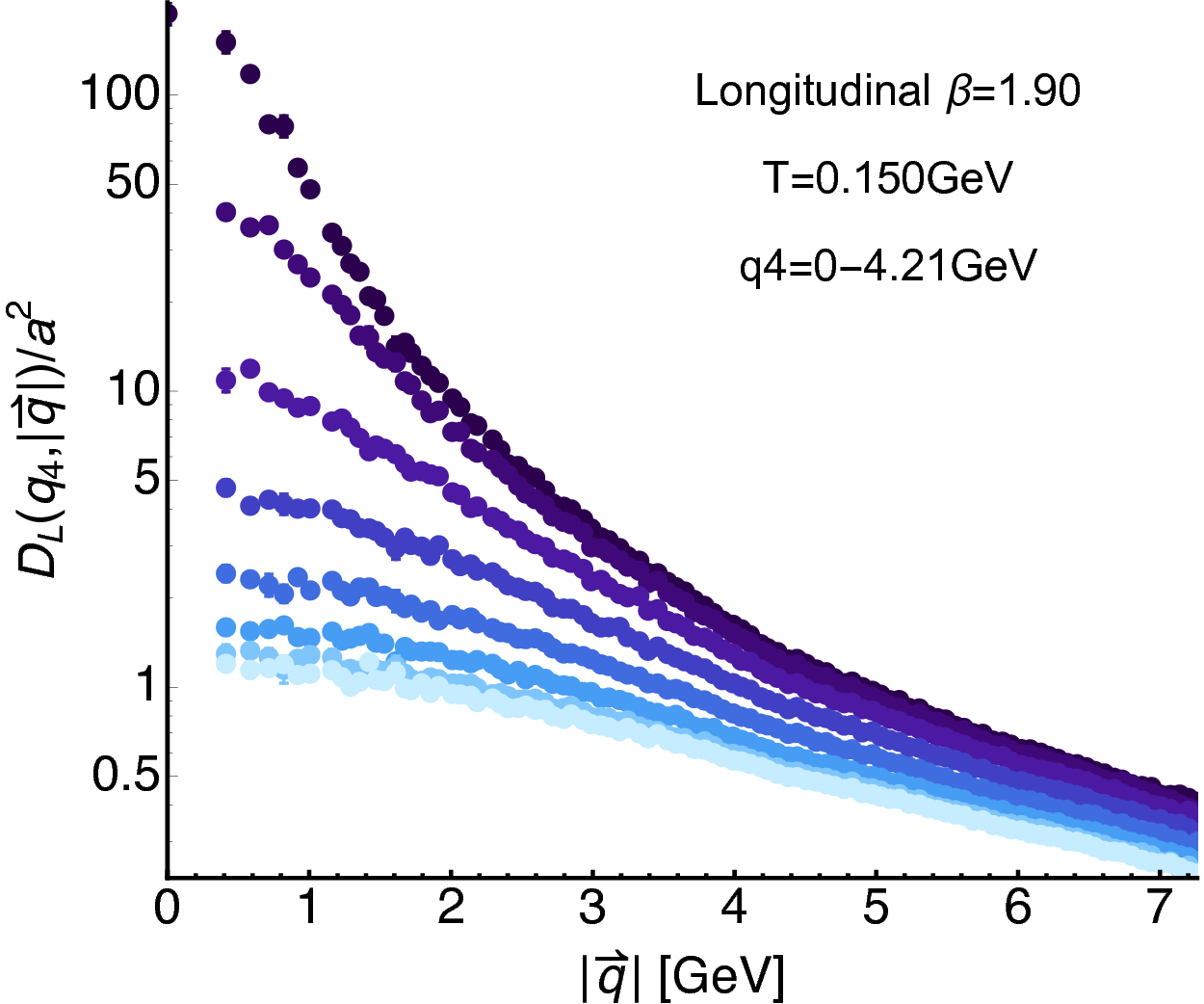}\vspace{0.1cm}\hspace{2.5cm}
\includegraphics[scale=0.125]{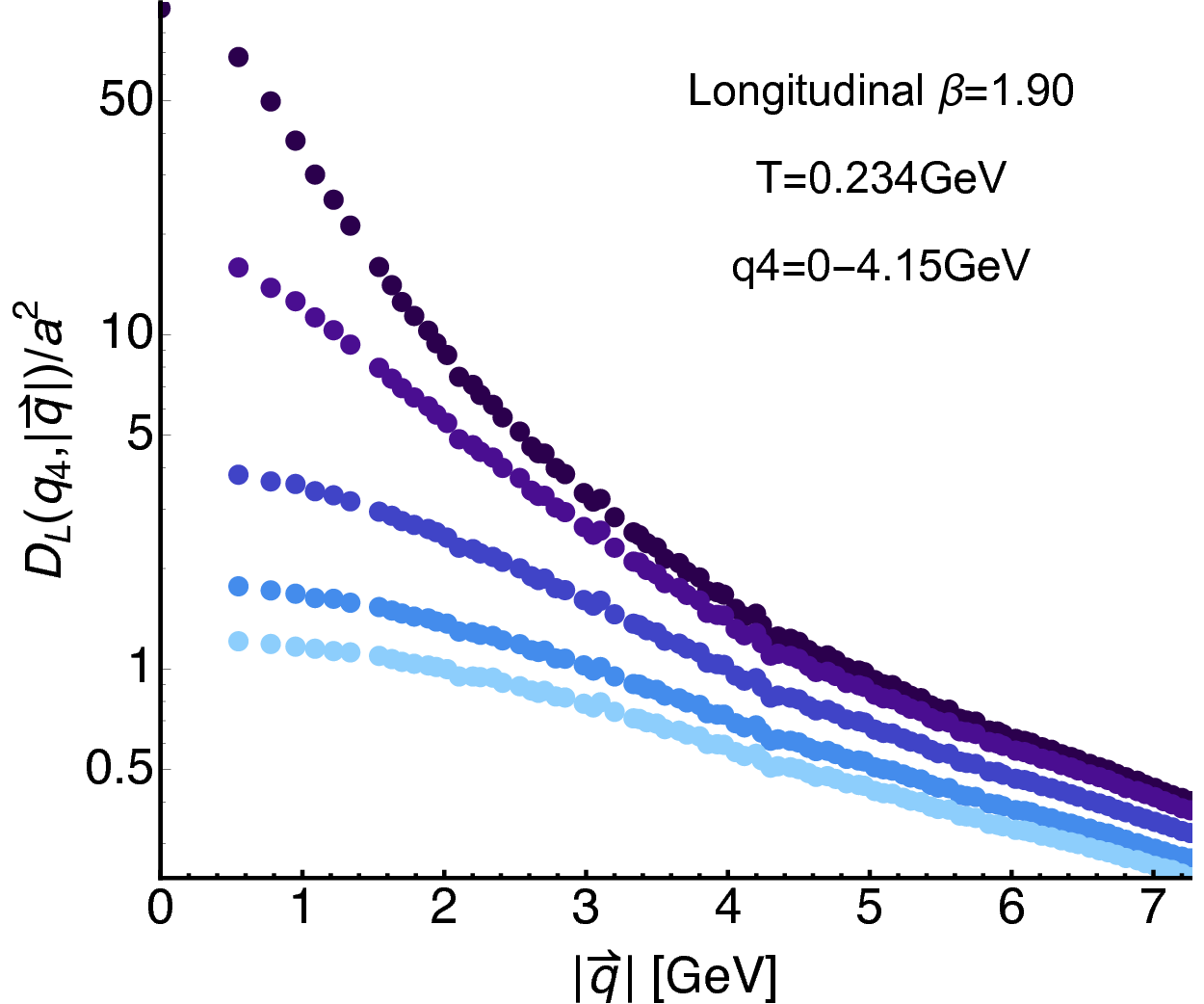}\vspace{0.1cm}\\
\includegraphics[scale=0.125]{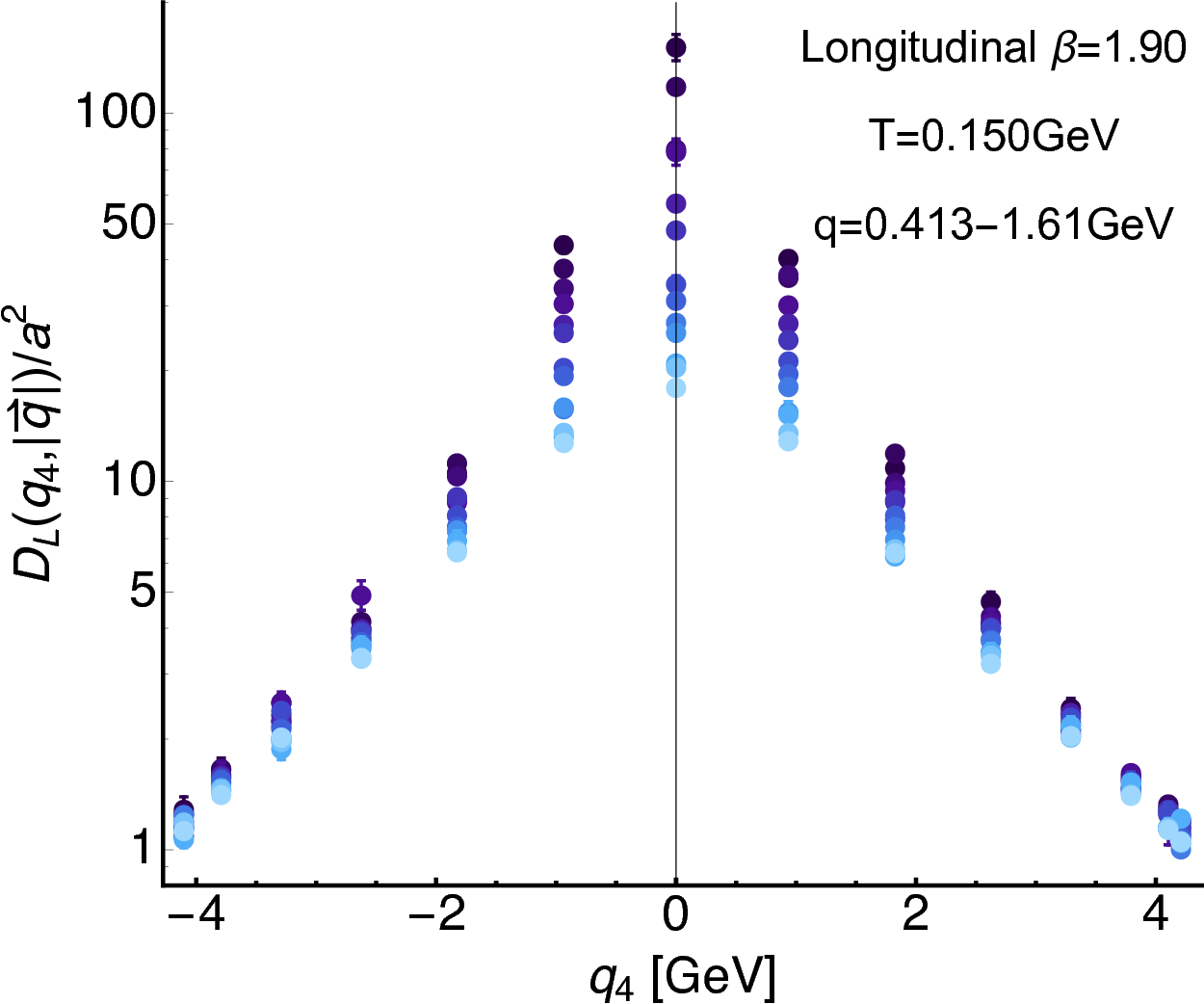}\hspace{2.5cm}
\includegraphics[scale=0.125]{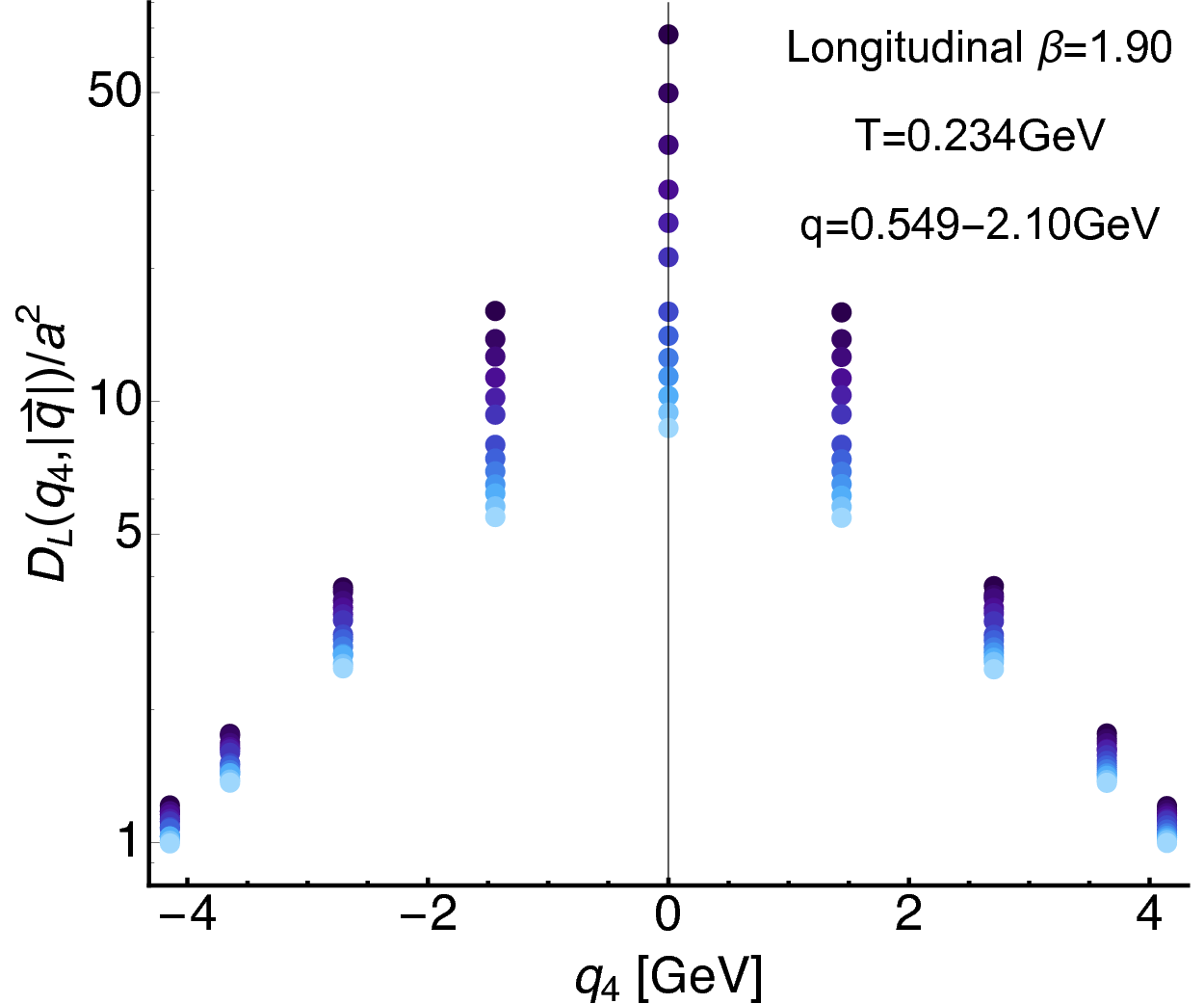}
\caption{The longitudinal gluon propagators at $\beta=1.90$ evaluated
  at both finite Matsubara frequency $q_4$ and spatial momentum
  $|\vec{q}|$.  The top row shows the $|\vec{q}|$ dependence at fixed
  $q_4$, while the bottom row we present the $q_4$ dependence for the
  fourteen lowest $|\vec{q}|$ values that will be used in the spectral
  reconstruction subsequently. The color coding assigns darkest colors
  to the lowest value of the corresponding parameter.  }
\end{figure*}

\begin{figure*}
\includegraphics[scale=0.125]{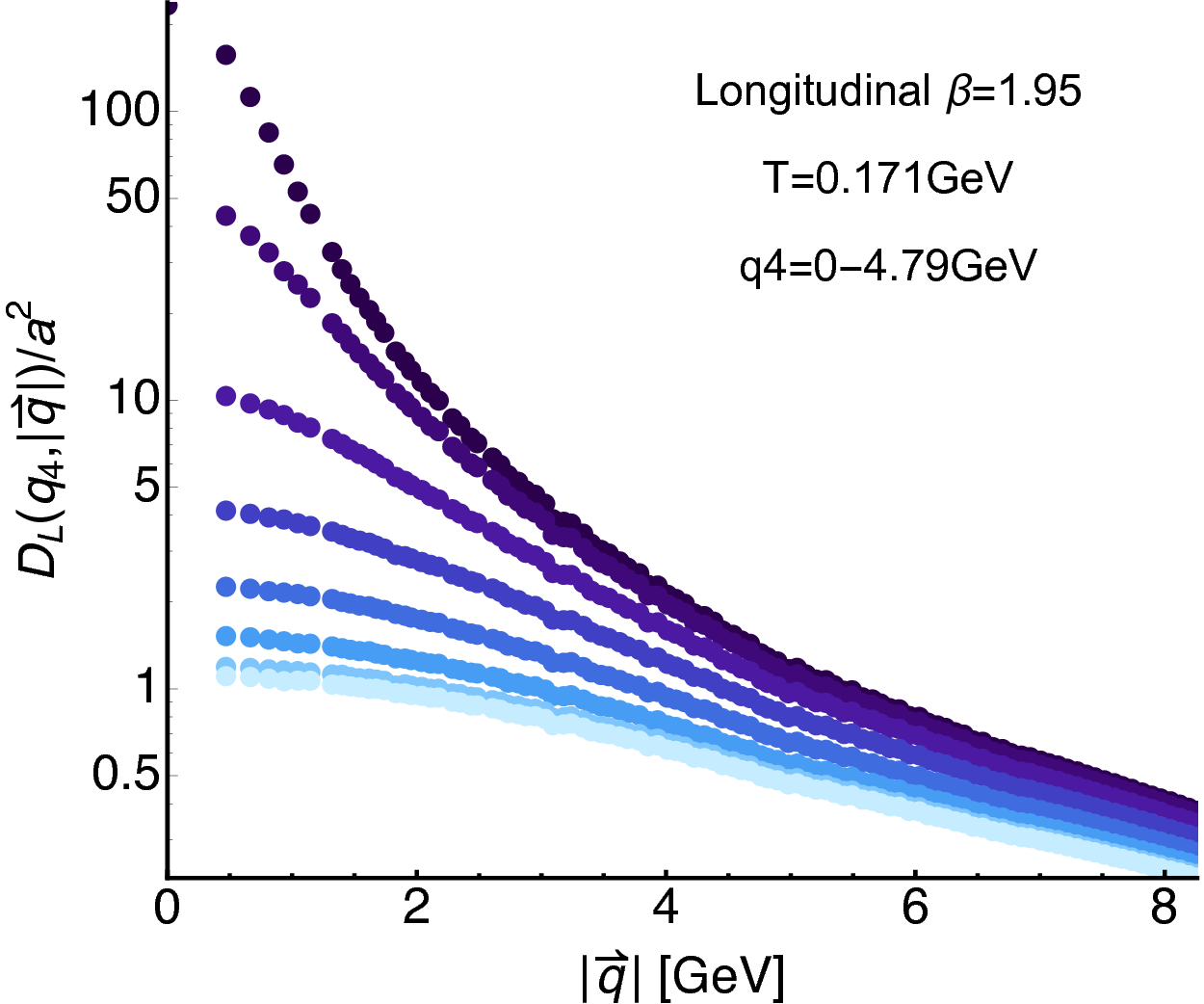}\vspace{0.1cm}\hspace{2.5cm}
\includegraphics[scale=0.125]{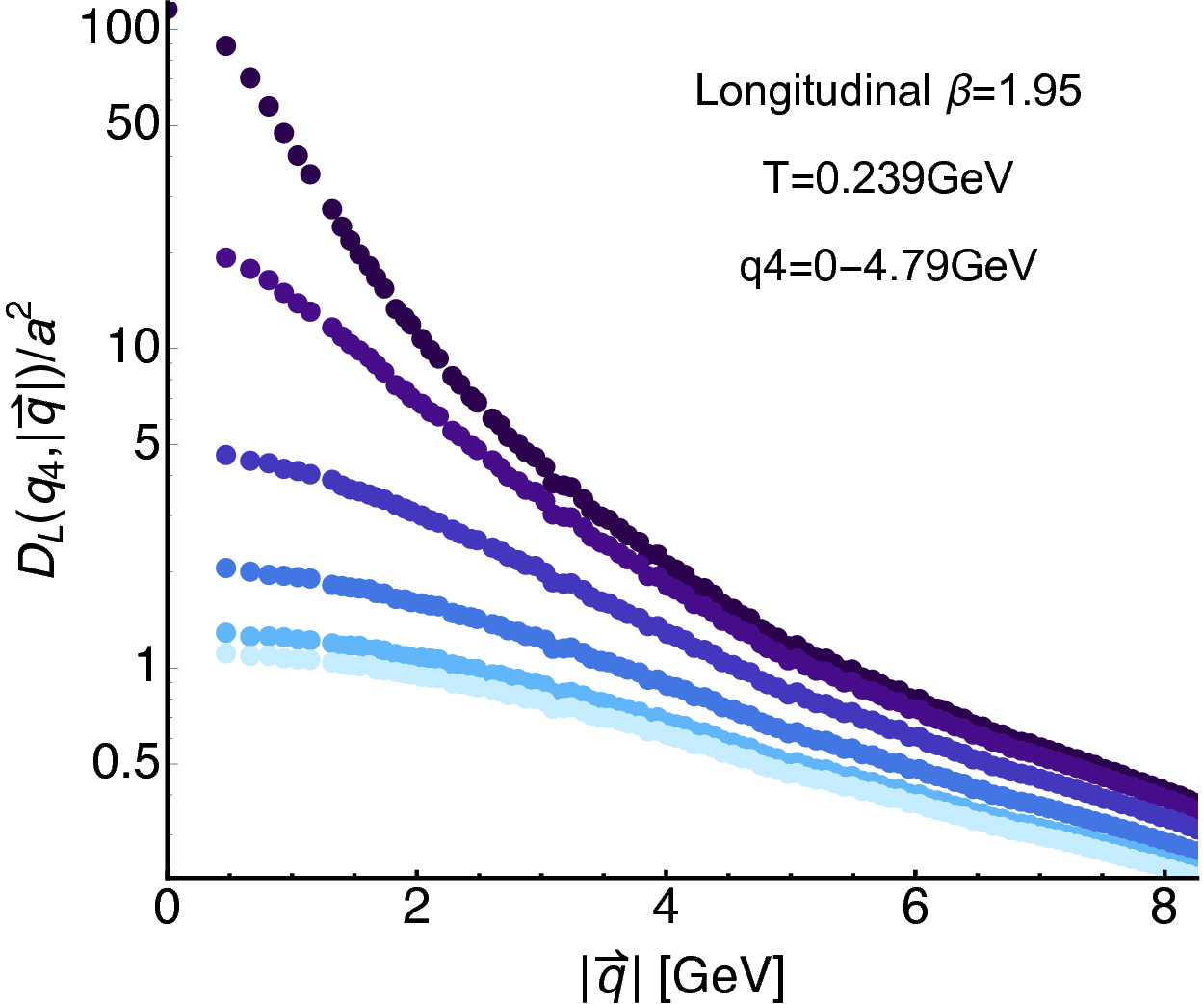}\vspace{0.1cm}\\
\includegraphics[scale=0.125]{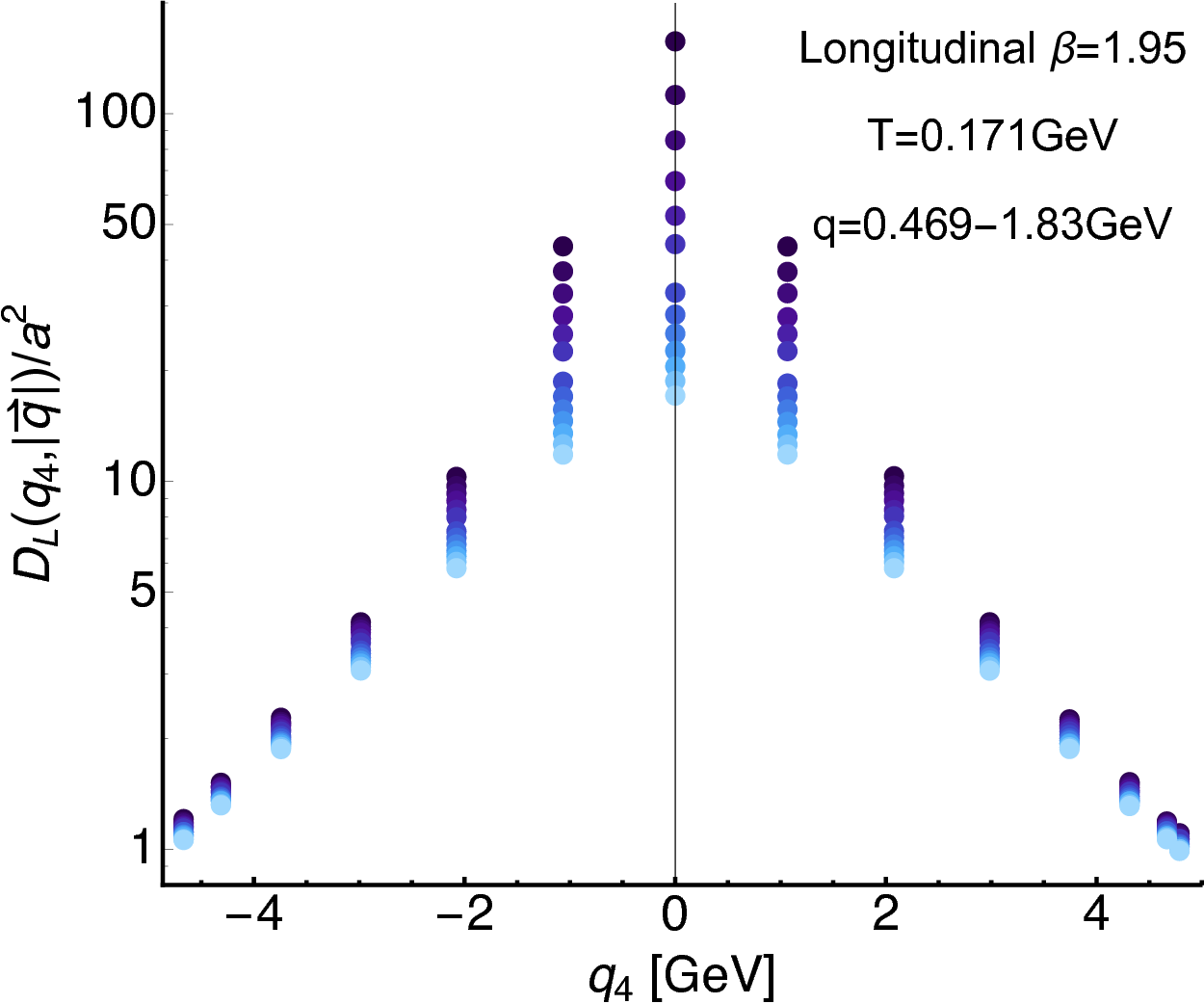}\hspace{2.5cm}
\includegraphics[scale=0.125]{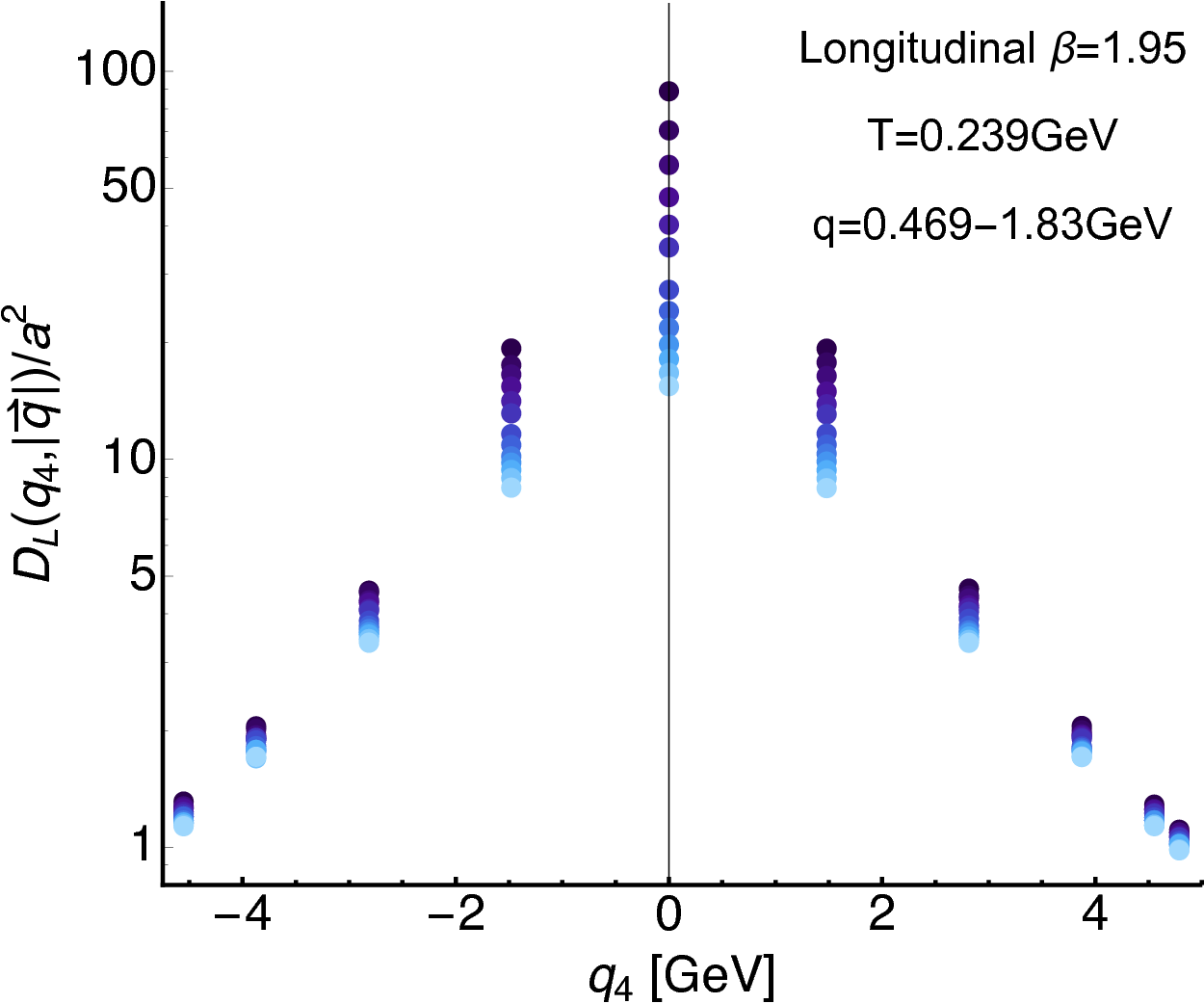}
\caption{The longitudinal gluon propagators at $\beta=1.95$ evaluated
  at both finite Matsubara frequency $q_4$ and spatial momentum
  $|\vec{q}|$.  The top row shows the $|\vec{q}|$ dependence at fixed
  $q_4$, while the bottom row we present the $q_4$ dependence for the
  fourteen lowest $|\vec{q}|$ values that will be used in the spectral
  reconstruction subsequently. The color coding assigns darkest colors
  to the lowest value of the corresponding parameter.  }
\end{figure*}

\begin{figure*}
\includegraphics[scale=0.125]{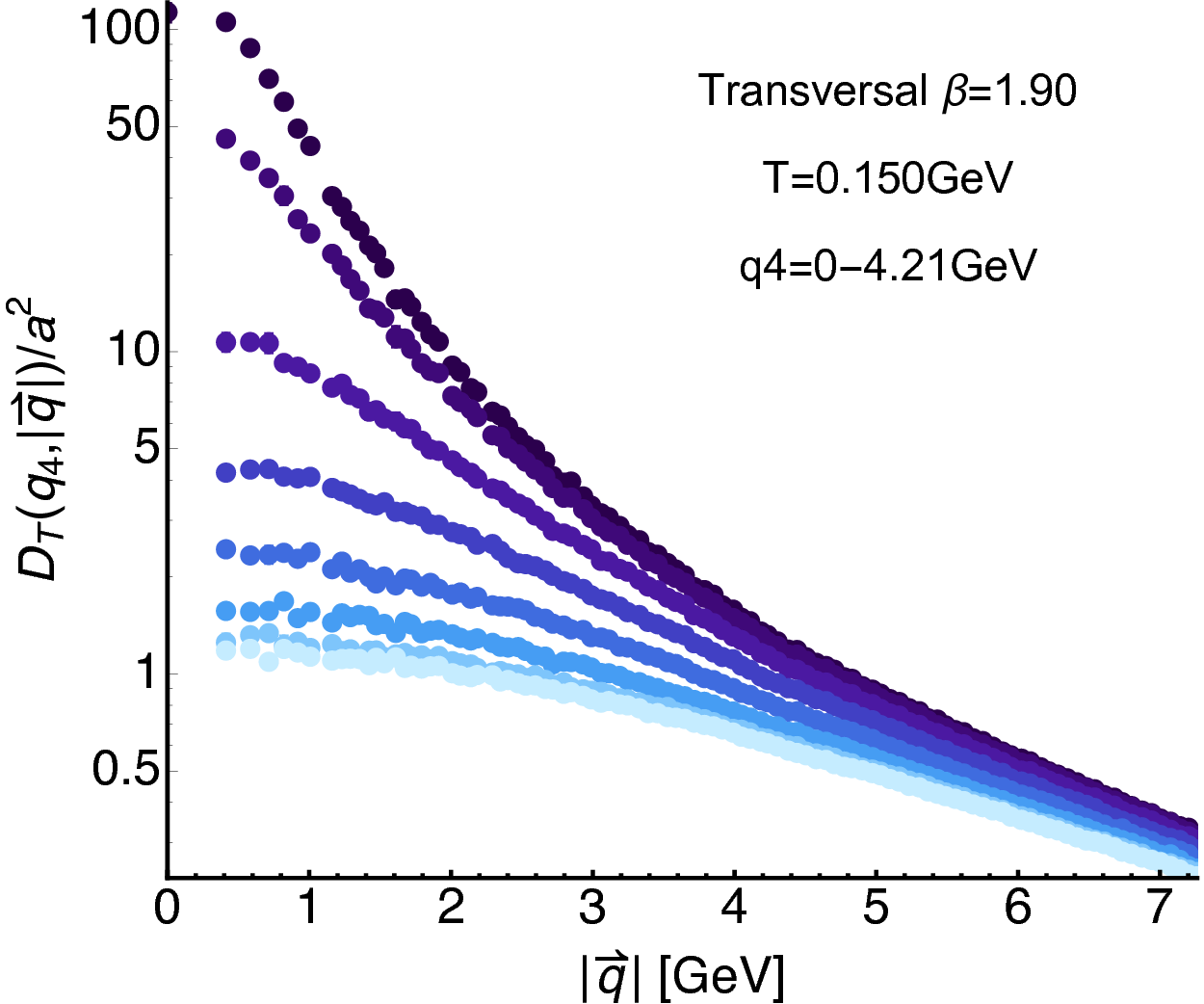}\vspace{0.1cm}\hspace{2.5cm}
\includegraphics[scale=0.125]{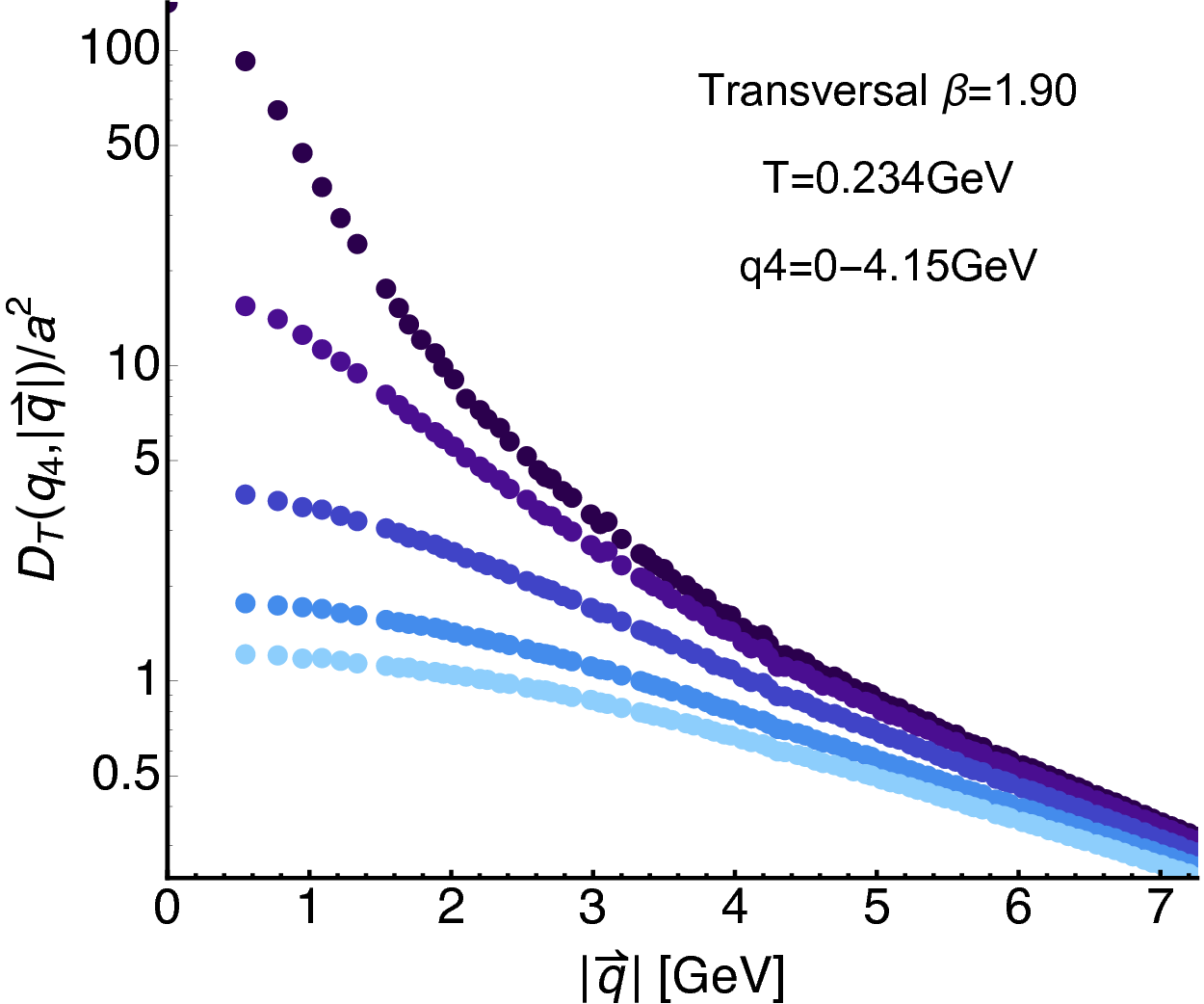}\vspace{0.1cm}\\
\includegraphics[scale=0.125]{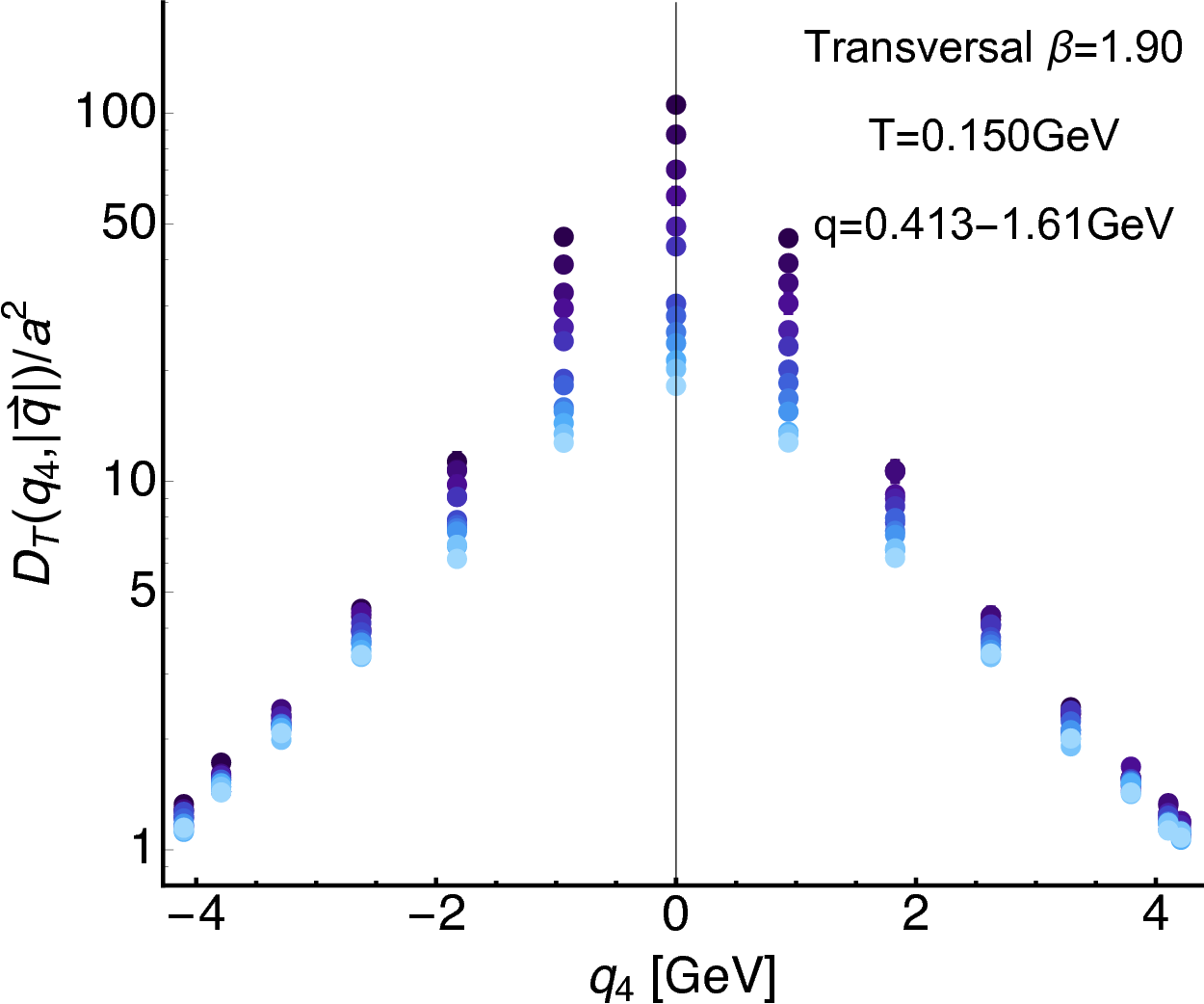}\hspace{2.5cm}
\includegraphics[scale=0.125]{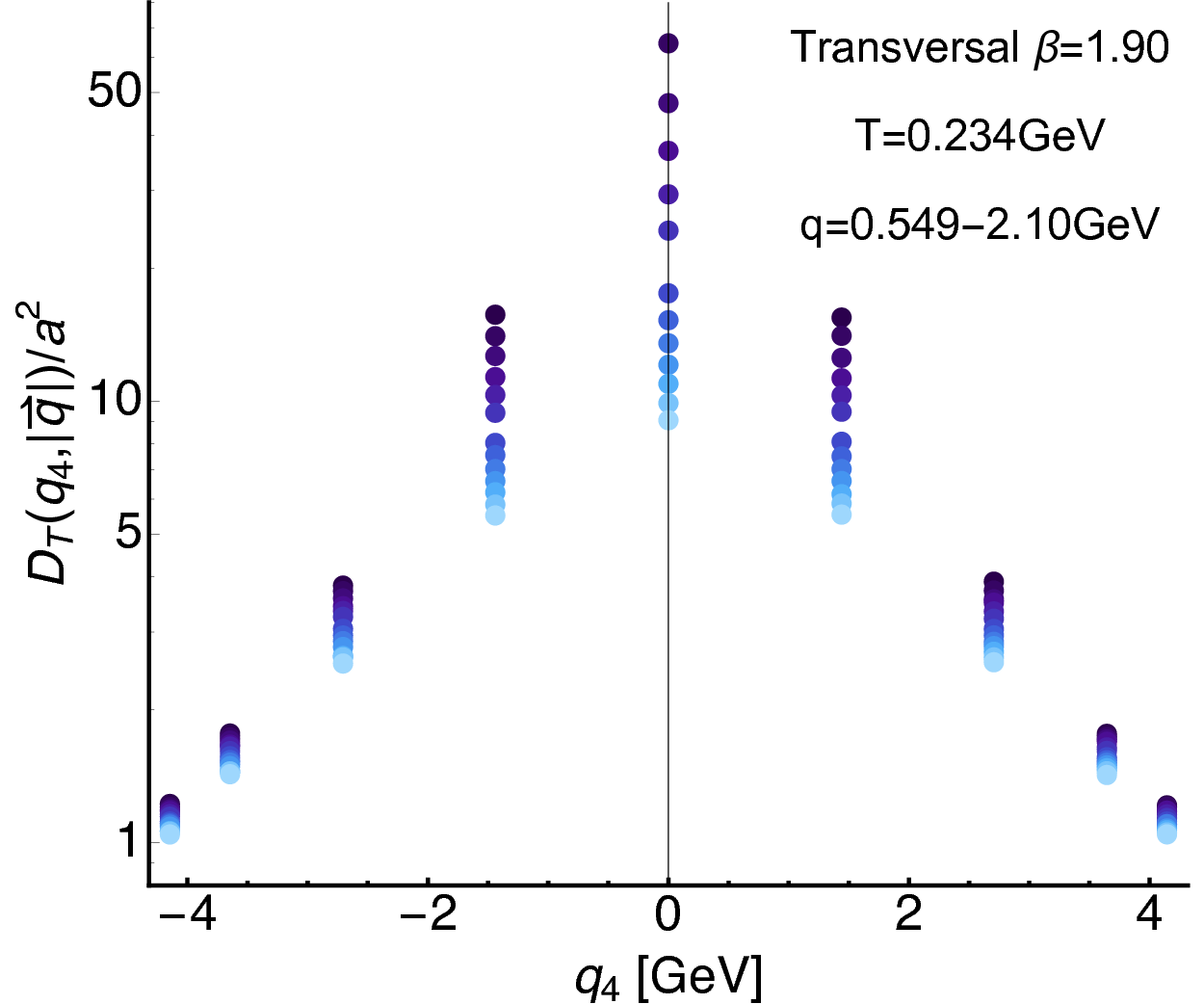}
\caption{The transversal gluon propagators at $\beta=1.90$ evaluated at both finite Matsubara frequency $q_4$ and spatial momentum $|\vec{q}|$.  The top row shows the $|\vec{q}|$ dependence at fixed $q_4$, while the bottom row we present the $q_4$ dependence for the fourteen lowest $|\vec{q}|$ values that will be used in the spectral reconstruction subsequently. The color coding assigns darkest colors to the lowest value of the corresponding parameter. }
\end{figure*}

\begin{figure*}
\includegraphics[scale=0.125]{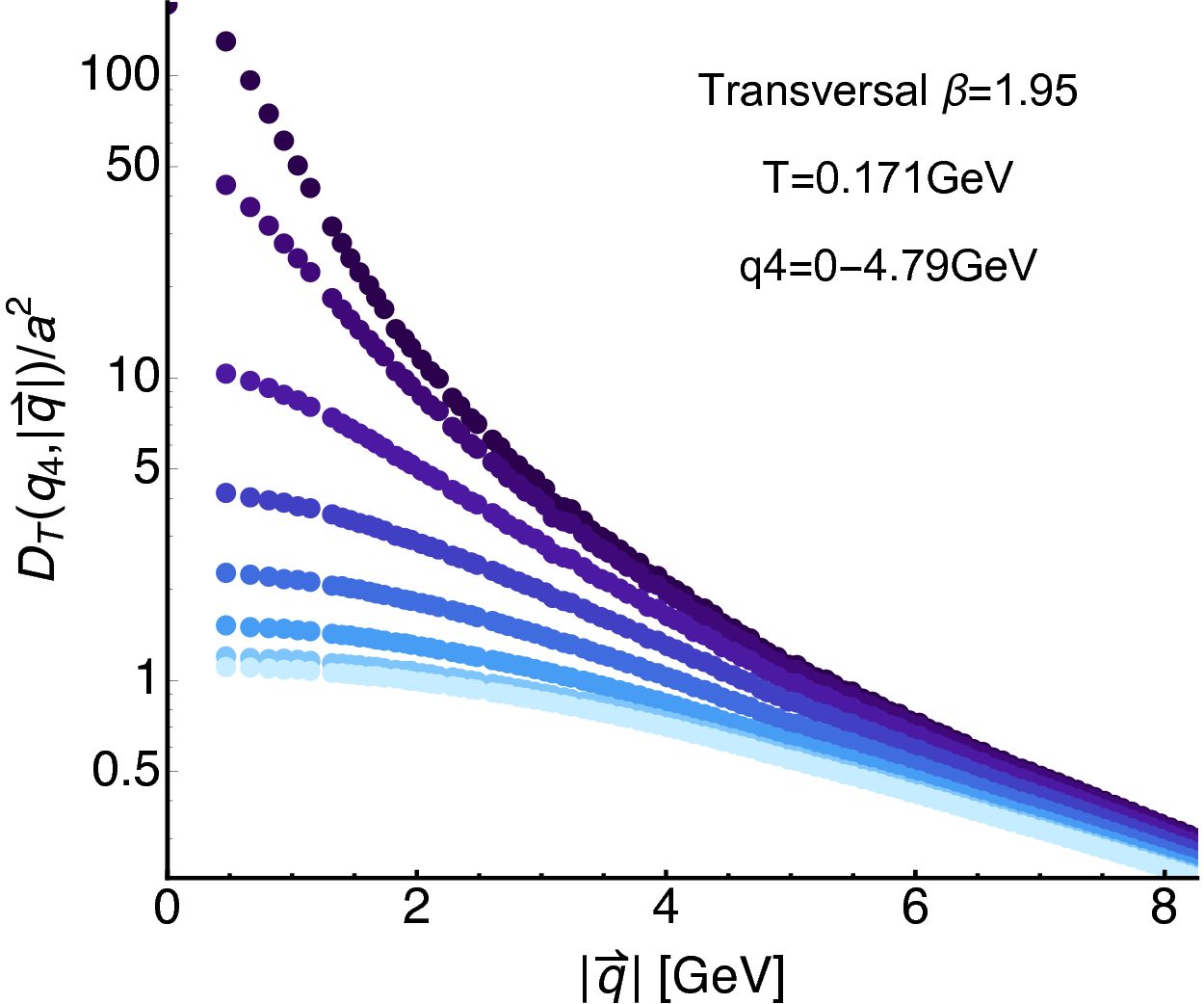}\vspace{0.1cm}\hspace{2.5cm}
\includegraphics[scale=0.125]{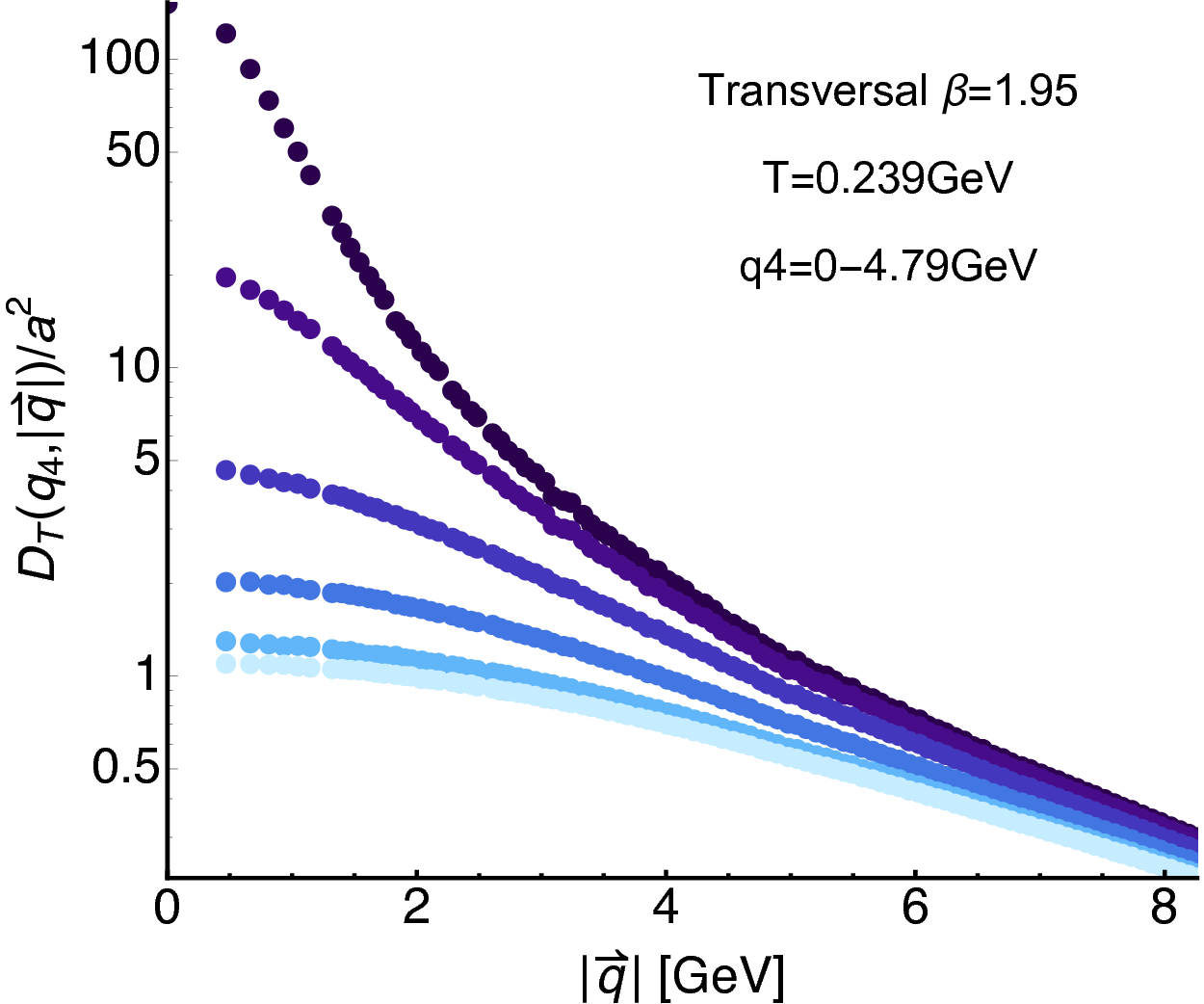}\vspace{0.1cm}\\
\includegraphics[scale=0.125]{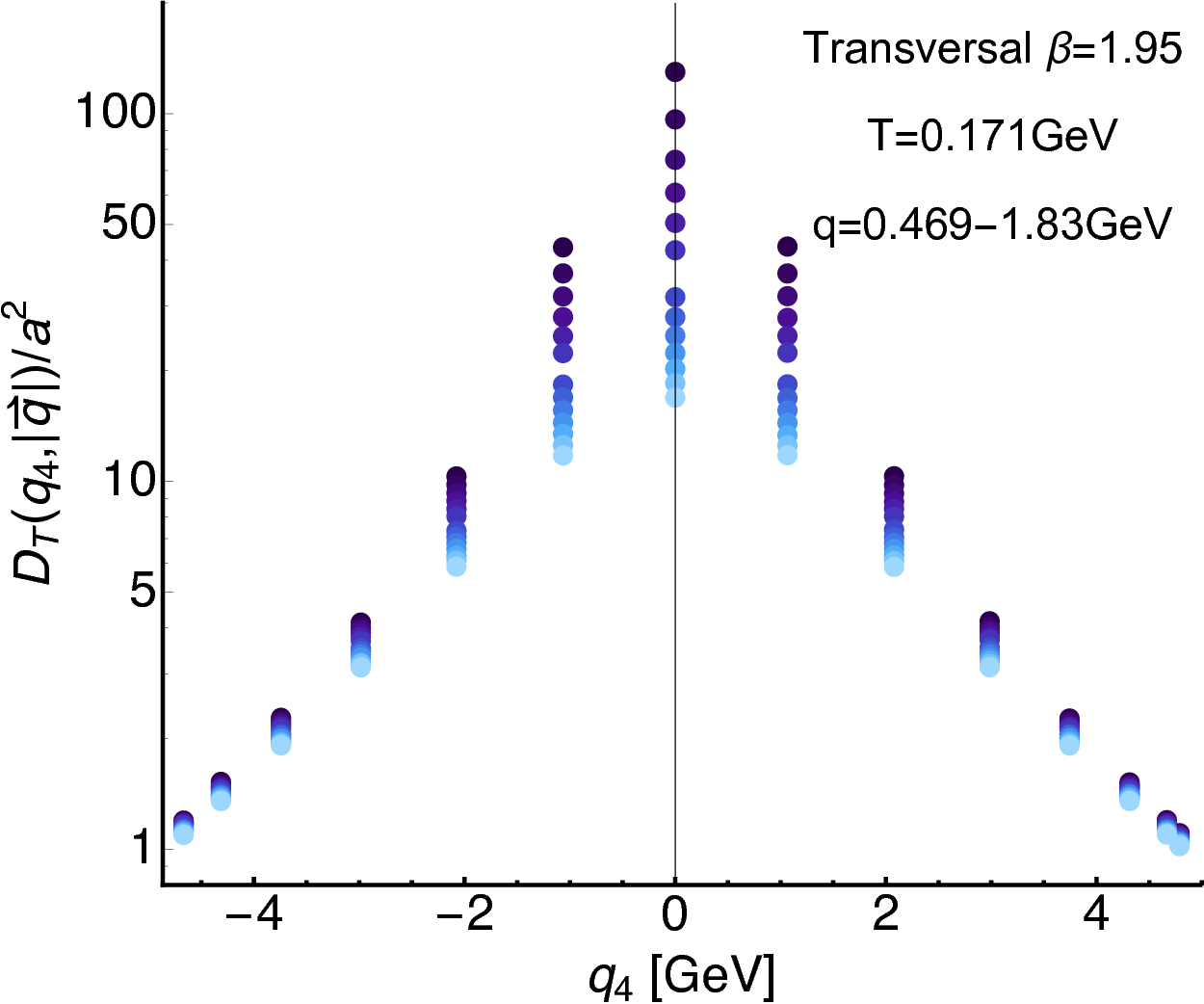}\hspace{2.5cm}
\includegraphics[scale=0.125]{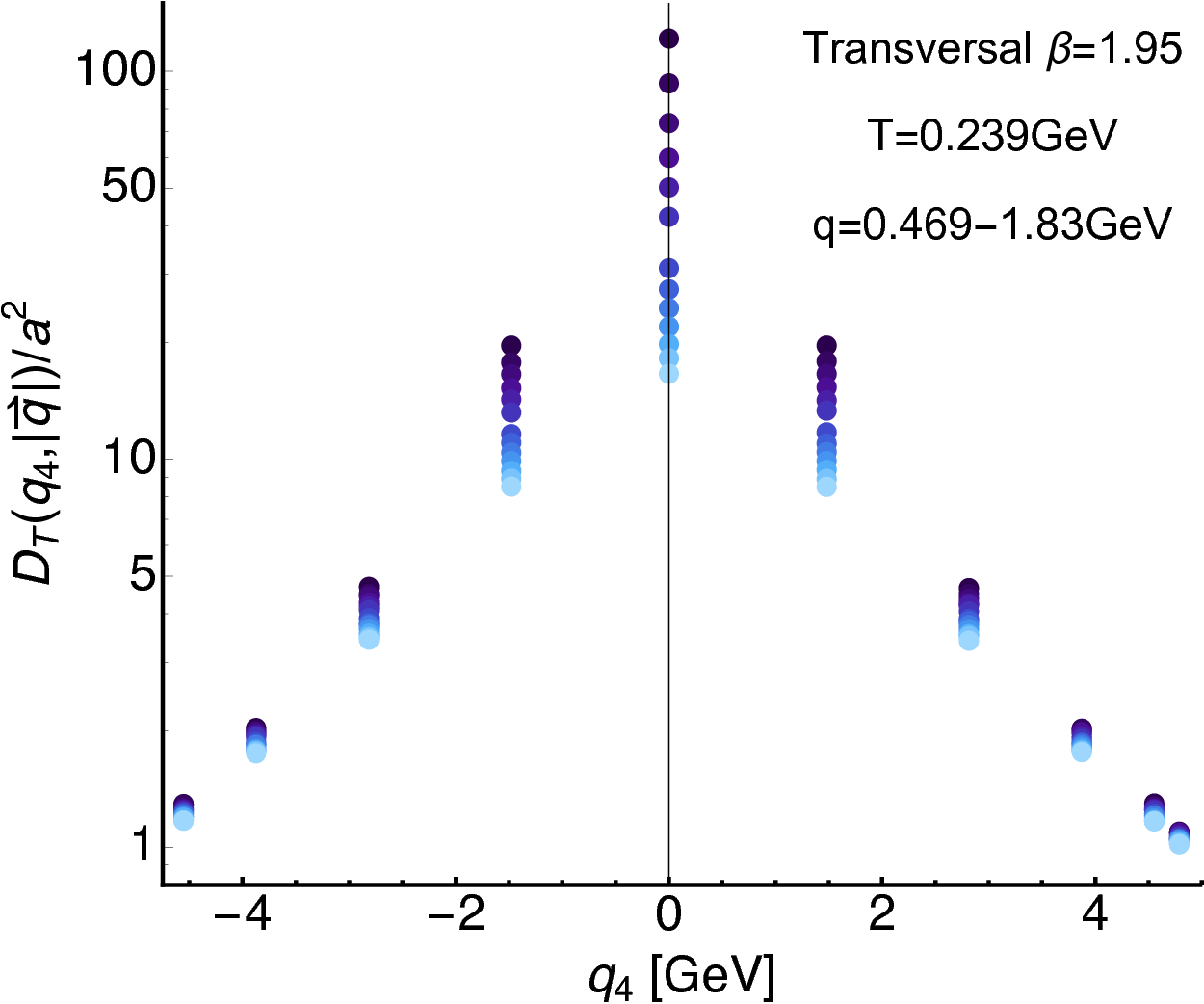}
\caption{The transversal gluon propagators at $\beta=1.95$ evaluated at both finite Matsubara frequency $q_4$ and spatial momentum $|\vec{q}|$.  The top row shows the $|\vec{q}|$ dependence at fixed $q_4$, while the bottom row we present the $q_4$ dependence for the fourteen lowest $|\vec{q}|$ values that will be used in the spectral reconstruction subsequently. The color coding assigns darkest colors to the lowest value of the corresponding parameter. }
\end{figure*}

\begin{figure*}
\includegraphics[scale=0.15]{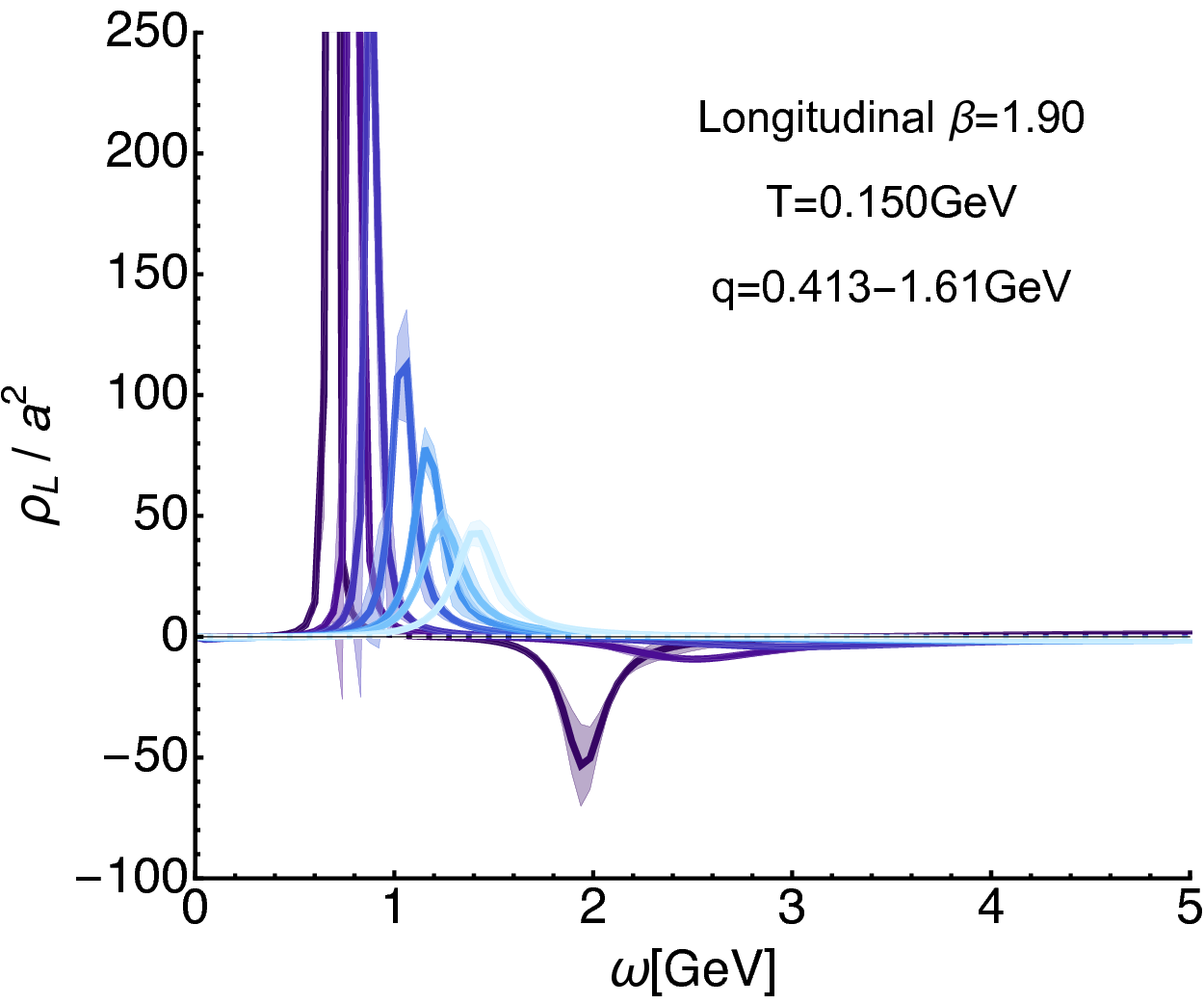}\hspace{0.5cm}\hspace{2.5cm}
\includegraphics[scale=0.15]{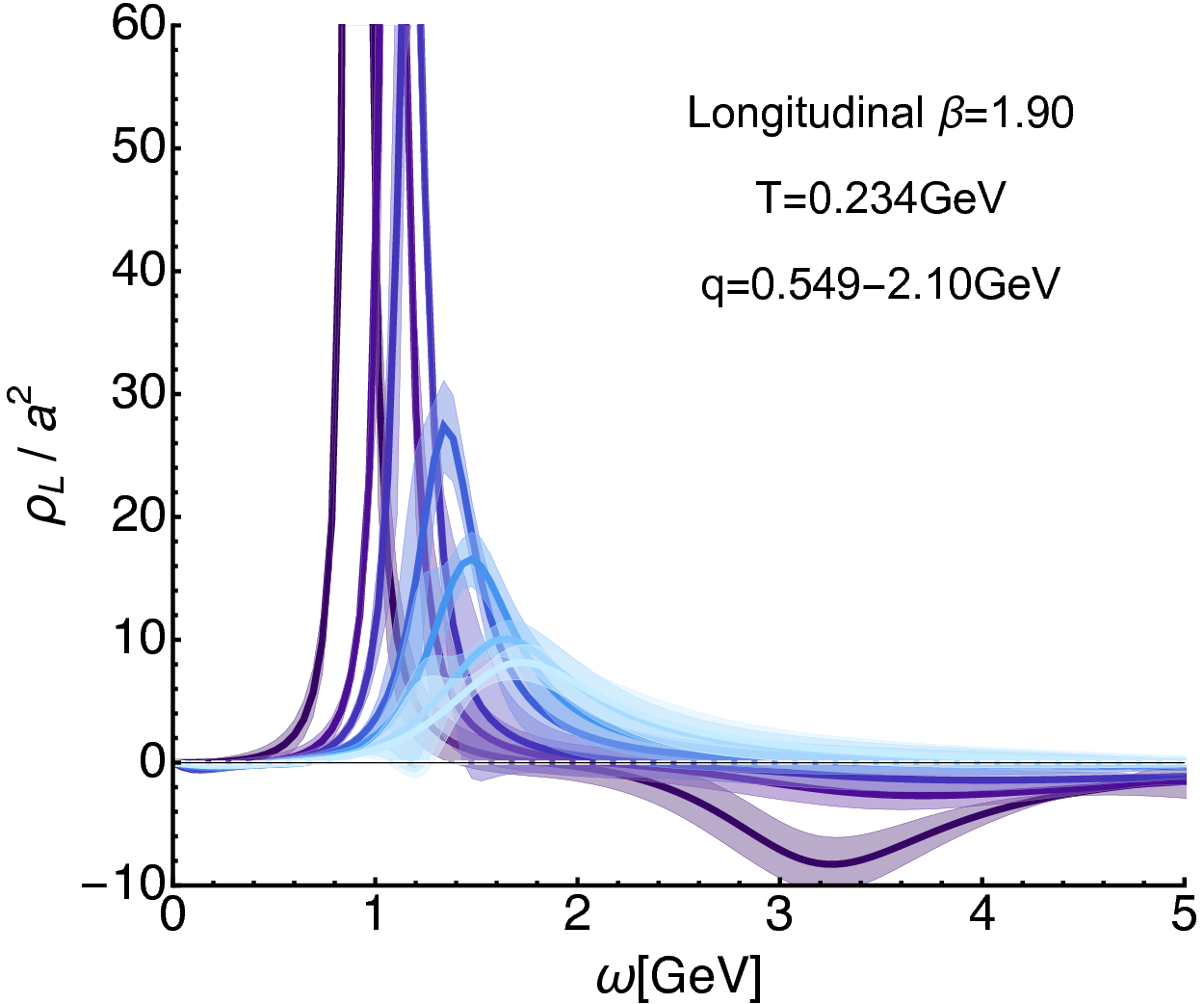}
\includegraphics[scale=0.15]{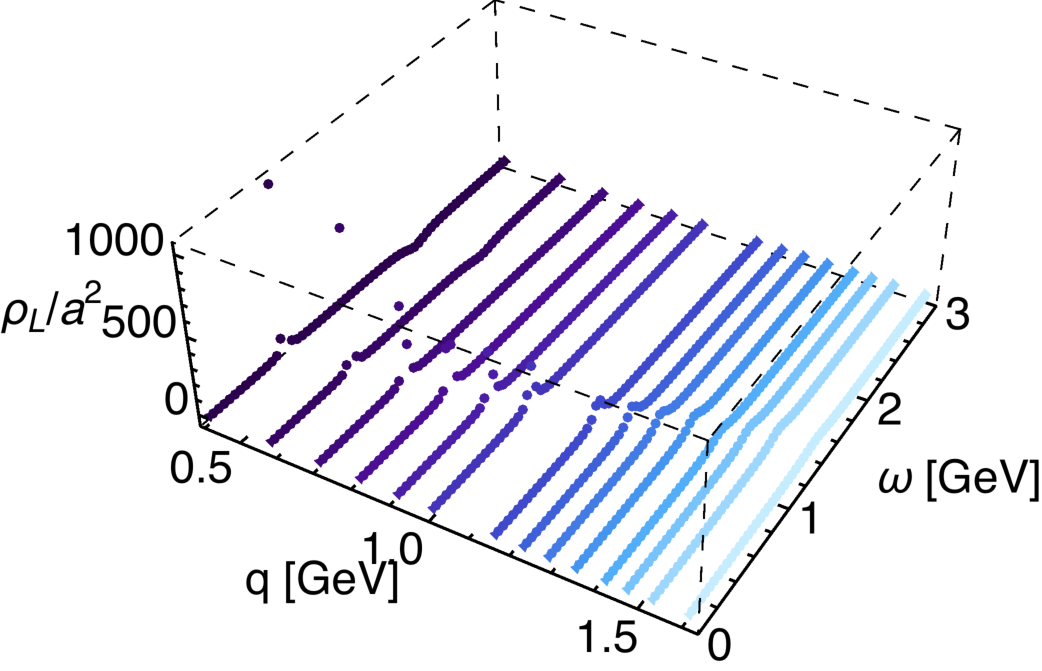}\hspace{0.5cm}\hspace{2.5cm}
\includegraphics[scale=0.15]{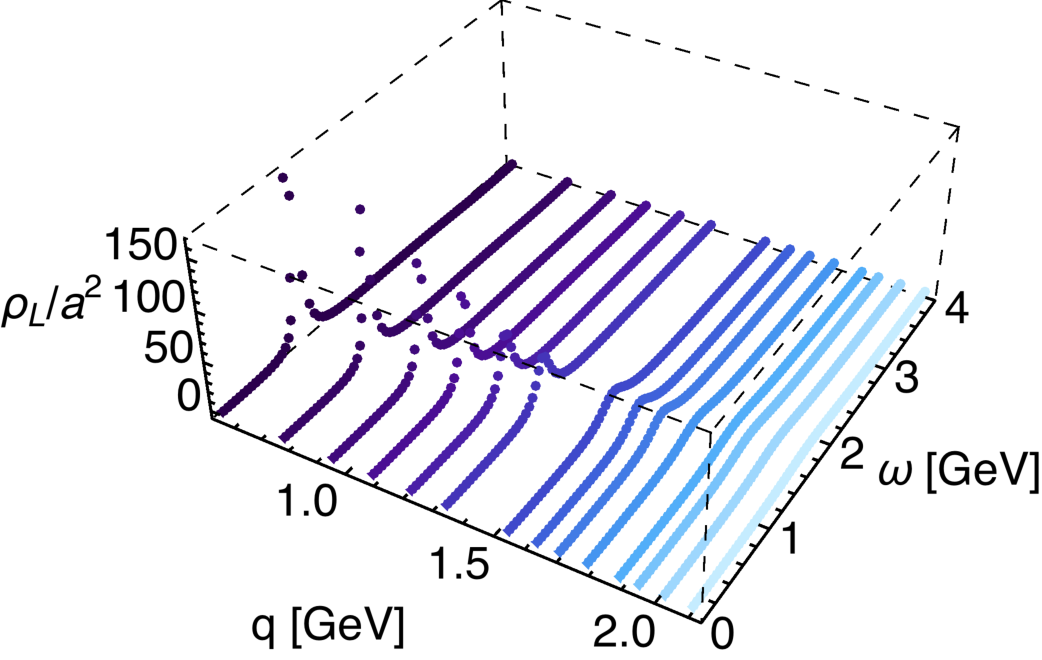}
\caption{(top) Reconstructed longitudinal gluon spectra at $\beta=1.90 $ for (left) $T=150$MeV and (right) $T=234$MeV. The seven different curves denote half of the fourteen lowest spatial momenta at which correlator data is available. One can clearly see at well defined lowest lying positive peak with a subsequent trough, which dies out towards higher frequencies. (bottom) Three dimensional visualization of the momentum dependence of the reconstructed longitudinal gluon spectra at $\beta=1.95$ for (left) $T=150$MeV and (right) $T=234$MeV.}\label{Fig:SpectraLong190}
\end{figure*}

\begin{figure*}
\includegraphics[scale=0.15]{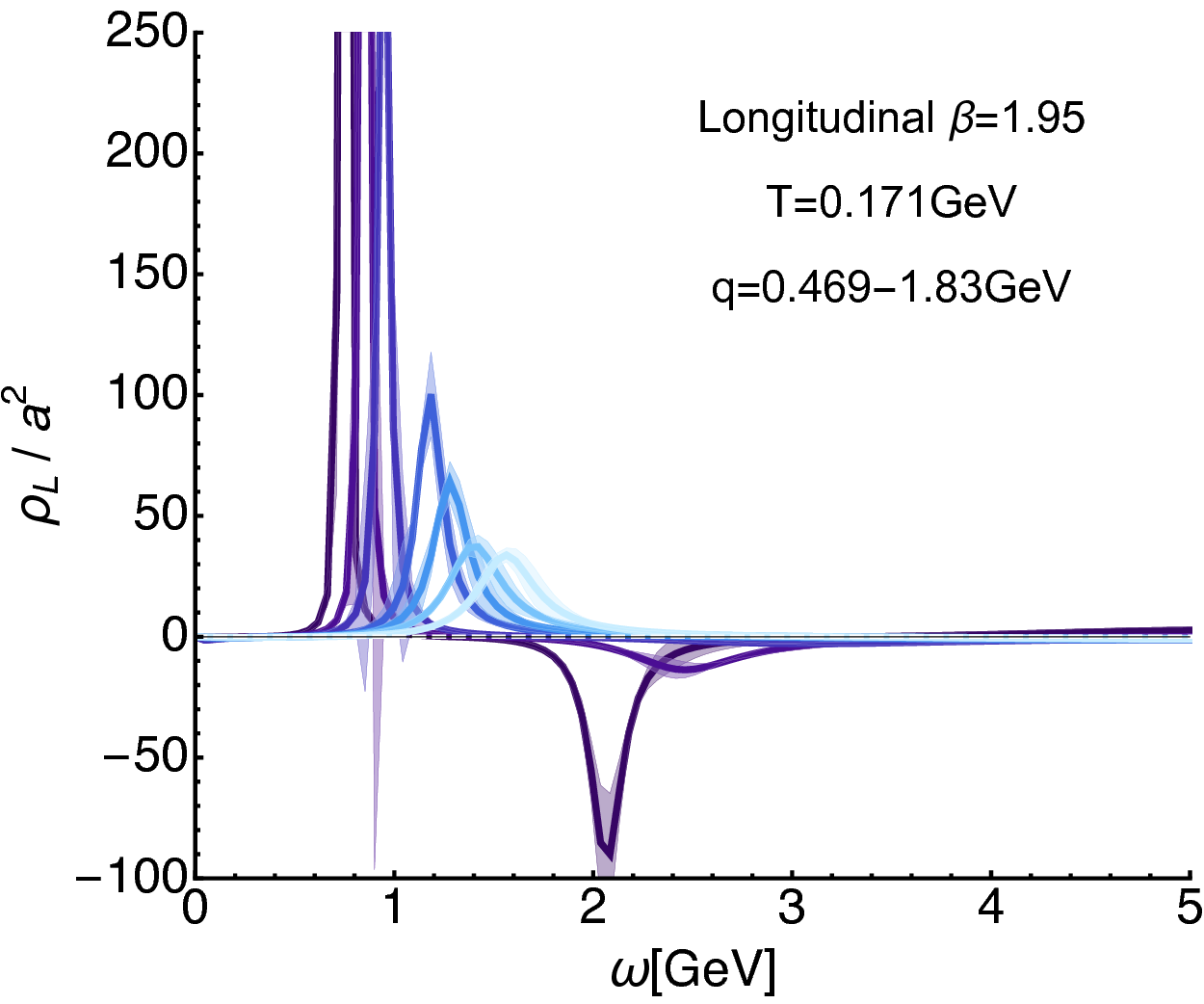}\hspace{0.5cm}\hspace{2cm}
\includegraphics[scale=0.15]{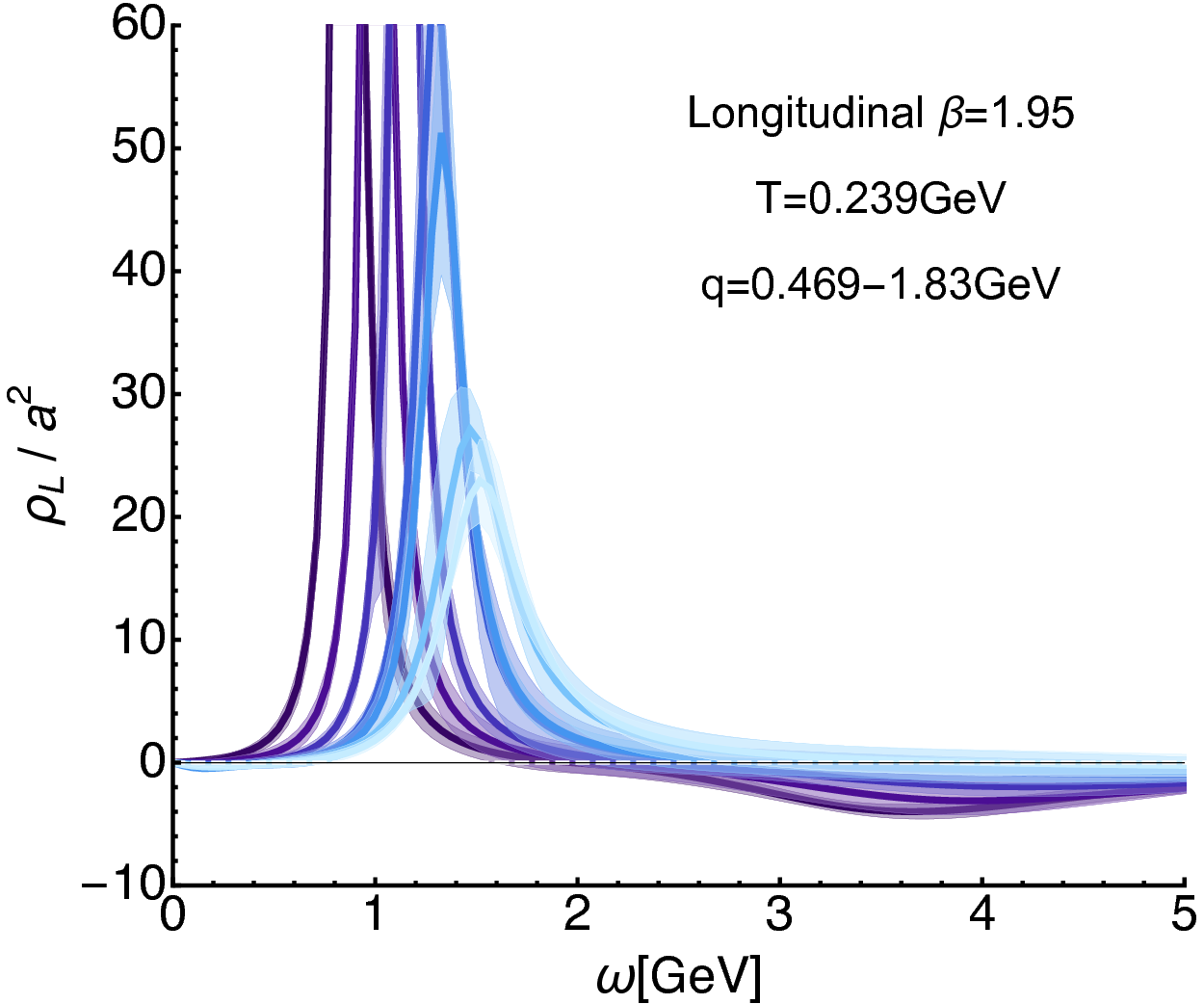}
\includegraphics[scale=0.17]{B190LongLowT3dMomenta.png}\hspace{0.5cm}\hspace{2cm}
\includegraphics[scale=0.17]{B190LongHighT3dMomenta.png}
\caption{(top) Reconstructed longitudinal gluon spectra at $\beta=1.95 $ for (left) $T=171$MeV and (right) $T=239$MeV. The seven different curves denote half of the fourteen lowest spatial momenta at which correlator data is available. (bottom) Three dimensional visualization of the momentum dependence of the reconstructed longitudinal gluon spectra at $\beta=1.95$ for (left) $T=171$MeV and (right) $T=239$MeV. }
\end{figure*}

\begin{figure*}
\includegraphics[scale=0.17]{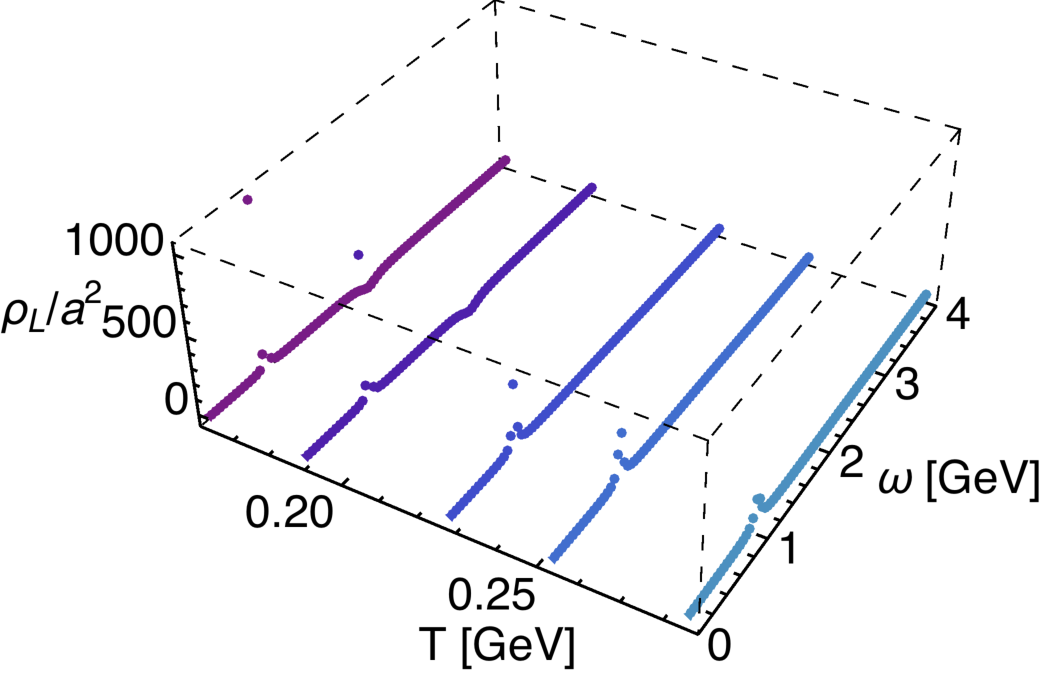}\hspace{0.5cm}\hspace{2cm}
\includegraphics[scale=0.17]{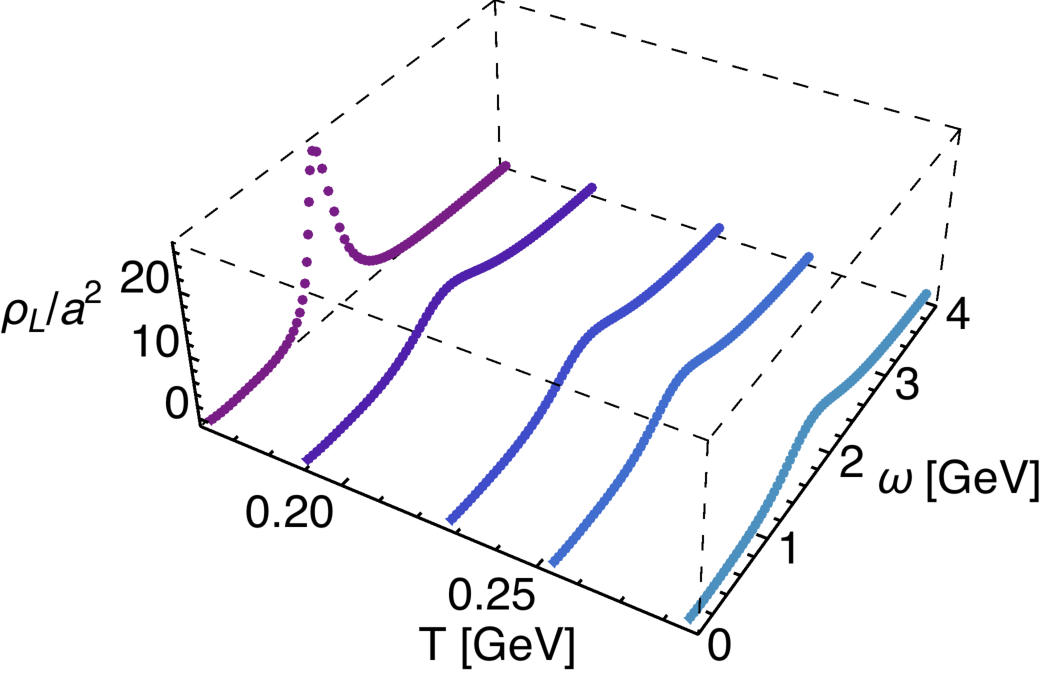}
\caption{Three dimensional visualization of the temperature dependence of the reconstructed longitudinal gluon spectra on the $\beta=1.90$ ensembles for (left) the lowest and (right) highest available spatial momentum. The five different curves denote the five lowest temperatures at which the reconstruction was possible.}
\end{figure*}

\begin{figure*}
\includegraphics[scale=0.17]{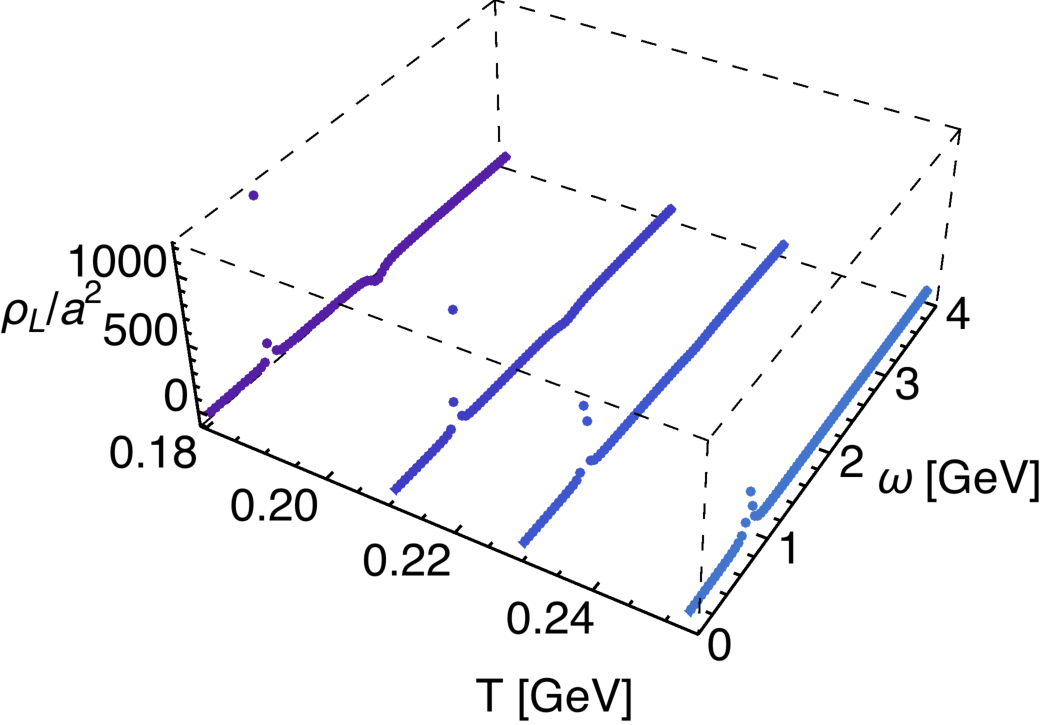}\hspace{0.5cm}\hspace{2cm}
\includegraphics[scale=0.17]{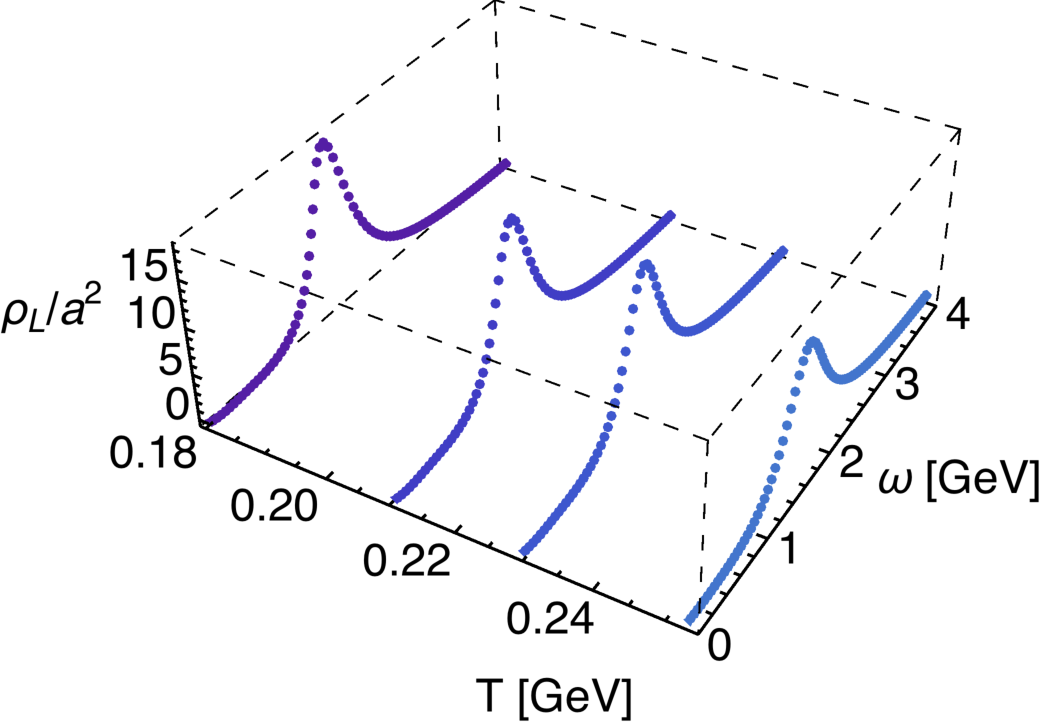}
\caption{Three dimensional visualization of the temperature dependence of the reconstructed longitudinal gluon spectra on the $\beta=1.95$ ensembles for (left) the lowest and (right) highest available spatial momentum. The four different curves denote the four lowest temperatures at which the reconstruction was possible.}
\end{figure*}

\begin{figure*}
\includegraphics[scale=0.26]{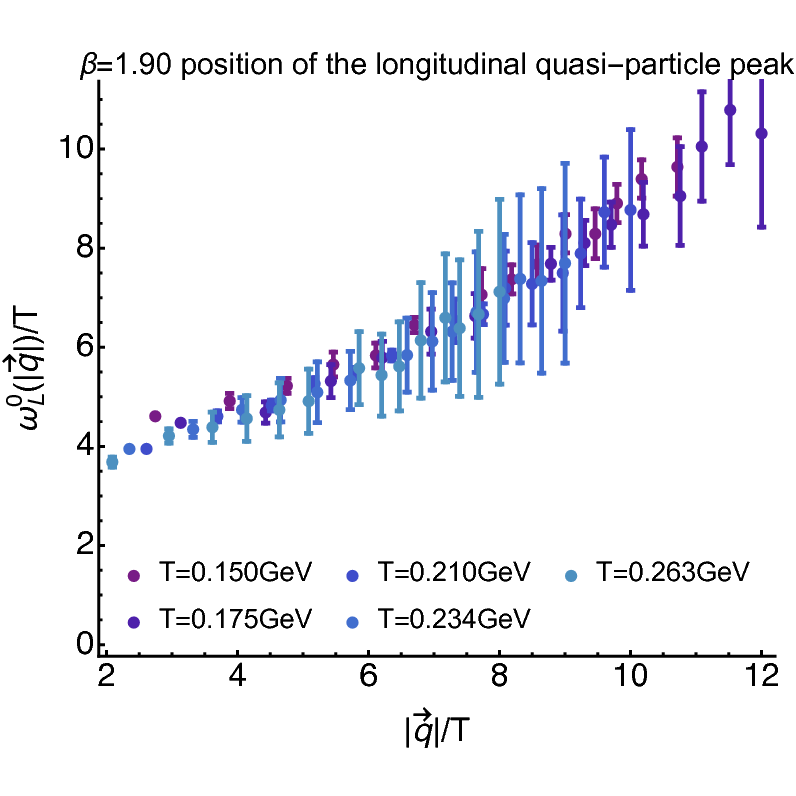}\hspace{0.1cm}\hspace{2cm}
\includegraphics[scale=0.25]{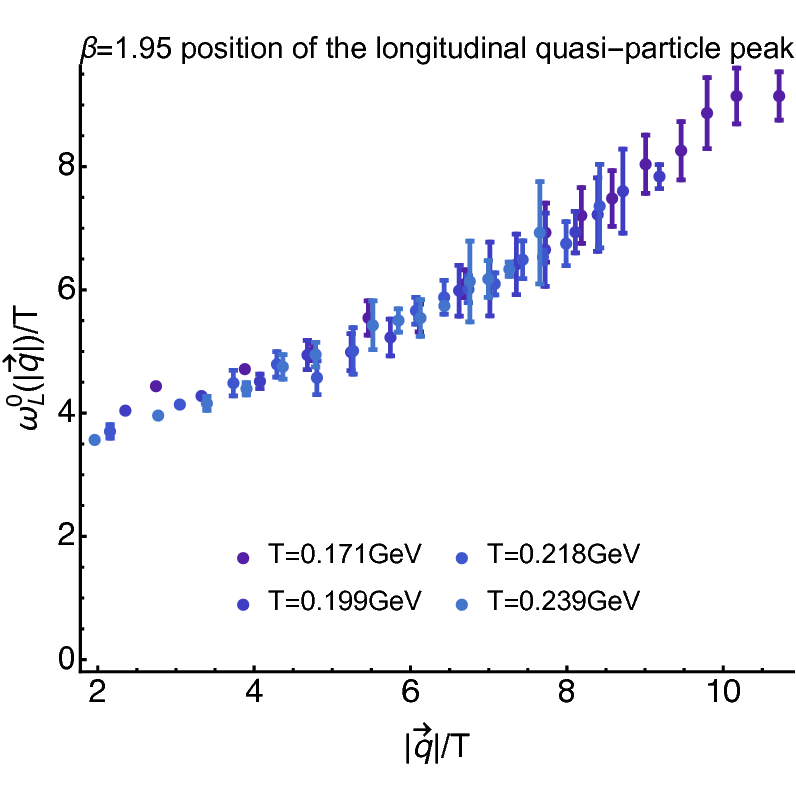}\hspace{0.1cm}
\caption{Momentum dependence of the longitudinal quasi-particle peak position at (left) $\beta=1.90$ and (right) $\beta=1.95$. The peak position for small momenta takes on a non-zero value. The reconstructions at the two lowest temperatures (within the hadronic phase) seem to exhibit a slightly larger intercept than the curves in the deconfined phase.} 
\end{figure*}

\begin{figure*}
\includegraphics[scale=0.14]{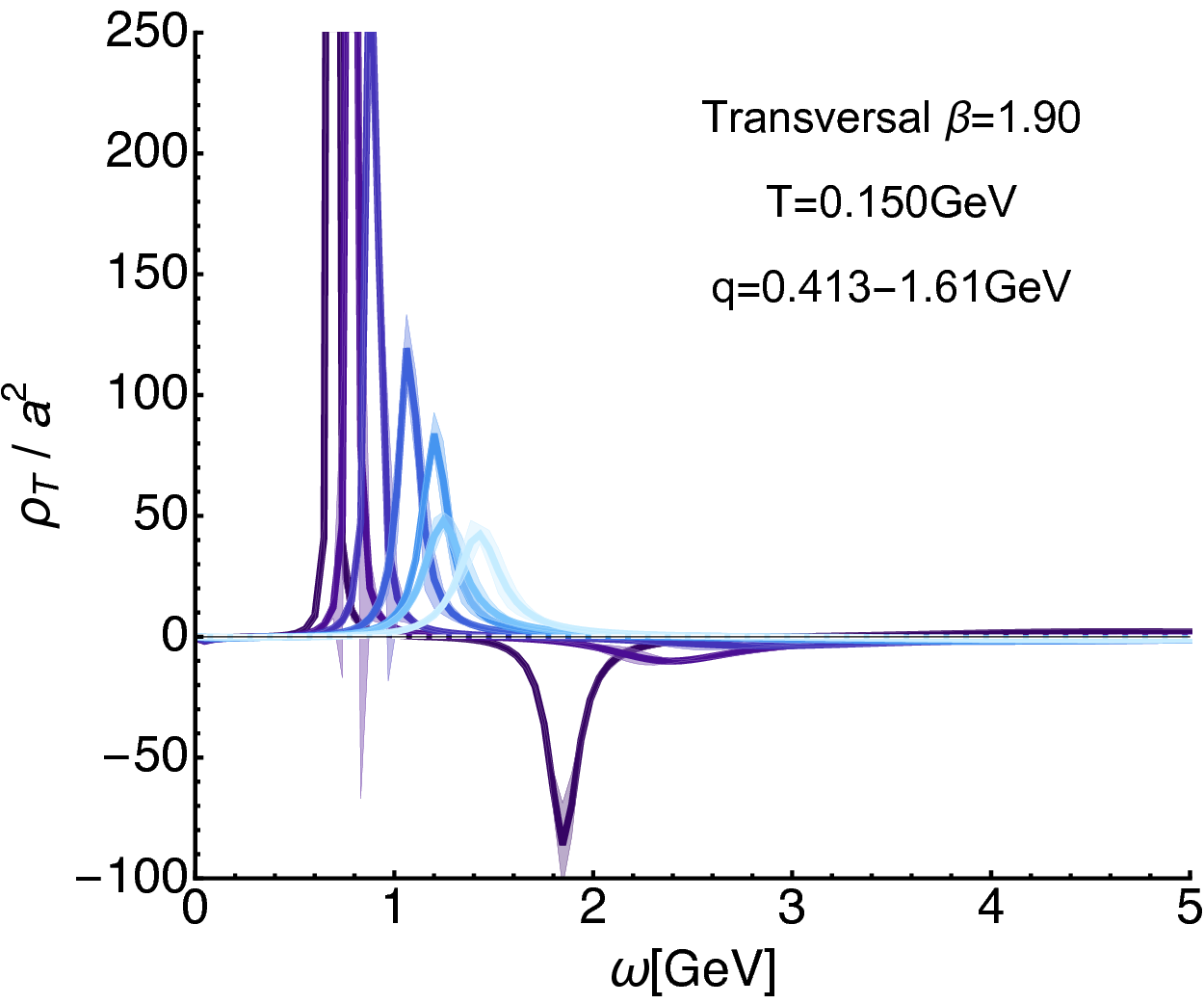}\hspace{0.5cm}\hspace{2cm}
\includegraphics[scale=0.14]{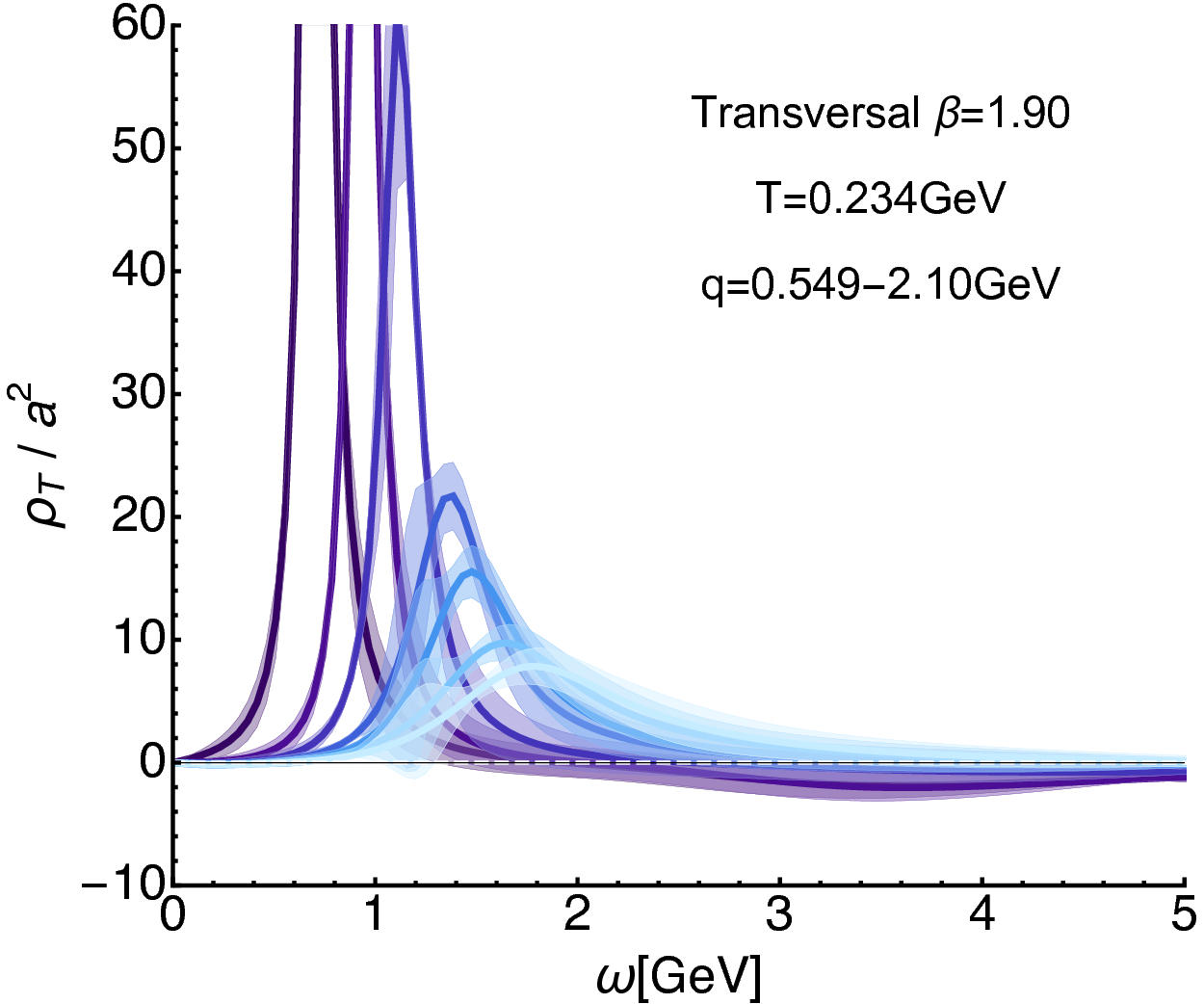}
\includegraphics[scale=0.16]{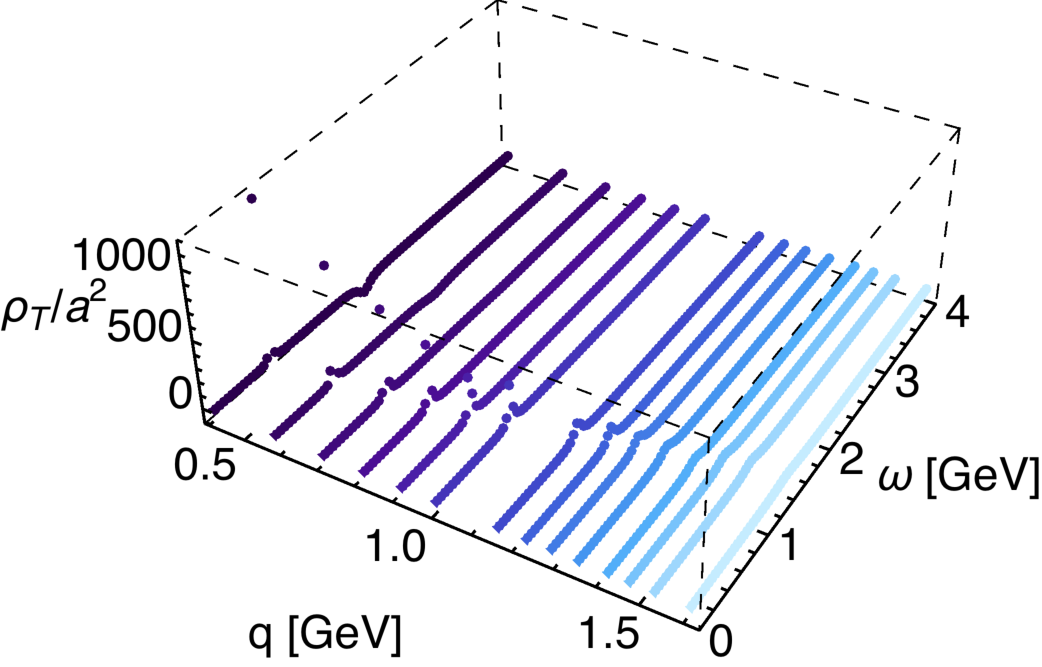}\hspace{0.5cm}\hspace{2cm}
\includegraphics[scale=0.16]{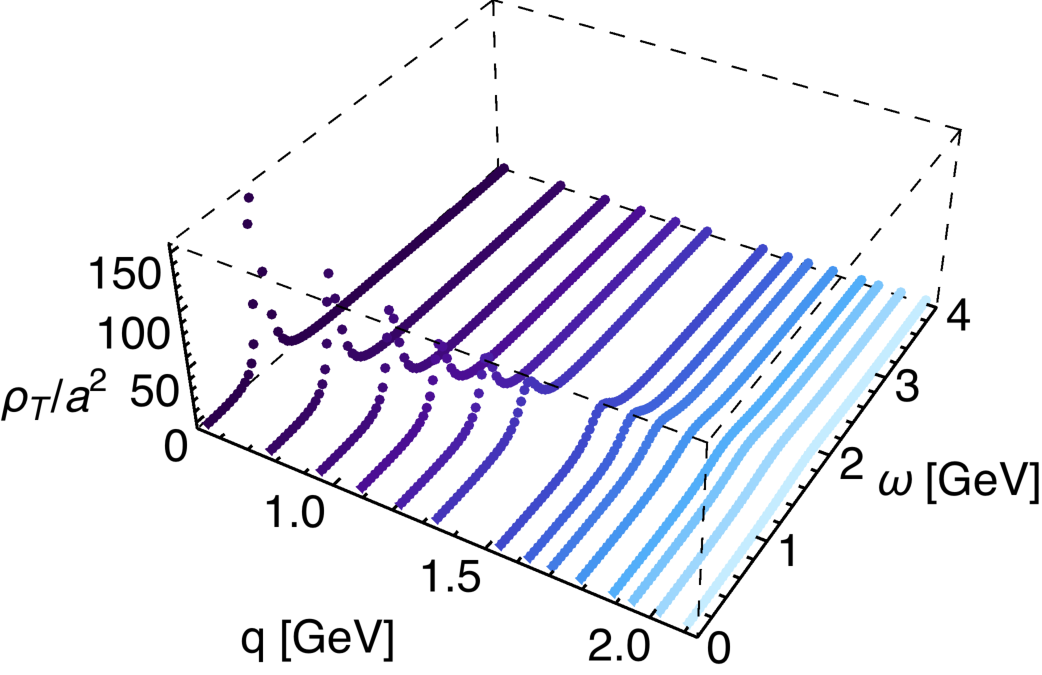}
\caption{(top) Reconstructed transversal gluon spectra at $\beta=1.90$ for (left) $T=150$MeV and (right) $T=234$MeV. The seven different curves denote half of the fourteen lowest spatial momenta at which correlator data is available.  One can clearly see at well defined lowest lying positive peak with a subsequent trough, which dies out towards higher frequencies. (bottom) Three dimensional visualization of the momentum dependence of the reconstructed transversal gluon spectra at $\beta=1.90$ for (left) $T=150$MeV and (right) $T=234$MeV. }\label{Fig:SpectraTrans190}
\end{figure*}

\begin{figure*}
\includegraphics[scale=0.14]{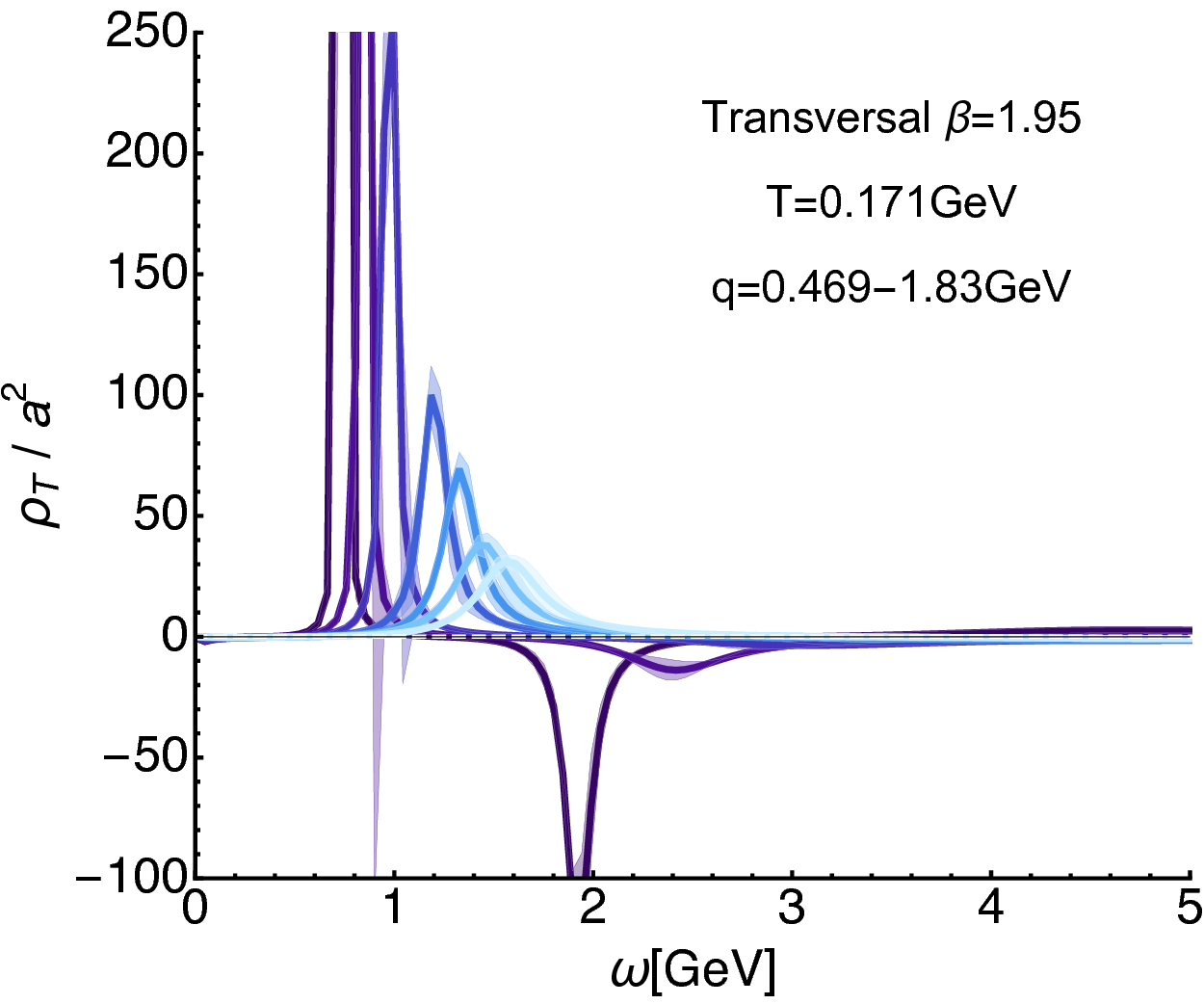}\hspace{0.5cm}\hspace{2cm}
\includegraphics[scale=0.14]{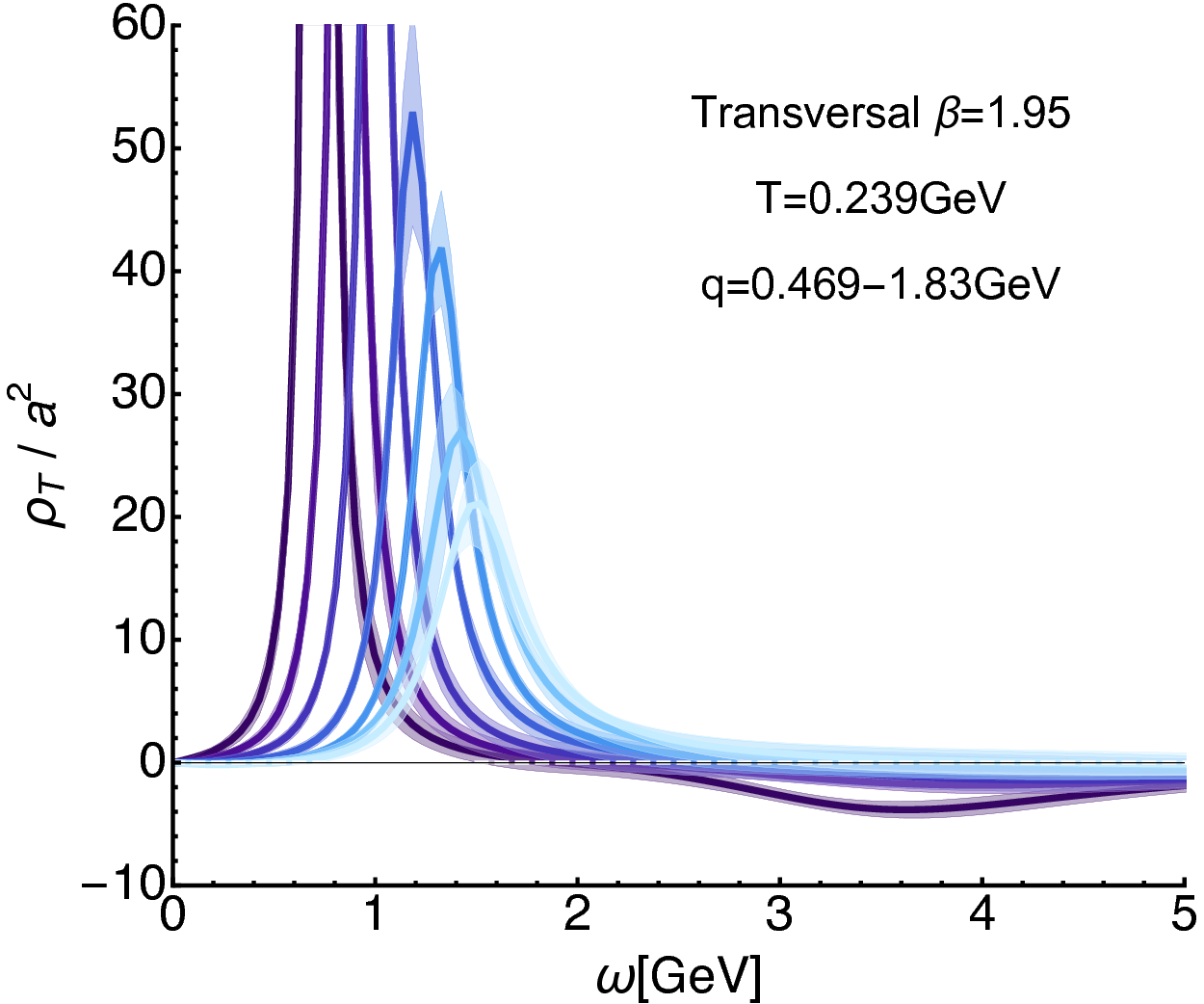}
\includegraphics[scale=0.16]{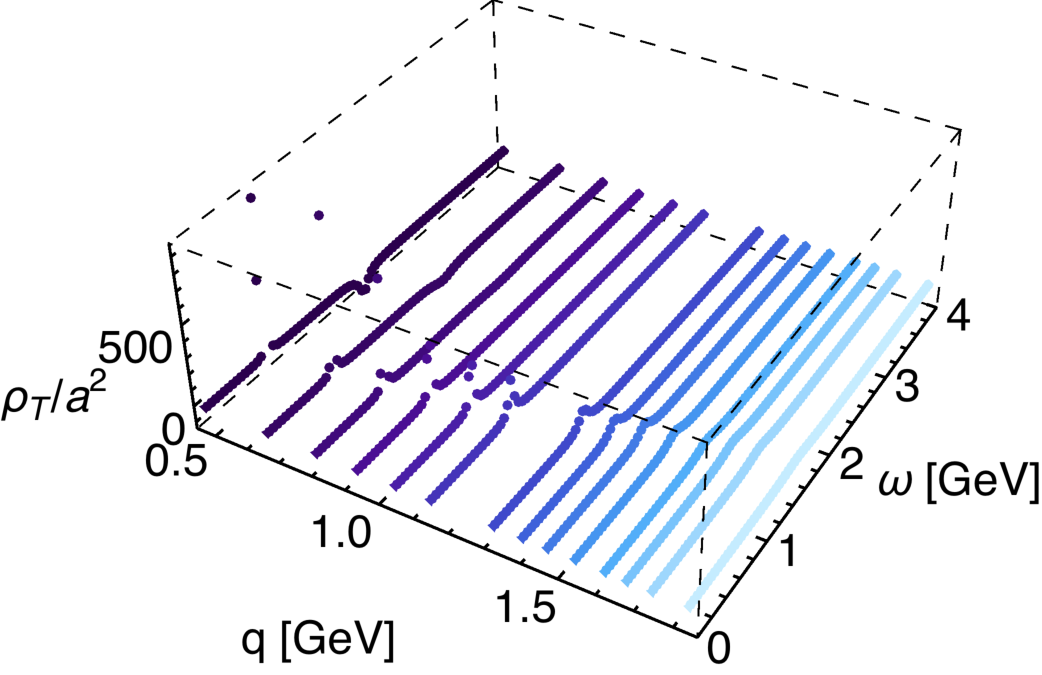}\hspace{0.5cm}\hspace{2cm}
\includegraphics[scale=0.16]{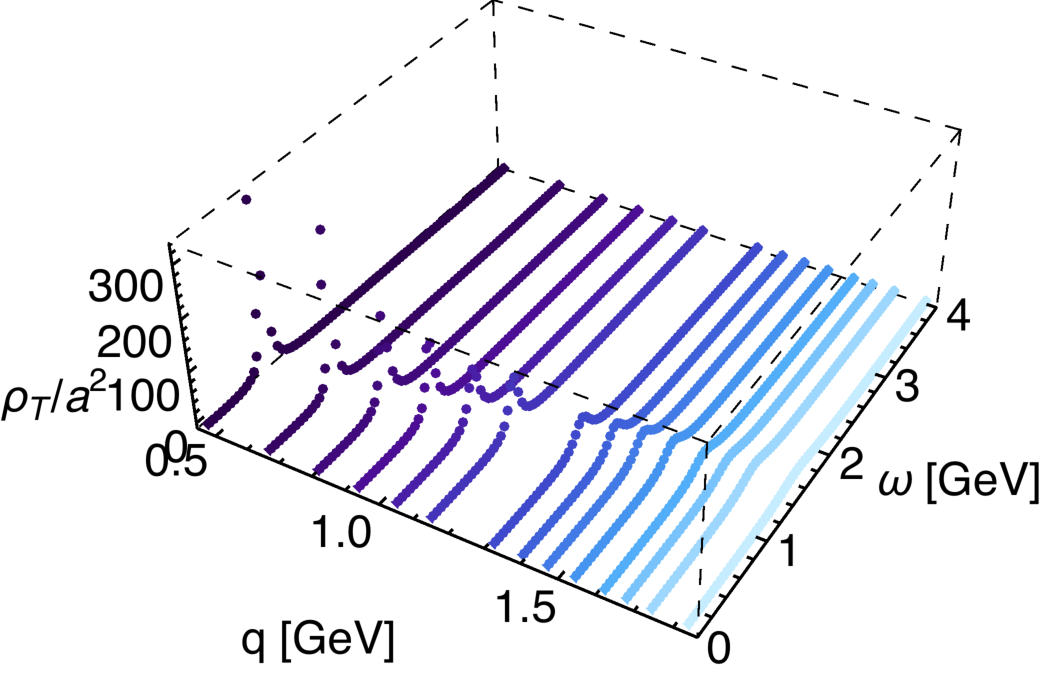}
\caption{(top) Reconstructed transversal gluon spectra at $\beta=1.95 $ for (left) $T=171$MeV and (right) $T=239$MeV. The seven different curves denote half of the fourteen lowest spatial momenta at which correlator data is available. (bottom) Three dimensional visualization of the momentum dependence of the reconstructed transversal gluon spectra on the $\beta=1.95$ ensembles for (left) $T=171$MeV and (right) $T=239$MeV. }
\end{figure*}

\begin{figure*}
\includegraphics[scale=0.17]{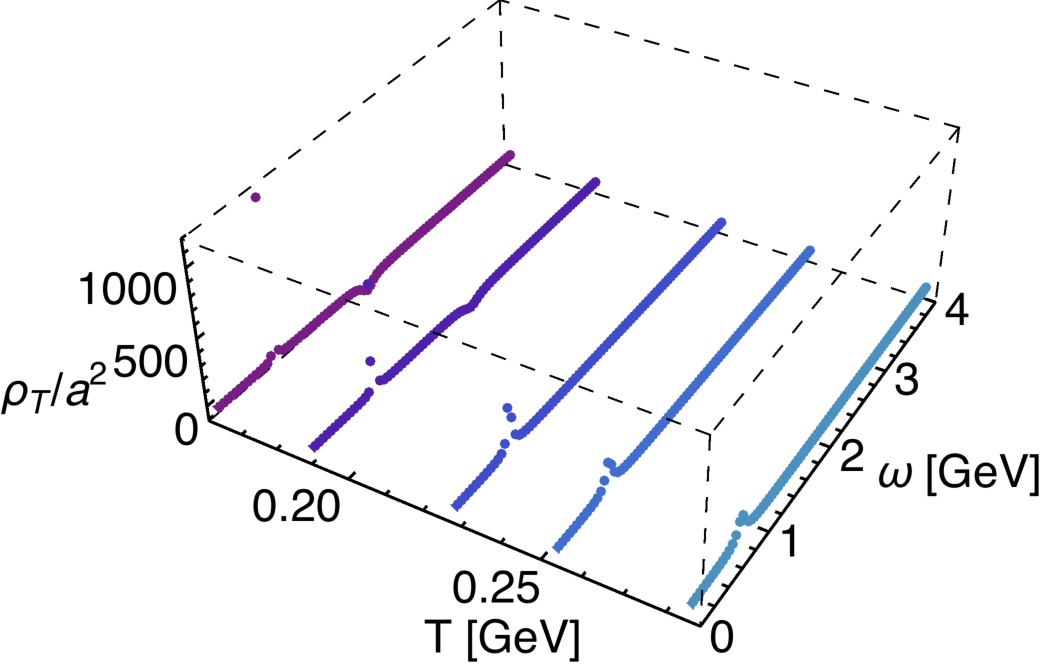}\hspace{0.5cm}\hspace{2cm}
\includegraphics[scale=0.17]{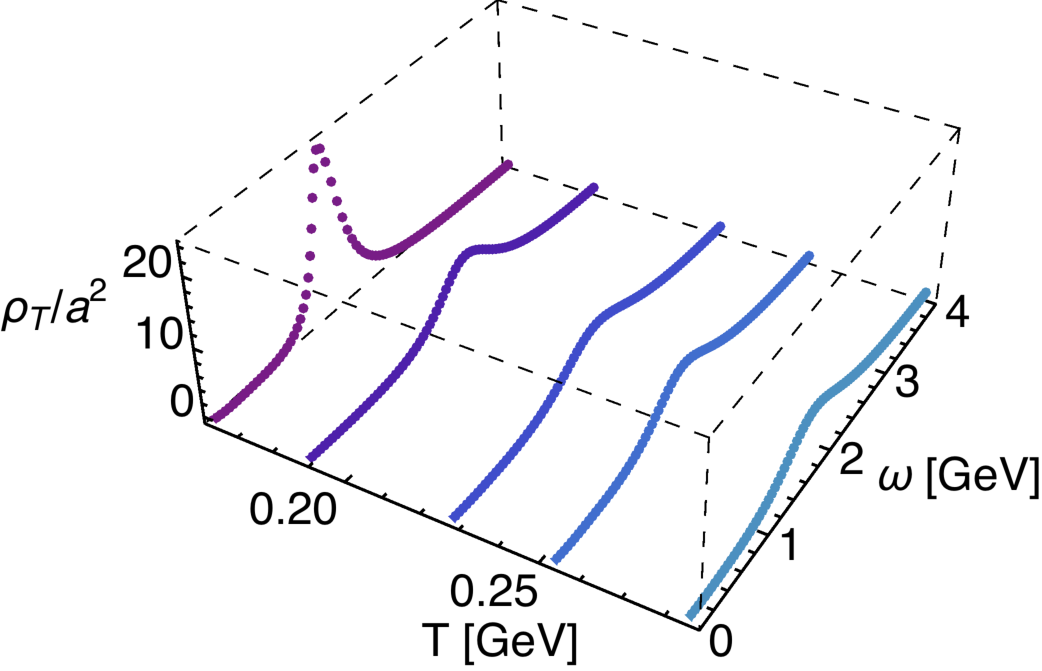}
\caption{Three dimensional visualization of the temperature dependence of the reconstructed transversal gluon spectra on the $\beta=1.90$ ensembles for (left) the lowest and (right) highest available spatial momentum. The five different curves denote the five lowest temperatures at which the reconstruction was possible.}
\end{figure*}

\begin{figure*}
\includegraphics[scale=0.17]{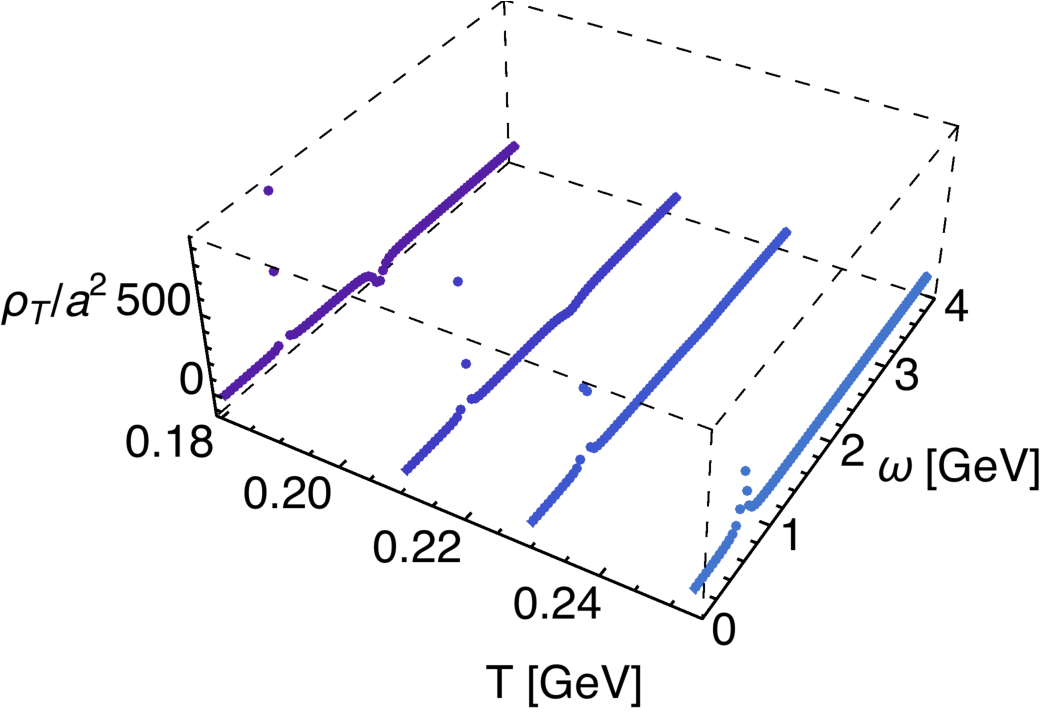}\hspace{0.5cm}\hspace{2cm}
\includegraphics[scale=0.17]{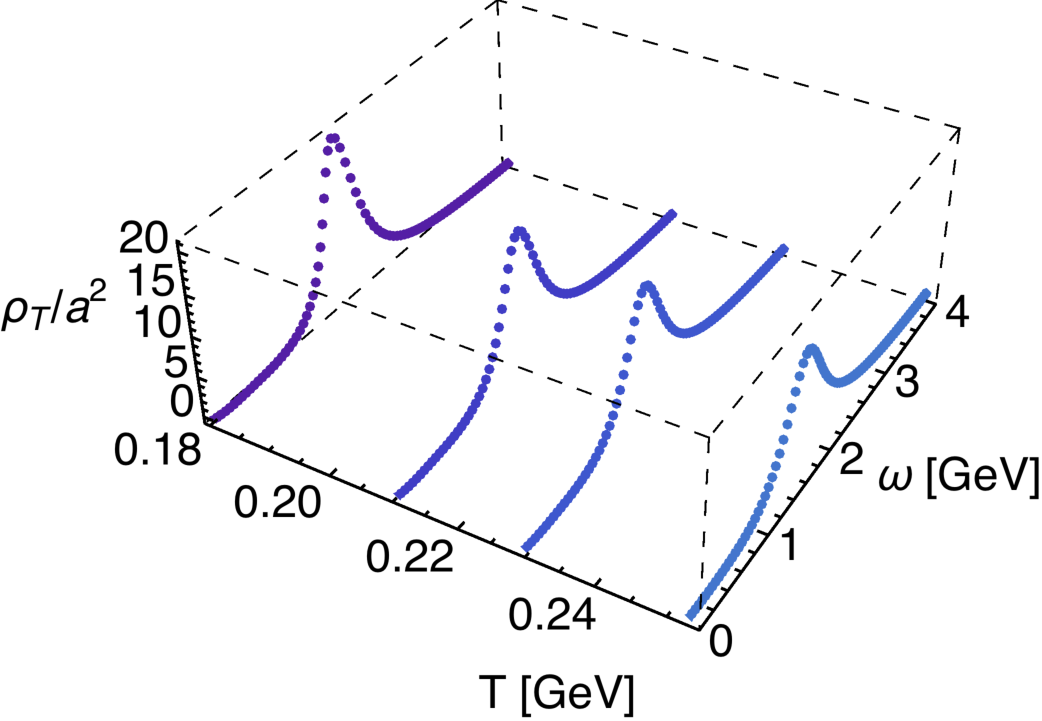}
\caption{Three dimensional visualization of the temperature dependence of the reconstructed transversal gluon spectra on the $\beta=1.95$ ensembles for (left) the lowest and (right) highest available spatial momentum. The four different curves denote the four lowest temperatures at which the reconstruction was possible.}
\end{figure*}

\begin{figure*}
\includegraphics[scale=0.21]{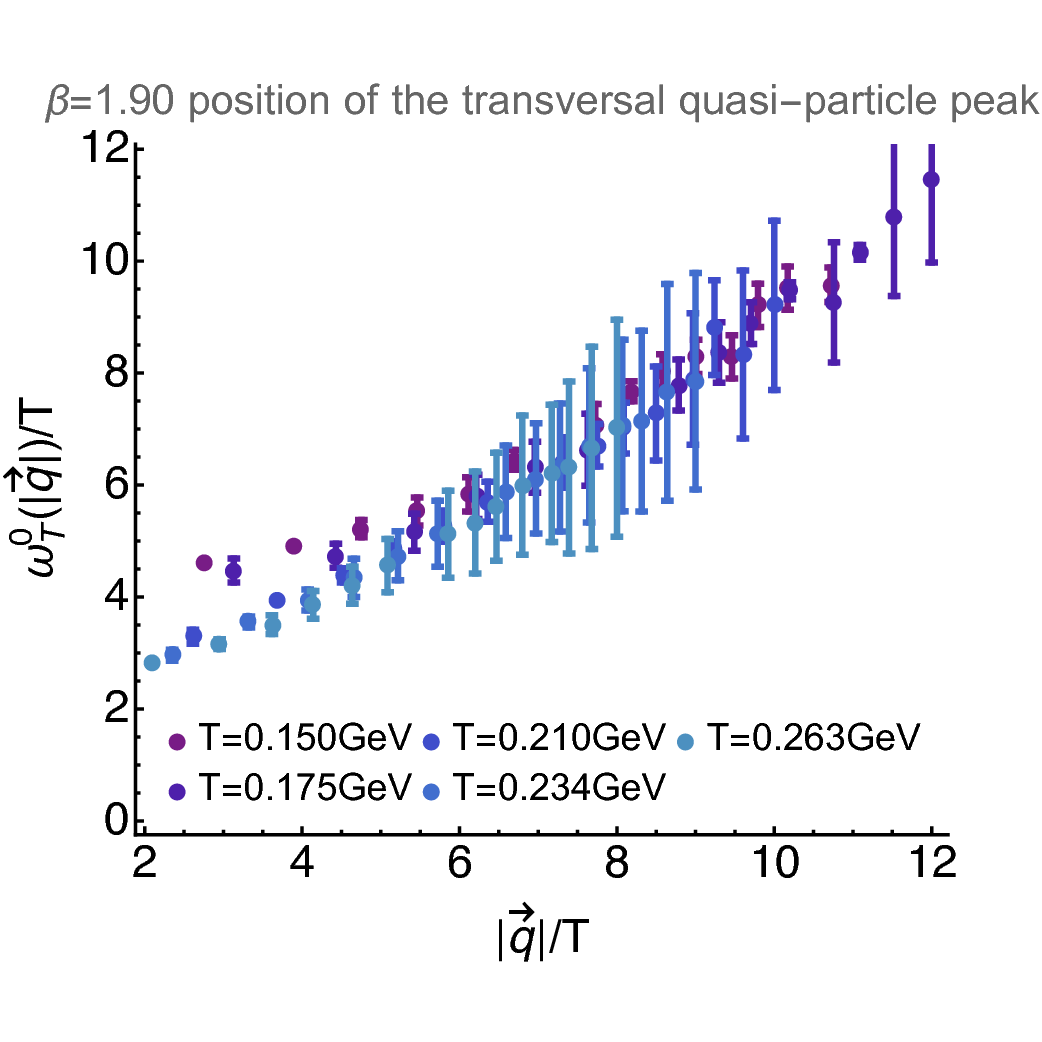}\hspace{0.1cm}\hspace{2cm}
\includegraphics[scale=0.28]{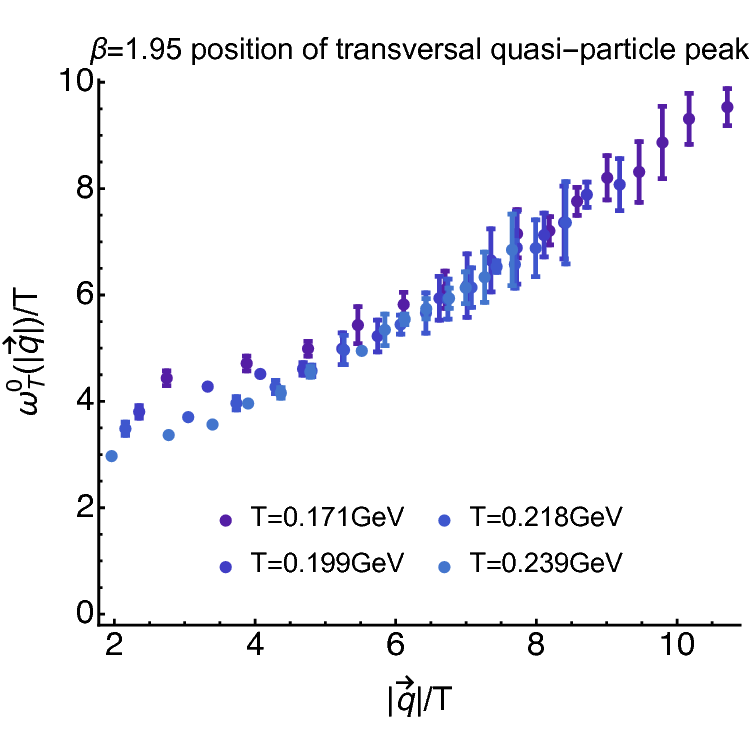}\hspace{0.1cm}
\caption{Momentum dependence of the transversal quasi-particle peak position at (left) $\beta=1.90$ and (right) $\beta=1.95$. The peak position for small momenta takes on a non-zero value. The reconstructions at the two lowest temperatures (within the hadronic phase) seem to exhibit a slightly larger intercept than the curves in the deconfined phase.} 
\end{figure*}

\FloatBarrier
\end{widetext}

\end{document}